\documentclass[apj]{emulateapj}

\usepackage{natbib}
\usepackage[bookmarks,breaklinks]{hyperref}
   \hypersetup{
     colorlinks,
     citecolor=blue,
     filecolor=magenta,
     linkcolor=blue,
     urlcolor=cyan
}
\usepackage{multirow}

\shorttitle{Single-epoch black hole mass beyond the local Universe}
\shortauthors{Karouzos et al.}

\begin{document}


\title{Rest-frame UV single-epoch black hole mass estimates of low-luminosity AGN at intermediate redshifts}


\author{Marios Karouzos$^{1}$, Jong-Hak Woo$^{1}$, Kenta Matsuoka$^{2}$, Christopher S. Kochanek$^{3}$, Christopher A. Onken$^{4}$, Juna A. Kollmeier$^{5}$, Dawoo Park$^{1}$, Tohru Nagao$^{6}$, Sang Chul Kim$^{7,8}$}
\affil{$^{1}$Astronomy Program, Department of Physics \& Astronomy, Seoul National University, Gwanak-gu, Seoul, Republic of Korea}
\affil{$^{2}$Department of Astronomy, Kyoto University, Oiwake-cho, Sakyo-ku, Kyoto 606-8502, Japan}
\affil{$^{3}$Department of Astronomy and the Center for Cosmology and Astroparticle Physics, Ohio State University, Columbus, Ohio, 43210, USA}
\affil{$^{4}$Research School of Astronomy and Astrophysics, Australian National University, Canberra, ACT 2611, Australia}
\affil{$^{5}$Carnegie Observatories, 813 Santa Barbara Street, Pasadena, California, 91101, USA}
\affil{$^{6}$Research Center for Space and Cosmic Evolution, Ehime University, Bunkyo-cho 2-5, Matsuyama, Ehime 790-8577, Japan}
\affil{$^{7}$Korea Astronomy and Space Science Institute (KASI), Daejeon 305-348, Republic Korea}
\affil{$^{8}$Korea University of Science and Technology (UST), Daejeon 305-350, Republic Korea}
\email{mkarouzos@astro.snu.ac.kr}




\begin{abstract}
The ability to accurately derive black hole (BH) masses at progressively higher redshifts and over a wide range of continuum luminosities has become indispensable in the era of large-area extragalactic spectroscopic surveys. In this paper we present an extension of existing comparisons between rest-frame UV and optical virial BH mass estimators to intermediate redshifts and luminosities comparable to the local H$\beta$ reverberation mapped active galactic nuclei (AGN). We focus on the MgII, CIV, and CIII] broad emission lines and compare them to both H$\alpha$ and H$\beta$. We use newly acquired near-infrared spectra from the FMOS instrument on the Subaru telescope for 89 broad-lined AGN at redshifts between 0.3 and 3.5, complemented by  data from the AGES survey. We employ two different prescriptions for measuring the emission line widths and compare the results. We confirm that MgII shows a tight correlation with H$\alpha$ and H$\beta$, with a scatter of $\sim0.25$ dex. The CIV and CIII] estimators, while showing larger scatter, are viable virial mass estimators after accounting for a trend with the UV-to-optical luminosity ratio. We find an intrinsic scatter of $\sim0.37$ dex between Balmer and carbon virial estimators by combining our dataset with previous high redshift measurements. This updated comparison spans a total of 3 decades in BH mass. We calculate a virial factor for CIV/CIII] $\log{f_{\mathrm{CIV/CIII]}}}=0.87$ {with an estimated systematic uncertainty of $\sim0.4$ dex} and find excellent agreement between the local reverberation mapped AGN sample and our high-z sample. 
\end{abstract}


\keywords{galaxies: active, quasars: emission lines, quasars: supermassive black holes }

\section{Introduction}
\label{sec:intro}

The correlation of central black hole (BH) masses with the properties of their host galaxies (\citealt{Ferrarese2000,Gebhardt2000, Kormendy2013, Woo2013}) implies a connection between galaxy evolution and BH growth, motivating numerous investigations on galaxy formation scenarios (e.g., \citealt{Kauffmann2000,Croton2006,Robertson2006,Ciotti2007, Angles2013, DeGraf2014}). In order to observationally constrain the cosmic evolution and BH growth history \citep[e.g.,][]{Woo2006,Woo2008, Peng2006, Jahnke2009, Merloni2010, Bennert2010, Canalizo2013,Schramm2013, Busch2014, Park2015}, it is crucial to obtain accurate BH mass estimates using consistently calibrated methods. 

Direct dynamical BH mass measurements based on spatially resolved kinematics are limited to the
relatively local Universe. However, the BH growth history has been probed using active galactic nuclei (AGN), where
BH masses can be estimated using the reverberation mapping technique (e.g., \citealt{Bahcall1972,Peterson1993}) 
or the empirical single-epoch methods  (e.g., \citealt{Kaspi2000, Woo2002,McLure2002, Kollmeier2006, Kelly2013}).

Traditionally, the bright Balmer lines (H$\beta$ and H$\alpha$) are used in mass estimates,
but the need for BH mass determination at higher redshifts led to the investigation of rest-frame UV lines (MgII and CIV) as virial mass estimators. 
Their calibration against the reverberation mapped AGN (e.g., \citealt{McLure2002,Vestergaard2006,Park2013,Feng2014}), 
gave rise to a whole industry of mass estimation (e.g., \citealt{Vestergaard2009,Shen2011}), allowing 
detailed studies of the BH mass function, cosmic BH accretion history, and galaxy evolution in general.

It is generally accepted that beyond the Balmer lines, the MgII emission line provides equally good, if not better, mass estimates (e.g., \citealt{Marziani2013}). The wavelength of MgII in the UV (2802\AA) makes it a natural choice as a mass estimator for intermediate redshifts (e.g., \citealt{McGill2008}). It has been shown that MgII emission should arise co-spatially with H$\beta$ (e.g., \citealt{McLure2002,Shen2008}, but also see \citealt{Wang2009}). Currently, the best calibrations give a scatter of $\sim0.2$ dex relative to mass estimates using H$\beta$ (e.g., \citealt{Wang2009,Shen2012}).

CIV and, to a lesser extent, CIII] lines provide alternative mass estimates that 
have been calibrated against H$\beta$ (e.g., \citealt{Vestergaard2006}, \citealt{Netzer2007b, Assef2011, Ho2012,Shen2012,Park2013,Zuo2015}). 
CIV often shows a complex emission line profile, (e.g., \citealt{Assef2011,Park2013}), with a broad wing (potentially due to winds), a confounding additional emission component at $\sim$1600\AA\, (e.g., \citealt{Fine2010}), and absorption features.
Thus, it is challenging to use CIV as a virial estimator for these cases. 
Nevertheless, recent studies have tried to account for this complexity with partial success (e.g, \citealt{Runnoe2013,Denney2012,Denney2013}). Only a handful of studies have investigated CIII] virial mass estimates (e.g., \citealt{Shen2012}) and in some cases CIII] has been used for reverberation mapping (e.g., \citealt{Peterson1999,Metzroth2006}). 

In this paper, we investigate the consistency of UV (MgII, CIV, CIII]) and optical (H$\alpha$, H$\beta$) virial mass estimators, by extending the AGN luminosity and mass range to be comparable to the local H$\beta$ reverberation sample \citep{Vestergaard2006}
based on new near-IR observations. Our results provide an invaluable comparison of high and low redshift virial mass estimates at comparable AGN luminosity ranges.

In Sections \ref{sec:sample} and \ref{sec:method} we present the sample, data, and methodology. Section \ref{sec:method} includes Monte-Carlo simulations to constrain the uncertainties in our analysis. In Section \ref{sec:results} we show the comparison of MgII and CIV/CIII] virial estimators to the H$\alpha$/H$\beta$ estimators. Sections \ref{sec:discuss} and \ref{sec:conc} provide discussion and conclusions. Throughout the paper, we assume the cosmological parameters $H_{0}=71$ km s$^{-1}$ Mpc$^{-1}$, $\Omega_{M}=0.27$, and $\Omega_{\Lambda}=0.73$ (\citealt{Komatsu2011}). 

\section{Sample and Observations}
\label{sec:sample}

\subsection{Sample selection}

{For this study, we selected relatively low luminosity AGN compared to the samples in previous studies. 
We used AGN from the AGN and Galaxy Evolution Survey (AGES; \citealt{Kochanek2012}), which has a lower flux limit
(m$_{I}<22.5$) than other wide-area surveys such as the Sloan Digital Sky Survey (SDSS). We selected 89 broad-line AGN at redshifts 0.3 $<$ z $<$ 3.5 with at least one of the broad UV lines of interest (MgII, CIV, and CIII]) detected in the optical spectra.}

\begin{figure}[tb]
\begin{center}
\includegraphics[width=0.48\textwidth,angle=0]{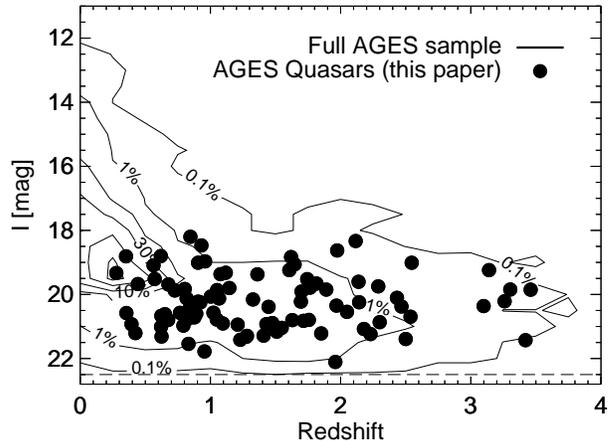}
\caption{Optical $I$-band magnitude and redshift distribution for the full AGES sample (\citealt{Kochanek2012}), shown with density contours at fractions of the maximum density (black curves), and for the sources in the sample presented in this paper (black circles). The nominal magnitude limit for AGN in the AGES sample is 22.5 and is shown with the dashed line.}
\label{fig:z}
\end{center}
\end{figure}

{We selected AGN in the AGES survey area by maximizing the number of AGN within the field of view of the FMOS instrument (see next sections for details). We consider this sample as a random subset of the full AGES quasar sample and therefore is representative of a complete flux-limited Type 1 quasar sample.}
{The redshift and $I$-band magnitude distribution of our sample is shown in Fig. \ref{fig:z} (also see Table \ref{tab:sample}), reaching down to optical luminosities of $3.1\times10^{43}$ erg s$^{-1}$. This is at least one order of magnitude deeper than previous single-epoch BH mass studies at similar redshifts (e.g., \citealt{Assef2011}) and comparable to reverberation mapped samples at low redshifts (e.g., \citealt{Vestergaard2006,Park2013,Woo2010,Woo2015}).}

\subsection{Near-infrared observations and data reduction}

We used the Fiber-fed Multi-Object Spectrograph (FMOS; \citealt{Kimura2010}) at the Subaru 8.2m telescope 
to obtain near-IR spectra (rest-frame optical spectra) of 89 high redshift and low luminosity Type 1 AGN.
FMOS consists of a fiber positioning system and two InfraRed Spectrographs (IRS1 and IRS2), with a fiber size of 1\farcs2 and a field of view of 30\arcmin\ diameter. 
The observations (Obs. ID: S10A-070) were performed over two nights (May 29-30, 2010) of bright time. In total three FMOS field configurations were used 
with total, on source, exposure times of 7200-9000 secs per configuration. 
The weather conditions during the two nights were good with mostly clear skies and seeing values between 0\farcs8 and 1\farcs0.
 
During our observing runs, we used 200 fibers with one detector IRS1 (the other detector IRS2 was not available) and the low-resolution mode covering wavelengths of 1.05-1.34$\mu$m (J band) and 1.43-1.77$\mu$m (H band) simultaneously. The spectral resolution is $\lambda/\Delta\lambda \sim 600$, corresponding to a velocity resolution (full width at half-maximum, FWHM) $\sim 500$ km s$^{-1}$. For optimal sky subtraction, we adopted a cross-beam switching (CBS) mode, which assigned two fibers offset by $60\arcsec$ to each target, where the paired fibers alternated between the sky and target spectra.

\begin{figure*}[tbp]
\begin{center}
\includegraphics[width=0.99\textwidth,angle=0]{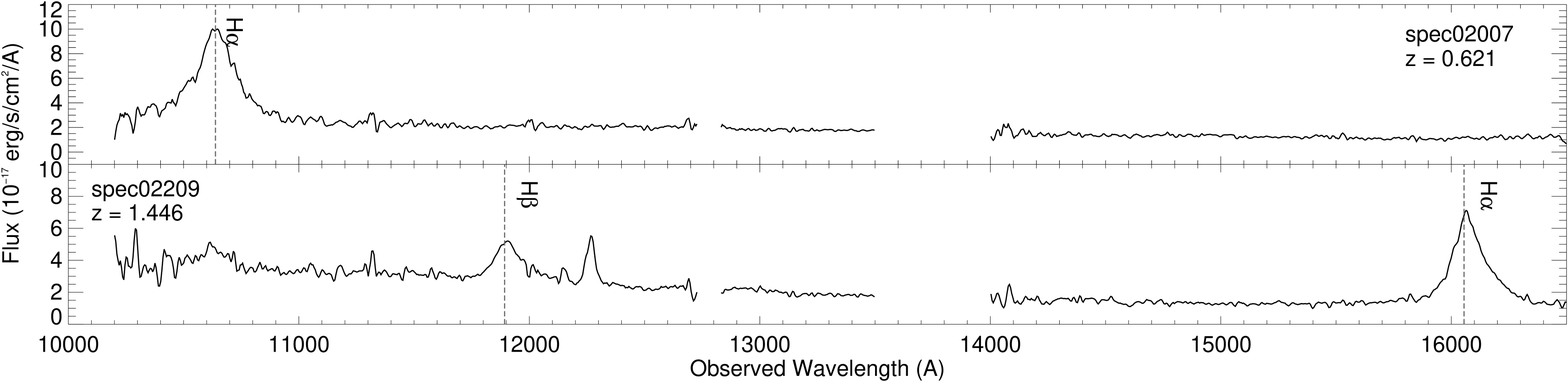}
\caption{Two examples of FMOS spectra for sources spec02007 and spec02209 at redshifts  0.621 and 1.446, respectively. Vertical dashed lines mark the emission lines of interest (H$\alpha$ and H$\beta$). The spectra have been smoothed with a 3 pixel boxcar filter for better visualization. The gap between the J and H bands is also visible. The small spectral gap around 12800\AA\, is masked due to high noise levels.}
\label{fig:FMOSspec}
\end{center}
\end{figure*}

We used the publicly available FMOS Image-based Reduction IRAF package software (\citealt{Iwamuro2012}) for the data reduction. Given the adopted CBS mode, sky subtraction was performed using the two different sky images. The difference in the bias across the four readout channels was corrected by making the average over each quadrant equal. The data were flat-fielded using dome flats. Bad pixels were masked throughout the reduction process. The distortion correction and the removal of residual airglow lines were done in additional steps. Individual images were combined into an average image and a noise image. We performed the wavelength calibration using the arc lines from a Th-Ar lamp. Flux calibration was performed using the spectra of bright stars obtained simultaneously with the science targets. Final one-dimensional science and error spectra were extracted for each fiber spectrum. In Fig. \ref{fig:FMOSspec} we show two examples of FMOS spectra {and all the FMOS spectra used in our analysis are shown in Appendix  \ref{sec:fmos_spec}}.

\LongTables
\begin{center}
\begin{deluxetable*}{c c c c c c c c c c c c c c c }
\tabletypesize{\footnotesize}
\tablecolumns{15}
\tablewidth{0pt}
\tablecaption{The FMOS-AGES sample of Type 1 quasars. \label{tab:sample}}
\tablehead{\colhead{ID} & \colhead{RA} & \colhead{DEC} & \colhead{z} & \colhead{m$_I$} & \colhead{$\log \lambda L_I$} & \colhead{Code06} & \colhead{SN$_{\mathrm{opt}}$} & \colhead{m$_{\mathrm{Ks}}$} & \colhead{$\log \lambda L_{\mathrm{Ks}}$} & \colhead{CIV} & \colhead{CIII]} & \colhead{MgII} & \colhead{H$\beta$} & \colhead{H$\alpha$}\\
\colhead{ } & \colhead{[hh:mm:ss]} & \colhead{[dd:mm:ss]} & \colhead{ } & \colhead{[AB]} & \colhead{[W]} & \colhead{ } & \colhead{ } & \colhead{[AB]} & \colhead{[W]} & \multicolumn{5}{c}{ }}
\startdata
spec00409 & 14:28:02.00 & +33:23:50.0 &    3.14 &   19.24 &   39.42 &           64 &      13 &   19.06 &   39.02& $\bullet$ & $\circ$ & $\circ$ & $\circ$ & $\circ$ \\ 
spec02247 & 14:36:27.36 & +35:41:16.1 &    2.12 &   18.33 &   39.36 &           64 &       7 &   18.68 &   38.76& $\bullet$ & $\bullet$ & $\circ$ & $\circ$ & $\circ$ \\ 
spec02019 & 14:38:13.85 & +35:23:41.8 &    2.55 &   19.01 &   39.29 &          112 &       8 &   18.64 &   38.97& $\bullet$ & $\bullet$ & $\circ$ & $\circ$ & $\circ$ \\ 
spec02091 & 14:38:14.56 & +35:29:26.5 &    3.46 &   19.85 &   39.27 &           16 &       6 &   19.30 &   39.03& $\bullet$ & $\circ$ & $\circ$ & $\circ$ & $\circ$ \\ 
spec01678 & 14:35:30.18 & +34:59:25.0 &    3.30 &   19.85 &   39.23 &           64 &      12 &   17.92 &   39.53& $\bullet$ & $\bullet$ & $\circ$ & $\circ$ & $\circ$ \\ 
spec00521 & 14:28:21.06 & +33:34:11.2 &    1.97 &   18.63 &   39.17 &          112 &      13 &   17.92 &   38.99& $\bullet$ & $\bullet$ & $\bullet$ & $\bullet$ & $\circ$ \\ 
spec01788 & 14:34:22.49 & +35:06:48.0 &    3.26 &   20.21 &   39.07 &           64 &      21 &   18.66 &   39.22& $\bullet$ & $\bullet$ & $\circ$ & $\circ$ & $\circ$ \\ 
spec02037 & 14:38:08.16 & +35:25:09.1 &    3.10 &   20.36 &   38.95 &          112 &       3 &   18.98 &   39.04& $\bullet$ & $\circ$ & $\circ$ & $\circ$ & $\circ$ \\ 
spec02153 & 14:38:12.45 & +35:33:36.6 &    2.29 &   19.75 &   38.88 &           96 &       6 &   18.84 &   38.78& $\bullet$ & $\bullet$ & $\circ$ & $\circ$ & $\circ$ \\ 
spec01980 & 14:37:17.38 & +35:20:57.3 &    1.62 &   18.83 &   38.87 &           80 &       2 &   18.24 &   38.64& $\bullet$ & $\bullet$ & $\circ$ & $\circ$ & $\circ$ \\ 
spec00716 & 14:27:04.74 & +33:48:36.7 &    2.14 &   19.60 &   38.86 &           96 &      15 &   18.07 &   39.01& $\bullet$ & $\bullet$ & $\circ$ & $\circ$ & $\circ$ \\ 
spec02170 & 14:37:58.70 & +35:34:46.5 &    2.43 &   20.11 &   38.80 &          112 &      10 &   18.02 &   39.17& $\bullet$ & $\bullet$ & $\circ$ & $\circ$ & $\circ$ \\ 
spec02251 & 14:37:12.95 & +35:41:22.6 &    1.65 &   19.06 &   38.80 &          112 &       2 &   18.48 &   38.57& $\circ$ & $\bullet$ & $\circ$ & $\circ$ & $\bullet$ \\
spec00600 & 14:27:30.56 & +33:40:37.3 &    2.47 &   20.38 &   38.71 &           64 &      35 &   18.86 &   38.85& $\bullet$ & $\bullet$ & $\circ$ & $\circ$ & $\circ$ \\ 
spec01812 & 14:34:57.45 & +35:08:29.4 &    1.61 &   19.24 &   38.70 &          116 &      17 &   17.26 &   39.03& $\bullet$ & $\bullet$ & $\bullet$ & $\bullet$ & $\bullet$ \\
spec02192 & 14:37:45.58 & +35:36:01.5 &    1.74 &   19.52 &   38.68 &           80 &      11 &   18.10 &   38.78& $\bullet$ & $\bullet$ & $\bullet$ & $\circ$ & $\circ$ \\ 
spec00427 & 14:27:18.33 & +33:25:32.3 &    1.81 &   19.66 &   38.66 &           80 &       3 &   18.52 &   38.66& $\bullet$ & $\bullet$ & $\bullet$ & $\circ$ & $\circ$ \\ 
spec00540 & 14:27:20.84 & +33:35:34.5 &    1.89 &   19.85 &   38.63 &           80 &      17 &   18.85 &   38.57& $\bullet$ & $\bullet$ & $\bullet$ & $\circ$ & $\circ$ \\ 
spec00598 & 14:27:29.18 & +33:40:33.4 &    3.42 &   21.43 &   38.63 &           32 &      33 &   18.75 &   39.24& $\bullet$ & $\bullet$ & $\circ$ & $\circ$ & $\circ$ \\ 
spec00732 & 14:27:14.30 & +33:49:33.5 &    2.14 &   20.24 &   38.61 &           80 &       3 &   18.36 &   38.90& $\circ$ & $\bullet$ & $\circ$ & $\circ$ & $\circ$ \\ 
spec02277 & 14:36:36.70 & +35:42:48.6 &    2.54 &   20.70 &   38.61 &           96 &       6 &   18.32 &   39.10& $\bullet$ & $\circ$ & $\circ$ & $\circ$ & $\circ$ \\ 
spec00724 & 14:27:37.38 & +33:49:04.5 &    1.77 &   19.79 &   38.58 &          112 &      12 &   18.18 &   38.76& $\bullet$ & $\bullet$ & $\bullet$ & $\circ$ & $\circ$ \\ 
spec01713 & 14:35:16.88 & +35:01:43.4 &    1.70 &   19.91 &   38.49 &          112 &       9 &   17.77 &   38.89& $\bullet$ & $\bullet$ & $\bullet$ & $\circ$ & $\circ$ \\ 
spec00571 & 14:28:06.39 & +33:38:23.6 &    1.97 &   20.35 &   38.48 &          112 &       4 &   18.99 &   38.56& $\bullet$ & $\bullet$ & $\circ$ & $\circ$ & $\circ$ \\ 
spec00513 & 14:27:03.64 & +33:33:45.0 &    1.36 &   19.37 &   38.47 &          112 &      13 &   16.38 &   39.20& $\circ$ & $\bullet$ & $\bullet$ & $\bullet$ & $\bullet$ \\
spec01557 & 14:34:11.47 & +34:51:48.1 &    2.05 &   20.54 &   38.44 &          116 &       4 &   18.14 &   38.94& $\bullet$ & $\bullet$ & $\circ$ & $\circ$ & $\circ$ \\ 
spec02016 & 14:38:05.14 & +35:23:28.6 &    2.30 &   20.86 &   38.44 &           96 &       2 &   18.60 &   38.88& $\circ$ & $\bullet$ & $\circ$ & $\bullet$ & $\circ$ \\ 
spec02230 & 14:36:42.16 & +35:39:28.8 &    0.85 &   18.20 &   38.42 &          112 &      20 &   16.49 &   38.64& $\circ$ & $\circ$ & $\bullet$ & $\circ$ & $\bullet$ \\
spec00646 & 14:28:06.83 & +33:43:37.0 &    0.93 &   18.47 &   38.42 &          112 &       5 &   17.48 &   38.35& $\circ$ & $\circ$ & $\bullet$ & $\circ$ & $\bullet$ \\
spec00752 & 14:27:31.53 & +33:51:08.5 &    1.70 &   20.22 &   38.37 &          112 &      20 &   17.49 &   38.99& $\bullet$ & $\bullet$ & $\bullet$ & $\circ$ & $\circ$ \\ 
spec00601 & 14:27:48.41 & +33:40:38.2 &    2.50 &   21.39 &   38.32 &           96 &      11 &   18.97 &   38.82& $\circ$ & $\bullet$ & $\circ$ & $\circ$ & $\circ$ \\ 
spec01599 & 14:35:51.53 & +34:54:37.8 &    2.18 &   21.08 &   38.29 &           80 &       7 &   18.76 &   38.76& $\bullet$ & $\bullet$ & $\circ$ & $\circ$ & $\circ$ \\ 
spec01647 & 14:35:07.56 & +34:57:24.2 &    1.12 &   19.32 &   38.28 &          116 &      11 &   16.67 &   38.87& $\circ$ & $\bullet$ & $\bullet$ & $\bullet$ & $\circ$ \\ 
spec01754 & 14:35:27.81 & +35:04:54.6 &    2.23 &   21.24 &   38.25 &           64 &       6 &   18.65 &   38.83& $\circ$ & $\bullet$ & $\circ$ & $\circ$ & $\circ$ \\ 
spec00688 & 14:27:10.62 & +33:46:38.4 &    0.96 &   18.97 &   38.25 &          112 &      17 &   17.50 &   38.37& $\circ$ & $\bullet$ & $\bullet$ & $\circ$ & $\bullet$ \\
spec00591 & 14:27:06.58 & +33:39:44.3 &    1.08 &   19.37 &   38.22 &           16 &       9 &   18.39 &   38.14& $\circ$ & $\bullet$ & $\bullet$ & $\circ$ & $\circ$ \\ 
spec01670 & 14:34:46.53 & +34:58:54.5 &    1.76 &   20.80 &   38.18 &           80 &      10 &   18.42 &   38.66& $\bullet$ & $\bullet$ & $\bullet$ & $\circ$ & $\circ$ \\ 
spec00523 & 14:28:37.79 & +33:34:14.8 &    0.91 &   19.01 &   38.17 &          112 &      14 &   18.20 &   38.03& $\circ$ & $\circ$ & $\bullet$ & $\circ$ & $\bullet$ \\
spec00467 & 14:27:41.22 & +33:29:37.6 &    1.72 &   20.82 &   38.14 &          112 &       8 &   18.10 &   38.76& $\bullet$ & $\bullet$ & $\bullet$ & $\circ$ & $\circ$ \\ 
spec02209 & 14:36:17.84 & +35:37:26.4 &    1.45 &   20.39 &   38.13 &          116 &      15 &   16.71 &   39.13& $\bullet$ & $\bullet$ & $\bullet$ & $\bullet$ & $\bullet$ \\
spec01581 & 14:34:11.18 & +34:53:09.0 &    1.33 &   20.16 &   38.13 &           80 &       9 &   17.66 &   38.66& $\circ$ & $\bullet$ & $\bullet$ & $\bullet$ & $\circ$ \\ 
spec01752 & 14:35:28.38 & +35:04:32.7 &    1.15 &   19.80 &   38.11 &           80 &      27 &   17.42 &   38.60& $\circ$ & $\bullet$ & $\bullet$ & $\circ$ & $\circ$ \\ 
spec00658 & 14:27:30.41 & +33:44:28.6 &    1.63 &   20.80 &   38.09 &           80 &       7 &   19.42 &   38.18& $\bullet$ & $\bullet$ & $\bullet$ & $\circ$ & $\circ$ \\ 
spec01805 & 14:34:50.81 & +35:07:56.7 &    1.85 &   21.22 &   38.06 &          100 &       3 &   18.28 &   38.77& $\bullet$ & $\bullet$ & $\circ$ & $\circ$ & $\circ$ \\ 
spec01745 & 14:35:20.17 & +35:04:13.3 &    1.05 &   19.86 &   37.99 &          112 &      15 &   17.49 &   38.48& $\circ$ & $\bullet$ & $\bullet$ & $\circ$ & $\circ$ \\ 
spec01652 & 14:34:14.53 & +34:57:43.6 &    1.48 &   20.89 &   37.95 &          112 &       7 &   18.04 &   38.62& $\bullet$ & $\bullet$ & $\bullet$ & $\circ$ & $\circ$ \\ 
spec01717 & 14:34:07.79 & +35:01:47.1 &    1.55 &   21.04 &   37.94 &           96 &       6 &   18.53 &   38.48& $\circ$ & $\bullet$ & $\bullet$ & $\circ$ & $\circ$ \\ 
spec02104 & 14:36:47.29 & +35:30:43.2 &    1.48 &   20.95 &   37.93 &           80 &       6 &   18.22 &   38.56& $\bullet$ & $\bullet$ & $\bullet$ & $\circ$ & $\bullet$ \\
spec00584 & 14:28:16.25 & +33:39:11.0 &    1.07 &   20.12 &   37.90 &           80 &      11 &   19.15 &   37.83& $\circ$ & $\bullet$ & $\bullet$ & $\circ$ & $\circ$ \\ 
spec00528 & 14:27:57.89 & +33:34:46.4 &    1.43 &   20.93 &   37.90 &          112 &       5 &   18.48 &   38.41& $\bullet$ & $\bullet$ & $\bullet$ & $\circ$ & $\circ$ \\ 
spec01555 & 14:33:44.04 & +34:51:43.3 &    1.51 &   21.16 &   37.87 &          112 &       2 &   18.42 &   38.50& $\circ$ & $\bullet$ & $\bullet$ & $\circ$ & $\circ$ \\ 
spec02185 & 14:37:48.10 & +35:35:31.6 &    1.00 &   20.07 &   37.86 &           48 &       5 &   17.38 &   38.47& $\circ$ & $\bullet$ & $\bullet$ & $\circ$ & $\bullet$ \\
spec02099 & 14:37:32.83 & +35:30:18.1 &    0.62 &   18.80 &   37.85 &       265856 &      16 &   16.58 &   38.27& $\circ$ & $\circ$ & $\circ$ & $\circ$ & $\bullet$ \\
spec00588 & 14:28:05.04 & +33:39:36.1 &    1.96 &   22.11 &   37.77 &          112 &      23 &   18.48 &   38.75& $\bullet$ & $\bullet$ & $\bullet$ & $\circ$ & $\circ$ \\ 
spec00739 & 14:28:09.12 & +33:50:12.0 &    1.41 &   21.29 &   37.74 &           80 &       4 &   18.04 &   38.57& $\circ$ & $\bullet$ & $\bullet$ & $\circ$ & $\circ$ \\ 
spec02044 & 14:38:01.13 & +35:25:34.2 &    0.80 &   19.83 &   37.71 &           80 &       1 &   18.66 &   37.72& $\circ$ & $\circ$ & $\bullet$ & $\circ$ & $\bullet$ \\
spec01637 & 14:34:53.77 & +34:56:38.4 &    1.21 &   20.94 &   37.71 &          116 &       6 &   17.97 &   38.44& $\circ$ & $\bullet$ & $\bullet$ & $\circ$ & $\circ$ \\ 
spec01529 & 14:35:34.44 & +34:49:07.3 &    0.92 &   20.25 &   37.70 &          112 &       6 &   17.05 &   38.51& $\circ$ & $\circ$ & $\bullet$ & $\circ$ & $\bullet$ \\
spec01971 & 14:37:30.13 & +35:20:15.8 &    0.90 &   20.22 &   37.69 &           96 &       3 &   17.73 &   38.22& $\circ$ & $\circ$ & $\bullet$ & $\circ$ & $\bullet$ \\
spec02047 & 14:36:41.30 & +35:25:37.0 &    1.02 &   20.57 &   37.68 &          112 &       6 &   17.61 &   38.40& $\circ$ & $\circ$ & $\bullet$ & $\circ$ & $\bullet$ \\
spec01723 & 14:34:30.49 & +35:02:10.7 &    1.28 &   21.29 &   37.63 &           80 &       7 &   17.82 &   38.56& $\circ$ & $\bullet$ & $\bullet$ & $\circ$ & $\circ$ \\ 
spec02138 & 14:37:52.92 & +35:32:51.5 &    0.56 &   19.09 &   37.63 &       266112 &       4 &   16.72 &   38.11& $\circ$ & $\circ$ & $\circ$ & $\circ$ & $\circ$ \\ 
spec01530 & 14:35:01.02 & +34:49:09.3 &    1.05 &   20.78 &   37.62 &           16 &       4 &   17.76 &   38.36& $\circ$ & $\circ$ & $\bullet$ & $\circ$ & $\bullet$ \\
spec00674 & 14:26:48.15 & +33:45:47.0 &    1.10 &   20.91 &   37.62 &          112 &      16 &   17.34 &   38.58& $\circ$ & $\bullet$ & $\bullet$ & $\bullet$ & $\circ$ \\ 
spec01597 & 14:35:02.29 & +34:54:31.7 &    1.28 &   21.32 &   37.62 &           64 &       2 &   18.12 &   38.43& $\circ$ & $\bullet$ & $\bullet$ & $\circ$ & $\circ$ \\ 
spec01430 & 14:34:33.00 & +34:42:35.6 &    0.82 &   20.12 &   37.62 &          116 &      10 &   17.13 &   38.35& $\circ$ & $\circ$ & $\bullet$ & $\circ$ & $\bullet$ \\
spec00679 & 14:27:58.86 & +33:45:19.3 &    0.68 &   19.69 &   37.59 &           64 &      11 &   99.00 &    5.40& $\circ$ & $\circ$ & $\bullet$ & $\circ$ & $\circ$ \\ 
spec01519 & 14:34:41.33 & +34:48:30.4 &    0.73 &   19.88 &   37.59 &          112 &       9 &   16.91 &   38.31& $\circ$ & $\circ$ & $\bullet$ & $\circ$ & $\bullet$ \\
spec02026 & 14:38:12.64 & +35:24:10.0 &    1.23 &   21.42 &   37.54 &          112 &       4 &   17.56 &   38.62& $\circ$ & $\bullet$ & $\bullet$ & $\circ$ & $\bullet$ \\
spec02142 & 14:36:15.42 & +35:33:00.0 &    0.89 &   20.60 &   37.52 &          100 &       8 &   17.40 &   38.33& $\circ$ & $\circ$ & $\bullet$ & $\circ$ & $\bullet$ \\
spec00533 & 14:28:42.73 & +33:35:09.0 &    0.84 &   20.44 &   37.52 &          112 &      14 &   17.89 &   38.07& $\circ$ & $\circ$ & $\bullet$ & $\circ$ & $\bullet$ \\
spec01716 & 14:35:04.83 & +35:01:44.7 &    0.57 &   19.52 &   37.48 &       262160 &       5 &   16.67 &   38.15& $\circ$ & $\circ$ & $\bullet$ & $\circ$ & $\circ$ \\ 
spec01501 & 14:33:53.39 & +34:47:18.2 &    0.88 &   20.72 &   37.46 &           80 &       3 &   18.05 &   38.05& $\circ$ & $\circ$ & $\bullet$ & $\circ$ & $\circ$ \\ 
spec01634 & 14:33:58.26 & +34:56:21.5 &    0.83 &   20.76 &   37.37 &          112 &       6 &   17.81 &   38.09& $\circ$ & $\circ$ & $\bullet$ & $\circ$ & $\bullet$ \\
spec02205 & 14:36:24.33 & +35:37:09.6 &    0.77 &   20.58 &   37.37 &          116 &      20 &   16.30 &   38.61& $\circ$ & $\circ$ & $\bullet$ & $\circ$ & $\bullet$ \\
spec01562 & 14:35:38.70 & +34:51:54.8 &    0.35 &   18.81 &   37.25 &          100 &       4 &   16.28 &   37.80& $\circ$ & $\circ$ & $\circ$ & $\circ$ & $\circ$ \\ 
spec00577 & 14:27:12.24 & +33:38:45.4 &    0.79 &   20.97 &   37.24 &           48 &       5 &   17.68 &   38.09& $\circ$ & $\circ$ & $\bullet$ & $\circ$ & $\bullet$ \\
spec01731 & 14:35:17.82 & +35:02:53.0 &    0.65 &   20.62 &   37.17 &           48 &       6 &   17.60 &   37.91& $\circ$ & $\circ$ & $\bullet$ & $\circ$ & $\circ$ \\ 
spec01651 & 14:34:34.18 & +34:57:42.1 &    0.44 &   19.68 &   37.14 &          112 &       6 &   16.90 &   37.79& $\circ$ & $\circ$ & $\circ$ & $\circ$ & $\circ$ \\ 
spec01547 & 14:33:57.31 & +34:50:58.5 &    0.68 &   20.80 &   37.14 &           16 &       2 &   17.64 &   37.94& $\circ$ & $\circ$ & $\bullet$ & $\circ$ & $\circ$ \\ 
spec00745 & 14:27:31.02 & +33:50:28.7 &    0.95 &   21.78 &   37.12 &          112 &       6 &   17.97 &   38.18& $\circ$ & $\circ$ & $\bullet$ & $\circ$ & $\circ$ \\ 
spec00424 & 14:26:40.12 & +33:25:07.6 &    0.62 &   20.67 &   37.11 &       262144 &      10 &   17.15 &   38.05& $\circ$ & $\circ$ & $\bullet$ & $\circ$ & $\circ$ \\ 
spec00547 & 14:26:46.77 & +33:36:00.0 &    0.83 &   21.54 &   37.07 &           96 &       3 &   16.77 &   38.51& $\circ$ & $\circ$ & $\circ$ & $\circ$ & $\bullet$ \\
spec02007 & 14:37:14.69 & +35:22:54.7 &    0.62 &   21.01 &   36.97 &          112 &      13 &   16.14 &   38.45& $\circ$ & $\circ$ & $\bullet$ & $\circ$ & $\bullet$ \\
spec01680 & 14:35:54.69 & +34:59:29.8 &    0.62 &   21.31 &   36.85 &           16 &       6 &   18.77 &   37.40& $\circ$ & $\circ$ & $\bullet$ & $\circ$ & $\circ$ \\ 
spec01729 & 14:34:24.65 & +35:02:42.0 &    0.28 &   19.33 &   36.80 &          112 &       7 &   17.14 &   37.21& $\circ$ & $\circ$ & $\circ$ & $\circ$ & $\circ$ \\ 
spec00602 & 14:28:28.57 & +33:40:51.0 &    0.35 &   20.58 &   36.55 &       782272 &      12 &   15.94 &   37.94& $\circ$ & $\circ$ & $\bullet$ & $\circ$ & $\circ$ \\ 
spec02080 & 14:37:45.00 & +35:28:24.0 &    0.39 &   20.93 &   36.52 &       267984 &       5 &   16.07 &   38.00& $\circ$ & $\circ$ & $\circ$ & $\circ$ & $\circ$ \\ 
spec02171 & 14:37:17.80 & +35:34:48.1 &    0.42 &   21.21 &   36.49 &          112 &      11 &   16.63 &   37.85& $\circ$ & $\circ$ & $\bullet$ & $\circ$ & $\circ$ \\ 
\enddata
\tablecomments{The sample of Type 1 quasars used in this Paper. We give the FMOS ID (Column 1), the coordinates (Columns 2 and 3), the redshift derived from the AGES survey (Column 4), {the $I$-band magnitude (Column 5) and monochromatic luminosity calculated at the effective wavelength 7467\AA\ } (Column 6), the selection code for the AGES survey (Column 7, see \citealt{Kochanek2012}), the mean signal-to-noise ratio for the AGES spectrum (Column 8), {the $K_{s}$-band magnitude (Column 9) and monochromatic luminosity calculated at the effective wavelength 21900\AA\ } (Column 10), and information about the detection of emission for the 5 broad emission lines studied here (detection:filled circle, non-detection:open circle).}
\end{deluxetable*}
\end{center}

\subsection{Optical data from the AGES survey}
AGES is a redshift survey in the Bo\"otes field (part of the NOAO Deep Wide-Field Survey; \citealt{Jannuzi1999}) that observed a total of $\sim$24,000 redshifts to a limiting magnitude of $I<20$ mag {for galaxies} and $<22.5$ for AGN, probing AGN luminosities $\sim10$ times fainter than SDSS. The optical spectroscopy (\citealt{Kochanek2012}) was acquired using the Hectospec instrument (\citealt{Fabricant1998,Fabricant2005,Roll1998}) at the 6.5m MMT telescope. The spectral coverage was 3700\AA-9200\AA\, with a spectral resolution of 6\AA\, (R$\sim$1000). 

\section{Methodology and analysis}
\label{sec:method}

Different methods have been used to fit AGN broad emission line profiles, ranging from fitting of single or multiple Gaussians, to more complicated profiles like Gauss-Hermite expansions (e.g., \citealt{vanderMarel1993, Cappellari2002,Woo2006}). The optimal model profile depends on the intrinsic line profile, the spectral resolution, and the S/N of the emission line. In the following, we describe the fitting method for each individual line used in our subsequent analysis.

\subsection{CIV $\lambda$ 1548, 1551\AA}
\begin{figure}[tb]
\begin{center}
\includegraphics[width=0.48\textwidth,angle=0]{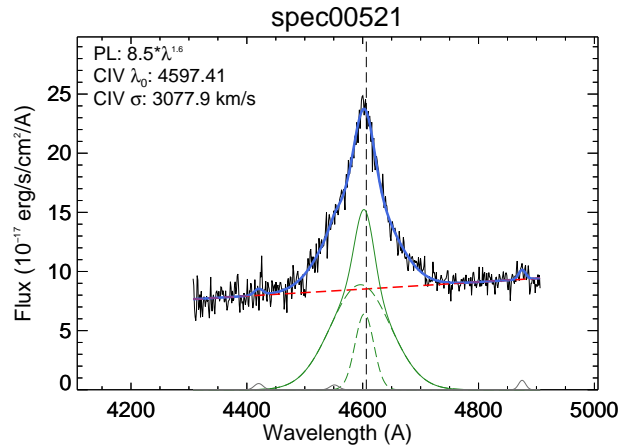}
\caption{Example of a fit to the CIV emission complex. The black line shows the AGES data while the blue line shows the total fit. The dashed vertical line shows the redshifted laboratory value of CIV. {The full} fit to all the narrow emission lines and the broad $\lambda$ 1600\AA\, feature (gray lines), together with the continuum (red dashed line), and the CIV emission line (green lines){, are also shown}. For CIV, the individual Gaussian components are plotted (green dashed lines). This spectrum, spec00521, has a visual quality flag of A.}
\label{fig:CIV}
\end{center}
\end{figure}

We simultaneously fit the continuum and the emission lines. For all the fits described here and in the following sections we employ the Levenberg-Marquardt least-squares algorithm (\citealt{Marquardt1963,More1978}), as implemented in the IDL procedure \textit{MPFIT} (\citealt{Markwardt2009}). While the continuum is fitted with a power law, we employ a combination of Gaussian profiles for the emission lines. We use a single Gaussian to fit the narrow emission lines (NIV], Si II, HeII, OIII]), and the broad feature at $\lambda$ 1600\AA, which we consider to be physically distinct from the CIV emission. We use a combination of two Gaussian profiles to fit the CIV emission\footnote{We do not fix each Gaussian to one of the components of the CIV doublet but instead allow the two Gaussians to vary within the limits discussed.}. We do not consider a separate additional narrow CIV component. This is for consistency with previous studies (e.g., \citealt{Assef2011, Park2013}). All emission line centers are allowed to shift within a range defined by the spectral resolution of the AGES data (R$\sim$1000, translating to $\sim$300 km s$^{-1}$). We do not subtract any Fe II mission, as it is considered relatively weak compared to the CIV flux and it is difficult to constrain given the data quality. In Fig. \ref{fig:CIV}, we show an example of a fit to the CIV line complex.

Out of 89 AGN in the sample, 34 objects have CIV in their optical spectra. The spectra of 11 AGN are too noisy for the full analysis. For these AGN we only fit the CIV emission line, either with a double Gaussian profile (9 sources) or a 4th order Gauss-Hermite profile (2 sources). Instead of physically interpreting the line profiles, we simply recover the best possible estimate of the width of the CIV line.

For each CIV fit (and all lines in the following) we calculate the first moment ($\lambda_{0}$), which represents the flux-weighted center, and the second moment ($\sigma_{\mathrm{CIV}}$) of the best-fit line profile, which represents the flux-weighted dispersion of the line:
\begin{equation}
\label{ math:smom}
\mathrm{\sigma}=\frac{\int\lambda^{2} f(\lambda)d\lambda}{\int f(\lambda)d\lambda}-\mathrm{\lambda_{0}}^{2}.
\end{equation}
We also calculate the FWHM of the best-fit profile. {Both quantities are corrected for the instrumental resolution.} The $\sigma$ width is more sensitive to any 
asymmetric deviations of the emission profile than the FWHM. In particular, $\sigma$ for carbon lines are sensitive to broad wings that can contain a substantial fraction of the total emission line flux (e.g., \citealt{Baskin2005,Nagao2006,Denney2009,Sluse2011,Denney2012,Marziani2012}). This also makes $\sigma$ more sensitive to bad or uncertain continuum fits in low S/N spectra. 

{Based on visual inspection, a quality flag is assigned to each fit, that ranges from A for the best fits to C for the poorest fits. These quality flags do not reflect the actual noise of the data but rather correspond to how well the best-fit model describes the emission line profile.} A visual quality flag F is assigned to sources for which the fit fails completely (usually due to extremely noisy data).
Of the 34 sources with CIV in their AGES spectra ($\sim40\%$), 23, 10, and one have fits with visual quality flags of A, B, and C, respectively. The FWHM of CIV for the 34 sources ranges from 826 to 10460 km s$^{-1}$. Similarly, the range of $\sigma_{\mathrm{CIV}}$ is from 440 to 7100 km s$^{-1}$.

\subsection{CIII] $\lambda$1907, 1909\AA}
\label{sec:ciii_method}
The blue side of the CIII] emission line is affected by the AlIII $\lambda$1857\AA\,  and SiIII $\lambda$1892\AA\, narrow emission lines. We perform a simultaneous fit to the CIII] complex with a total of four Gaussian components and a power-law continuum. The CIII] emission line itself is fitted using a double Gaussian profile.

\begin{figure}[tb]
\begin{center}
\includegraphics[width=0.48\textwidth,angle=0]{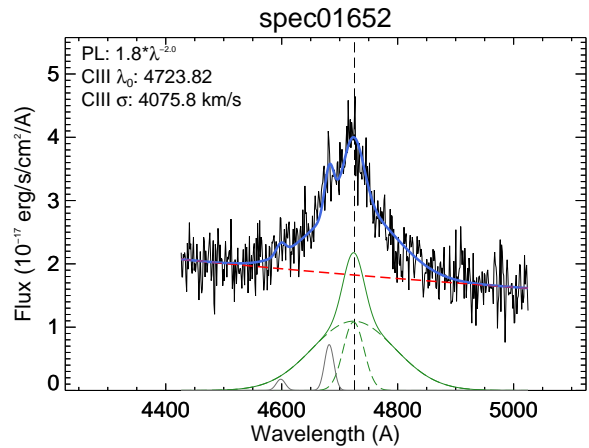}
\caption{Example of a fit to the CIII] emission complex. The black line shows the AGES data while the blue line shows the total fit. The dashed vertical line shows the redshifted laboratory value of CIII]. {The full} fit to the narrow AlIII and SiIII emission lines (gray lines), together with the continuum (red dashed line), and the CIII] emission line (green lines), {are shown}. For CIII], the individual Gaussian components are also plotted (green dashed lines). The spectrum, spec01652, has a visual quality flag A.}
\label{fig:CIII}
\end{center}
\end{figure}

Of the 89 sources, a total of 52 AGN have the CIII] emission complex in their optical spectra. For 11 sources, low S/N prevents us from performing a full fit. In these cases we only fit the CIII] line using either a double Gaussian or a 4th order Gauss-Hermite profile. In Fig. \ref{fig:CIII} we show an example of a CIII] fit.
Of the 52 sources with successful fits to the CIII] emission, 12, 25, and 15 are given a flag of A, B, and C, respectively. The FWHM of CIII] ranges from 1281 to 12000 km s$^{-1}$, while $\sigma_{\mathrm{CIII]}}$ ranges from 1100 to 15000 km s$^{-1}$. In total, 30 sources have fits for both the CIV and CIII] lines.

\subsection{MgII $\lambda$2796, 2803\AA}

For the MgII line we perform a simultaneous multi-component fit that includes the continuum (a power law), the FeII emission, and the MgII emission line. For the FeII emission, we use the FeII UV template from \citet{Tsuzuki2006}, which includes a careful treatment of the FeII emission at the edge and within the MgII emission region. Using the FeII emission template, we create a library of templates convolved with Gaussian profiles of varying velocity dispersions (500 to 6000 km s$^{-1}$, see also \citealt{McGill2008}). Finally, the MgII emission line is fit with a 4th order Gauss-Hermite profile. Through $\chi^{2}$ minimization, we find the best combination of the continuum, FeII pseudo-continuum, and MgII emission line.

\begin{figure}[tbp]
\begin{center}
\includegraphics[width=0.48\textwidth,angle=0]{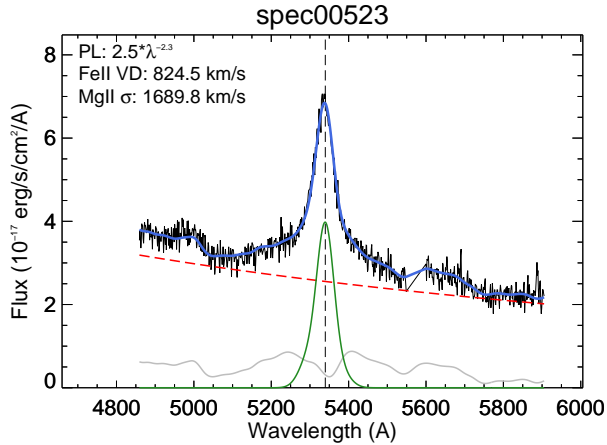}
\caption{Example of a fit to the MgII emission line. The black line shows the AGES data while the blue line shows the total fit. The dashed vertical line shows the redshifted laboratory value of MgII. {The simultaneous} fit to the continuum (power law; red dashed line), the FeII emission (from \citealt{Tsuzuki2006}; gray solid line), and the MgII emission line (Gauss-Hermite; green solid line) {are shown}. The continuum normalisation and slope, the FeII Gauss-convolved template velocity dispersion, and the $\sigma_{\mathrm{MgII}}$ in km/s {are also given}. The spectrum, spec00523, has a visual quality flag of A.}
\label{fig:MgII}
\end{center}
\end{figure}

Of 89 sources, a total of 59 had MgII lines that could be fitted ($\sim67\%$). Of these, 35 ,13, and 11 are assigned visual quality flags A, B, and C, respectively. For 20 sources with noisy spectra and/or weak MgII emission, a simple single Gaussian fit to the MgII emission line was performed\footnote{Of these 20, there were 8, 4, and 8 sources assigned visual quality flags A, B, and C, respectively.}. In Fig. \ref{fig:MgII} we show an example of a fit to a MgII line.
The FWHM of MgII for the 59 sources ranges from 750 to 12300 km s$^{-1}$ while the $\sigma_{\mathrm{MgII}}$ ranges from 700 to 9100 km s$^{-1}$.

\subsection{H$\beta$ $\lambda$4861\AA}

We fit the H$\beta$ emission line with a single Gaussian profile after subtracting a continuum of the form $\alpha$+$\beta\lambda$, using featureless continuum windows bracketing the line. Only 8 FMOS spectra have successful H$\beta$ measurements. In addition to the usual line properties, we calculate the flux of the H$\beta$ line as an alternative to the continuum luminosity. Using the same visual quality flags{, we assign a quality flag B to 5 sources, and 3 sources are assigned a quality flag C}. The FWHM of H$\beta$ for the 8 sources covers a range from 1200 to 3600 km s$^{-1}$, and $\sigma_{H\beta}$ ranges from 550 to 1500 km s$^{-1}$.\\

\subsection{H$\alpha$ $\lambda$6563\AA}
\label{sec:Ha}

We fit the H$\alpha$ broad emission line with either a 4th order Gauss-Hermite or a double-Gaussian profile, after continuum subtraction. As for H$\beta$, the continuum is assumed to be a linear function of wavelength and is fitted within featureless windows bracketing H$\alpha$. 

\begin{figure}[tbp]
\begin{center}
\includegraphics[width=0.48\textwidth,angle=0]{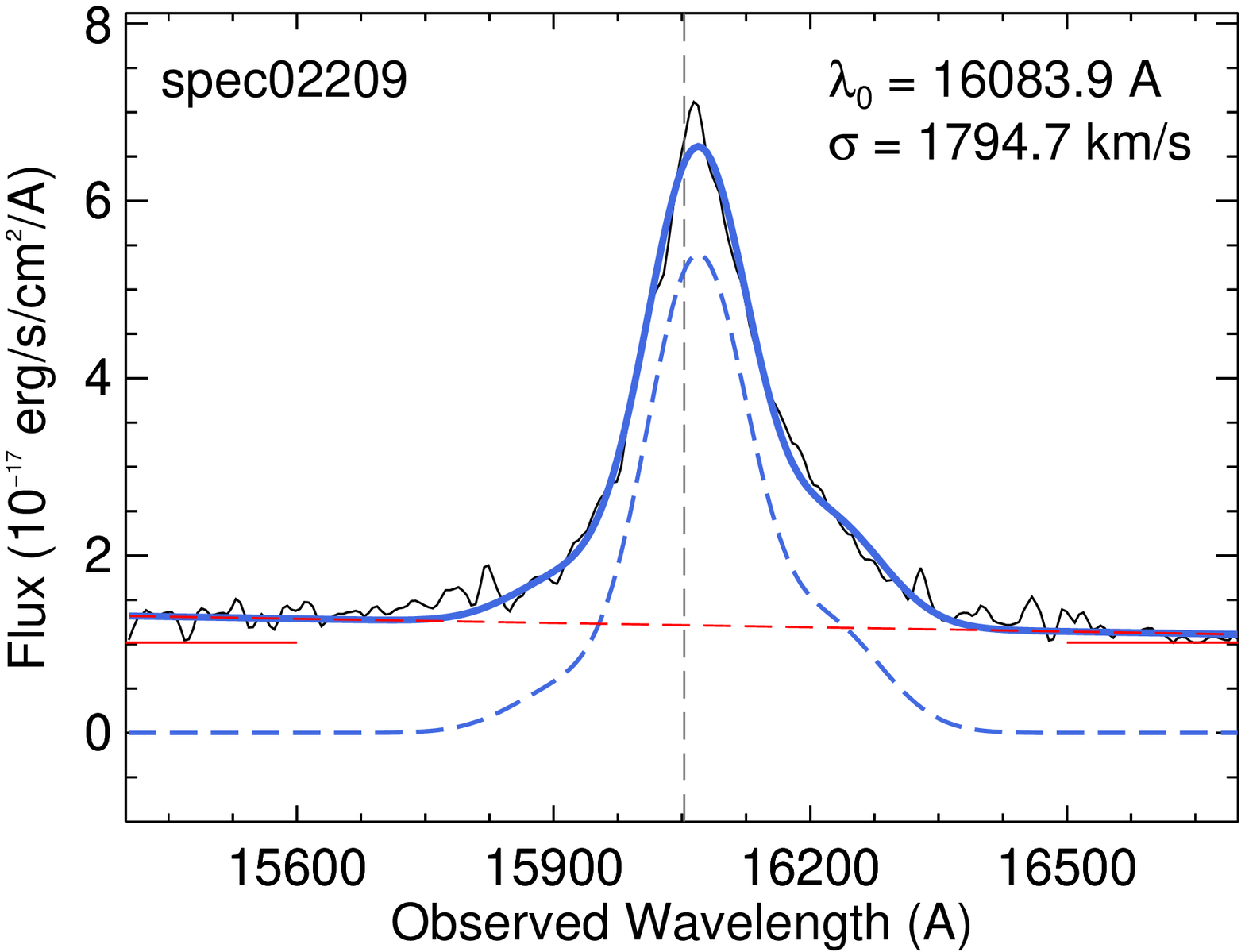}\\
\includegraphics[width=0.48\textwidth,angle=0]{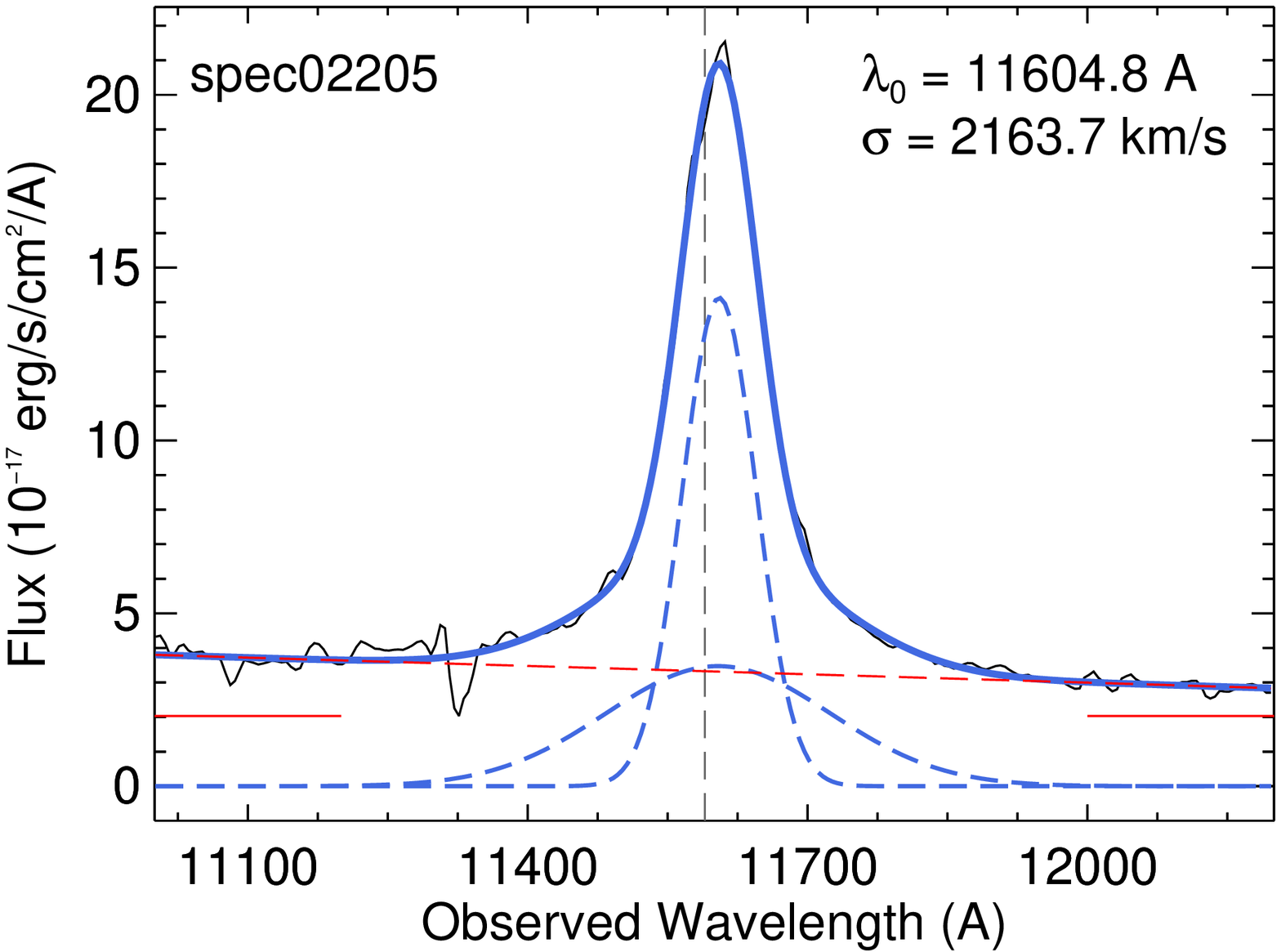}
\caption{Two example fits to the H$\alpha$ line using a 4th order Gauss-Hermite profile (top) and a double-Gaussian profile (bottom). The original spectrum (black solid line) is shown together with the wavelength windows, which are used for the continuum emission fitting (red solid horizontal lines), and the fitted continuum (red dashed line). The best-fit emission line model (blue solid line) is also shown. For the case of the multi-Gaussian fit, the individual Gaussian components are shown with blue dashed lines. spec02209 and spec02205 have visual quality flags B and A, respectively.}
\label{fig:Ha}
\end{center}
\end{figure}

Out of the 89 sources, 28 include H$\alpha$ in their FMOS spectra. Out of these, 13, 11, and 1 have a visual quality flag A, B, and C. Of the remaining 3, the fit failed for two and one shows a double-peaked emission line. As it is uncertain whether the double peak is due to low S/N or of intrinsic origin, we remove this last source from further analysis. In total, we use a double-Gaussian profile for fitting 8 out of the 25 sources, for which a Gauss-Hermite profile produces a poorer fit. In Fig. \ref{fig:Ha} we show two examples of the H$\alpha$ emission line fitting with both a 4th order Gauss-Hermite and a double-Gaussian profile.

The FWHM of H$\alpha$ for the 25 sources ranges from 1300 to 5670 km s$^{-1}$ while the $\sigma_{H\alpha}$ ranges from 600 to 3360 km s$^{-1}$.\\

\subsection{UV Continuum Luminosity}

We calculate the continuum {luminosities} at 3000\AA, in the proximity of the MgII emission line, and at 1350\AA\ and 1450\AA, near the CIV line (e.g., \citealt{Vestergaard2006,Park2013}). {For a typical quasar UV continuum slope of -0.59 (1450\AA\ to 2200\AA\ slope for SDSS quasars from \citealt{Davis2007}), this leads to a continuum luminosity difference between the two wavelengths of $\sim4\%$, smaller than typical absolute flux measurement uncertainties. We thus consider the luminosities at 1350\AA\, and 1450\AA\, to be interchangeable.} We focus on the former, where allowed by the wavelength coverage, but switch to the latter for intermediate redshift sources. We also calculate the continuum luminosity at 1800\AA, which is near the CIII] line. All luminosities have been calculated within $\sim$50\AA\ windows  and are not corrected for any intra- or extra-galactic extinction. The former is comparatively small and the latter is beyond the scope of this paper to calculate. A comparison of the three different rest-frame UV luminosities shows excellent agreement (scatter $\lesssim0.1$ dex and negligible offset).

\subsection{Optical Continuum}

Single-epoch BH mass estimates using the rest-frame optical Balmer lines have utilized both the continuum and emission line luminosities. However, the low continuum level and the fact that the 5100\AA\ region lies at the blue edge of the FMOS spectra for most of our high-redshift sources, leads to very large uncertainties in the determination of L$_{5100}$. Nevertheless, a comparison between continuum and H$\alpha$ line luminosities shows a good correlation, with an average ratio of $\sim$100. For the following we will use L$_{\mathrm{H\alpha}}$ instead of L$_{5100}$ to avoid large uncertainties for individual objects. For the few sources with only H$\beta$ measurements, we translate L$_{\mathrm{H\beta}}$ to L$_{\mathrm{H\alpha}}$, assuming a fixed ratio of 3. While we only have 3 sources with both H$\alpha$ and H$\beta$ emission, a comparison of their luminosities gives us a ratio of $3.3\pm0.3$, consistent with the expected value. 

\subsection{Linear fitting and statistics}
{The underlying assumption of this study (and that of all similar studies in the past) is that all broad emission lines are emitted by fast moving ionized gas in the vicinity of the supermassive BH and thus kinematically should reflect the BH's gravitational potential. Therefore, we expect a consistency among BH masses calculated using different broad emission lines. Deviations from this one-to-one relation should reflect measurement uncertainties, different geometries and stratification of the broad emission-line region (BLR), and additional (non-virialized) kinematic components.}

{We perform linear regression analyses using the \textit{fitexy} code, based on the linear regression algorithm introduced in \citet{Tremaine2002}, in order to compare BH mass estimates from various methods.  
The previous study by \citet{ParkD2012} provides a detailed analysis of the method \citep[see also][]{Park2015}. The \textit{fitexy} code allows for measurement errors on both the independent and measured variables.
For this study, both variables are measured and as such for each comparison we additionally perform reverse linear regression fits. In most cases the results are consistent with each-other.}

{The uncertainties in the fits are determined through a set of 100 Monte Carlo (MC) realizations, where a random subset of the full sample of measurements is used in the fit. For each fit we also calculate the intrinsic scatter of the data iteratively, adjusting it so that the reduced $\chi_\nu^2\approx1$. The intrinsic scatter is then defined as the required error-weighted reduced $\chi^{2}$ difference along the y-axis of the fit.}
{Finally, for each comparison we calculate Kendall's rank correlation coefficient, $\tau$, to quantify the degree of correlation in the data. {This is preferred for smaller samples, as the ones presented here, over Spearman's rank correlation. Kendall's $\tau$ results in smaller correlation coefficients than Spearman's rank correlation, with $\tau$ values $\gtrsim0.4$ implying a strong correlation between the compared quantities.}}

\subsection{Photometric and kinematic measurement uncertainties}
\label{sec:mc}
For the FMOS data, the error spectra provide the statistical noise per spectral pixel. Hence, we calculate uncertainties for the width and flux measurements of each emission line, based on MC simulations. For each object we produce a set of 1000 mock spectra by randomizing the flux using the estimated flux error. We take the standard deviation of the fits as the uncertainty. We adopt an iterative 4$\sigma$ clipping process to ensure that catastrophic fits are removed before calculating the uncertainty, particularly since this procedure corresponds to analyzing noisier spectra than the actual data. The clipping is stopped once the change in the values is less than 10\%. In practice, no more than 3 iterations are required for all 89 sources.

A similar procedure is followed for the AGES spectra. We calculate the noise in 5 different wavelength regions of the AGES spectra that are close to the emission lines of interest and free of emission or absorption lines. These values are then used to perform MC simulations as described above.

\subsection{Narrow Emission Components}

\begin{figure}[tbp]
\begin{center}
\includegraphics[width=0.48\textwidth,angle=0]{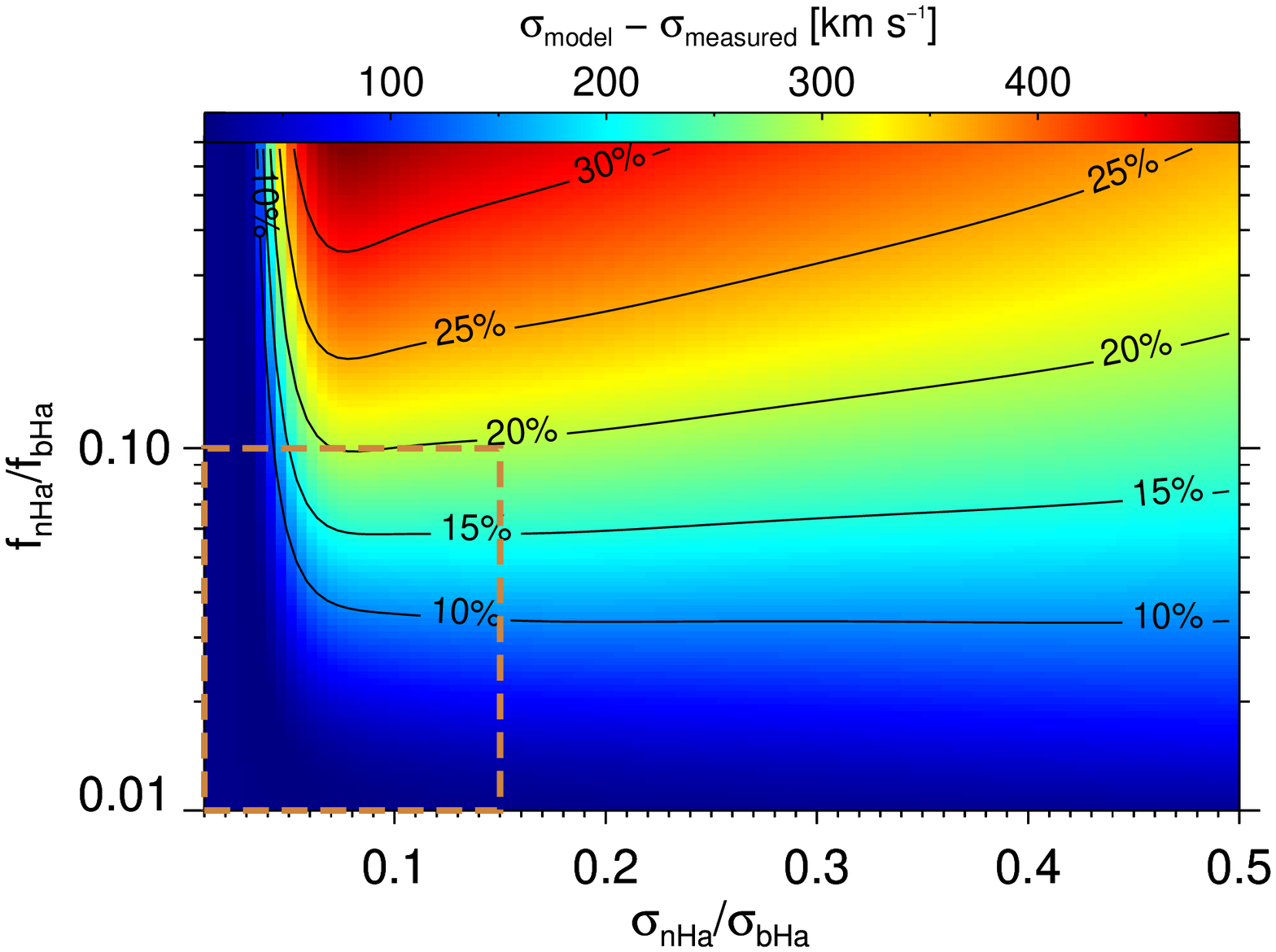}\\
\includegraphics[width=0.48\textwidth,angle=0]{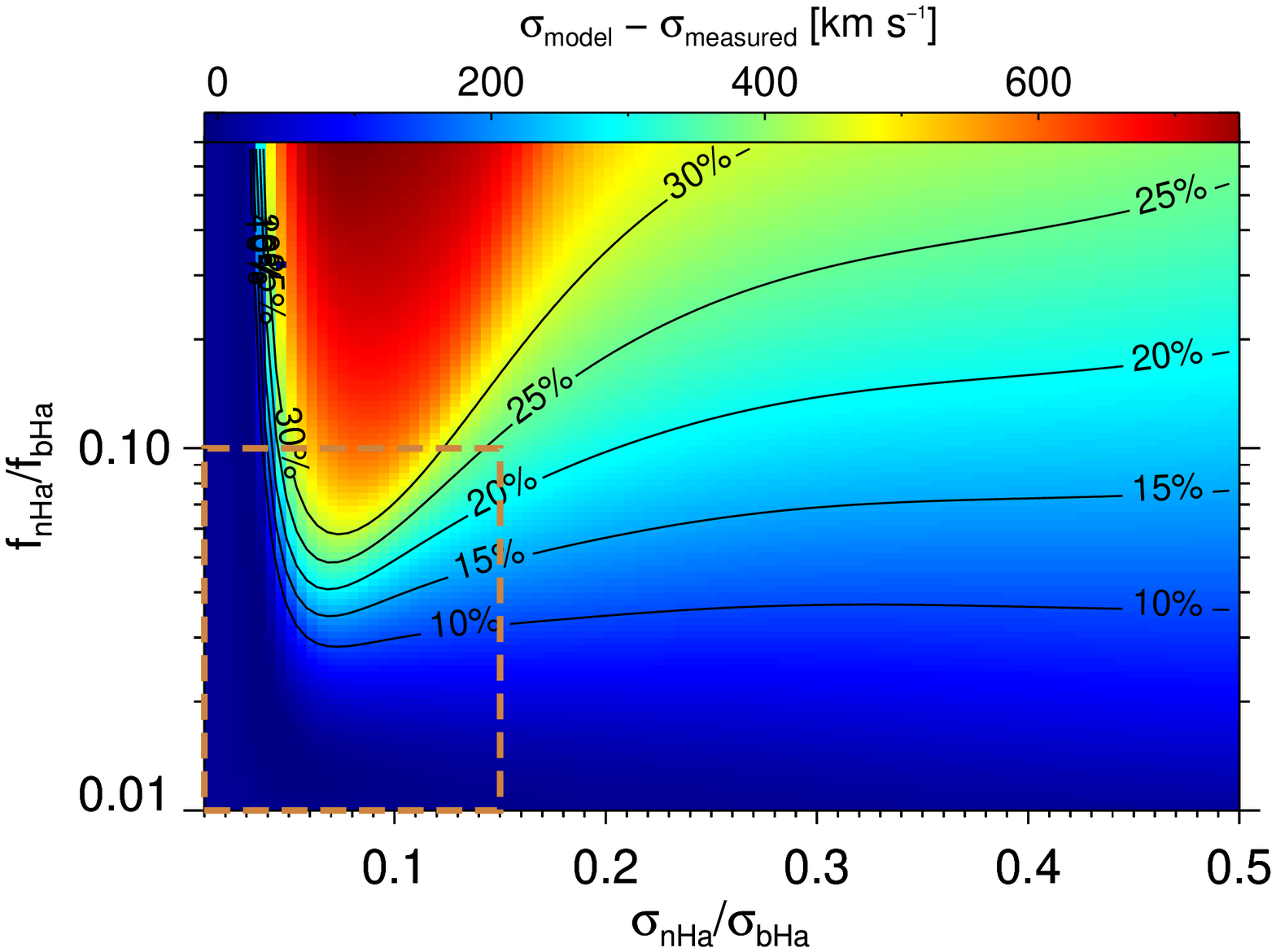}
\caption{Velocity difference (in color) between measured and true broad H$\alpha$ dispersion, as a function of the fractional values of narrow H$\alpha$ flux and dispersion, with respect to broad H$\alpha$. This is shown for spectral resolutions of R=600 (top) and R=1200 (bottom). We also show contours of fractional uncertainty. The dashed boxes show the parameter space where we expect the actual range of values of the narrow H$\alpha$ contribution to lie with respect to the broad H$\alpha$ component.}
\label{fig:sim_model}
\end{center}
\end{figure}

Here we perform a set of simulations to constrain the uncertainty due to the exclusion of a narrow component from our H$\alpha$ fits. We generate model spectra that include both broad and narrow H$\alpha$ emission components, as well as the narrow [NII] doublet. Next we convolve the spectra with a Gaussian kernel of dispersion equal to the FMOS resolution and measure the $\sigma$, FWHM, and flux of H$\alpha$ line. Keeping the broad H$\alpha$ emission component parameters constant, we repeat this over a grid of values for the H$\alpha$ narrow emission component flux and width fractions, with respect to the broad component. The difference between the input and output estimates of $\sigma$ for the broad component provides an estimate of the bias created by a narrow line component as a function of its flux and width. This is shown in Fig. \ref{fig:sim_model} for two different assumed instrumental resolutions.

For simulations at the R=600 resolution of FMOS (upper panel of Fig. \ref{fig:sim_model}), the uncertainties range from below 10\% for the weakest narrow H$\alpha$ components up to $\sim30\%$ for the strongest ones. We see a mild trend for smaller underestimates with increasing narrow H$\alpha$ width. The modest effects are a consequence of the low spectral resolution. If we repeat the simulations assuming R=1200, the uncertainties in $\sigma$ for the broad H$\alpha$ line are up to $\sim40\%$, twice those found for R=600.

We conclude that for the FMOS H$\alpha$ $\sigma$ measurements can be uncertain by up to 20\%, depending on the relative strength of the narrow H$\alpha$ component. Averaging over the possible parameter space of narrow H$\alpha$ (see dashed box in Fig. \ref{fig:sim_model}), we get an overall uncertainty of $\lesssim10\%$. While this error is systematic in nature, it is generally small compared to the statistical uncertainties.

\section{Results}
\label{sec:results}
\subsection{Comparing MgII and Balmer estimators}
\label{sec:MgIIHaHb}

In this section, we compare the line widths of the Balmer lines with those of MgII and then investigate the consistency of the virial products estimated from these lines. 
We combine H$\beta$ and H$\alpha$ measurements, given the small number of sources with measured H$\beta$ emission lines.
For this comparison we exclude highly uncertain measurements (i.e., fractional errors $>100\%$)\footnote{We note that the inclusion of these measurements does not change our results.} and present error-weighted results.



{If we compare the MgII and Balmer line widths we find that most sources are clustered around the 1:1 relation, with a Kendall's $\tau$ value of 0.52 (0.39) and a significance for rejecting the null hypothesis (no correlation) of 0.0006 (0.016) for the $\sigma$ (FWHM) measurements. {Since $\tau$ values above 0.4 are considered to imply strong correlations, the calculated $\tau$ for the FWHM measurements suggests a modest correlation, which can be understood as a result of the asynchronous observations and the intrinsic scatter of the relation.} Fits to the line width data give slopes that are consistent with unity, zero intercepts, and an intrinsic scatter $\sim0.2$ dex. Thus, although AGES and FMOS observations were not contemporaneous, the line widths of Balmer and MgII lines are roughly consistent.}

Next we examine the virial product
\begin{equation}
\left(\frac{VP_{x}}{M_\sun}\right)=\frac{(M_{\mathrm{BH}}/M_\sun)}{f_{x}}=\frac{33.65}{G}\cdot\sigma^{2}_{x}\cdot \lambda L^{0.533}_{x},
\end{equation}
{where $x$ denotes the line used for the calculation, $\sigma_{x}$ is the line width (FWHM or $\sigma$), {$\lambda L_{x}$ is either the monochromatic continuum luminosity measured at specific wavelengths near the emission line (e.g., 3000\AA\ for MgII) or the emission line luminosity (for the Balmer lines) in erg s$^{-1}$ units}, $f_{x}$ is the virial factor, and G is the gravitational constant. To derive the BLR size, we utilize the latest BLR size-luminosity calibration from \citet{Bentz2013}, from which the power index 0.533 and the normalization $33.65$ are taken. Given the uncertainty in the virial factor, especially for less often used lines like CIV and CIII], we primarily focus on the VP rather than the actual BH mass estimates\footnote{{We use the same size-luminosity power law index and normalization from \citet{Bentz2013} for both Balmer and carbon VPs. While it is uncertain that the same relation holds for different ionization species, for consistency we assume it does.}}.}

{While there are alternate choices for the luminosity and velocity power law indices (for a compilation see e.g., \citealt{McGill2008, Park2013}), fixing them to the values given above allows us to perform self-consistent comparisons between the different VP estimates and does not affect our results. }


\begin{figure}[tbp]
\begin{center}
\includegraphics[width=0.48\textwidth,angle=0]{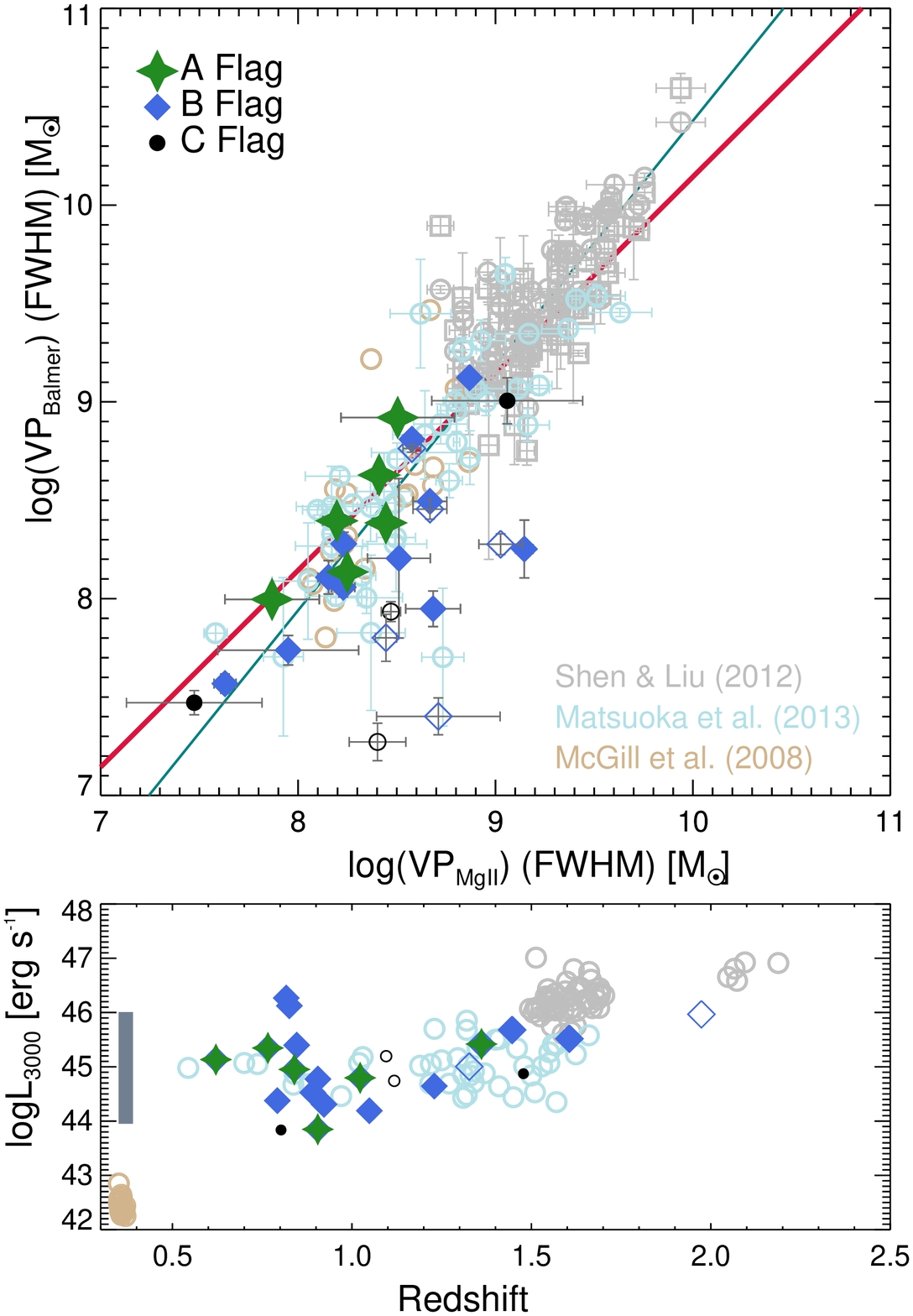}
\caption{Top: Virial products for the MgII and H$\alpha$/H$\beta$ broad emission lines for this (green stars, blue squares, and black circles) and previous studies (\citealt{McGill2008,Shen2012,Matsuoka2013}; light brown, gray, and light blue open circles, respectively). Green stars indicate sources where both lines have flag A, while blue diamonds are cases with AB or BB flags. Black circles denote cases where either of the lines has a visual quality flag C. For our sample, filled and open symbols denote VPs using the H$\alpha$ and H$\beta$ line, respectively. The red line shows a linear fit with a fixed slope of one to all VPs. We obtain an intercept $0.14\pm0.03$ dex and an intrinsic scatter $0.31\pm0.03$ dex. Considering only H$\alpha$, the intrinsic scatter reduces to $0.25\pm0.02$ dex. The teal line shows the forward linear regression fit to the combined sample, allowing the slope to vary. Bottom: Continuum luminosity at 3000\AA\, as a function of redshift for the different samples presented in the top panel, using the same notation. We also show the redshift and luminosity range of the SDSS sources with MgII and H$\alpha$ measurements from the sample of \citet{Shen2011} with the gray stripe at low redshift.}
\label{fig:MgIIHa_VP_lit}
\end{center}
\end{figure}

In Fig. \ref{fig:MgIIHa_VP_lit} we show the FWHM-based VPs of our sample combined with values from previous studies beyond the local Universe: the intermediate redshift sample of \citet{McGill2008}, the SDSS  high-redshift sample from \citet{Shen2012}, and the high-z sample from \citet{Matsuoka2013}. The combined sample spans a large range of redshifts (0.5 $<$ z $<$ 2.3) and luminosities ($10^{43}$ $<$ L$_{3000}$ $<$ $10^{48}$ erg s$^{-1}$), as shown in the lower panel of Fig. \ref{fig:MgIIHa_VP_lit}. 
The luminosity of our sample is much lower {than} those of \citet{Shen2012}, while at the same time it expands the luminosity coverage at redshifts 0.5 $<$ z $<$ 1, compared to the sample of \citet{Matsuoka2013}. {When we perform a linear fit to the VPs {(assuming MgII to be the independent measurement)}, the resulting slope is 1.24$\pm$0.04. This significantly super-linear slope is driven by the H$\beta$ VPs, which are systematically lower that their MgII counterparts. For only H$\alpha$, we obtain a slope 1.19$\pm$0.04 that is still significantly ($\sim$5$\sigma$) steeper than the 1:1 relation, and has an intrinsic scatter  $\sim0.23\pm0.03$ dex. {Reverse linear regression fitting provides a sub-linear slope of 0.73$\pm$0.03 and an intrinsic scatter of 0.18$\pm$0.02 dex}. For all VPs, we derive an intercept of $0.14\pm0.03$ dex and an intrinsic scatter of $0.31\pm0.03$ dex around the 1:1 relation. When considering only the H$\alpha$ VPs, the intrinsic scatter reduces to $0.25\pm0.02$ dex.}

\begin{figure}[tbp]
\begin{center}
\includegraphics[width=0.48\textwidth,angle=0]{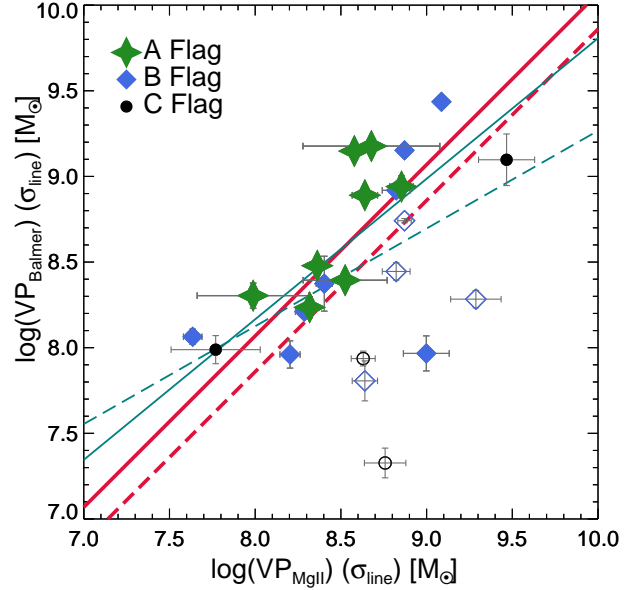}
\caption{Virial products for the MgII and H$\alpha$/H$\beta$ broad emission lines for our sample (green stars, blue squares, and black circles) based on the measured second moment ($\sigma_{line}$) of the emission lines. Filled and open symbols denote VPs using the H$\alpha$ and H$\beta$ lines, respectively. The dashed and solid red lines show linear fits with a fixed slope of one to all and only H$\alpha$ VPs, respectively. Similarly, the teal lines show the forward linear regression fits with free slope.}
\label{fig:MgIIHa_VP_SMOM}
\end{center}
\end{figure}

{In Fig. \ref{fig:MgIIHa_VP_SMOM}, we show the $\sigma$-based VPs for our sample. Second moment measurements for other large samples at similar redshifts do not exist and we thus probe limited dynamic range of VPs. We find a reasonable agreement between the VPs with a zero intercept. Considering all VPs, we obtain a significant intrinsic scatter $0.53\pm0.08$ dex, while a linear regression fit with a free slope gives a sub-linear slope $0.57\pm0.19$. Considering only H$\alpha$ VPs, we obtain an intrinsic scatter $0.36\pm0.11$ dex around the 1:1 relation, while a free-slope regression fit gives a slope of $0.82\pm0.26$, consistent with one. The deviations of the slope from unity are again driven by the results for H$\beta$.}

\subsection{The kinematics of CIV and CIII]}
\label{sec:civciii}

\begin{figure}[tbp]
\begin{center}
\includegraphics[width=0.48\textwidth,angle=0]{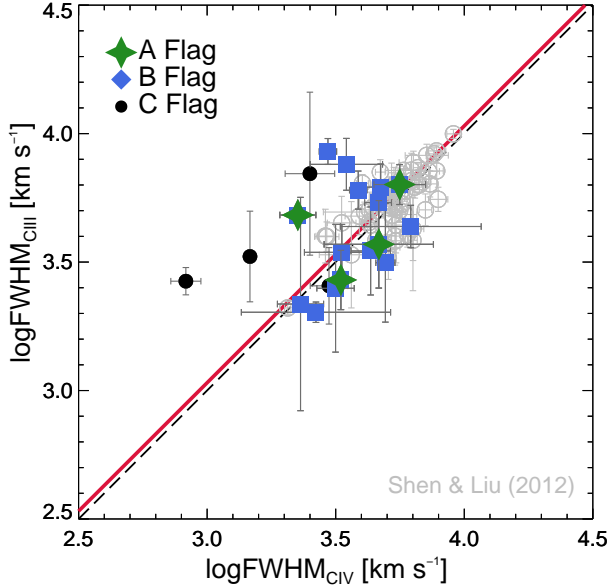}
\caption{Comparison between the FWHM of the CIV and CIII] emission lines. Sources are divided into three classes according to their visual quality flags, following the notation of Fig. \ref{fig:MgIIHa_VP_lit}. We increase the FWHM range by including the data from \citet{Shen2012} (gray open circles). The dashed black line shows the 1:1 relation. The red line shows a linear fit with a fixed slope of unity to the combined sample. We calculate an intercept $0.03\pm0.01$ dex and an intrinsic scatter $0.10\pm0.10$ dex around the 1:1 relation.}
\label{fig:CIVCIII_comp}
\end{center}
\end{figure}

We next turn to comparisons between the CIII]/CIV lines and the H$\alpha$/H$\beta$ lines. First, we investigate whether CIII] is a viable alternative to CIV as a single-epoch BH mass estimator by comparing the FWHM of CIV and CIII] for our sample combined with the SDSS sample of \citet{Shen2012} (Fig. \ref{fig:CIVCIII_comp}). For CIII] we derive a negligible intercept of $0.03\pm0.01$ dex and an intrinsic scatter of $0.1\pm0.1$ dex around the 1:1 relation, suggesting that the two lines trace similar kinematics. {We calculate Kendall's $\tau$=0.43 with a very high confidence (p$<$0.00001). The comparison for $\sigma$ (not shown) is much noisier (intrinsic scatter $0.25\pm0.04$ dex), partly due to the lack of $\sigma$ estimates in the literature, the small number and limited dynamic range of our sample, and $\sigma$ being more sensitive to low S/N than the FWHM.} The scatter around the 1:1 relation for both width measures and the different sub-sets of sources are given in Table \ref{tab:CIVCIII}.


\begin{deluxetable}{c| c c}
\tablecolumns{3}
\tablewidth{0pt}
\tablecaption{CIV/CIII] intrinsic scatter \label{tab:CIVCIII}}
\tablehead{ \colhead{} &	\colhead{$\sigma$}	&	\colhead{FWHM}\\
	\colhead{}	&	\colhead{[dex]} & \colhead{[dex]}	}
\startdata
A+B+C (All)			&	0.35		&	0.20		\\
A+B					&	0.28		&	0.19		\\
A+B+Shen \& Liu (2012)	&	\nodata	&	0.10		\\
\enddata
\tablecomments{{Intrinsic scatter (i.e., scatter beyond the individual measurement uncertainties)} around the 1:1 relation (in dex) for the CIV and CIII] emission lines using either the $\sigma$ or the FWHM. For the data from \citet{Shen2012} no $\sigma$ measurements are available.}
\end{deluxetable}%

\subsection{Balmer vs. carbon VPs}
\label{sec:CIVCIIIHaHb}

\begin{figure}[tbp]
\begin{center}
\includegraphics[width=0.23\textwidth,angle=0]{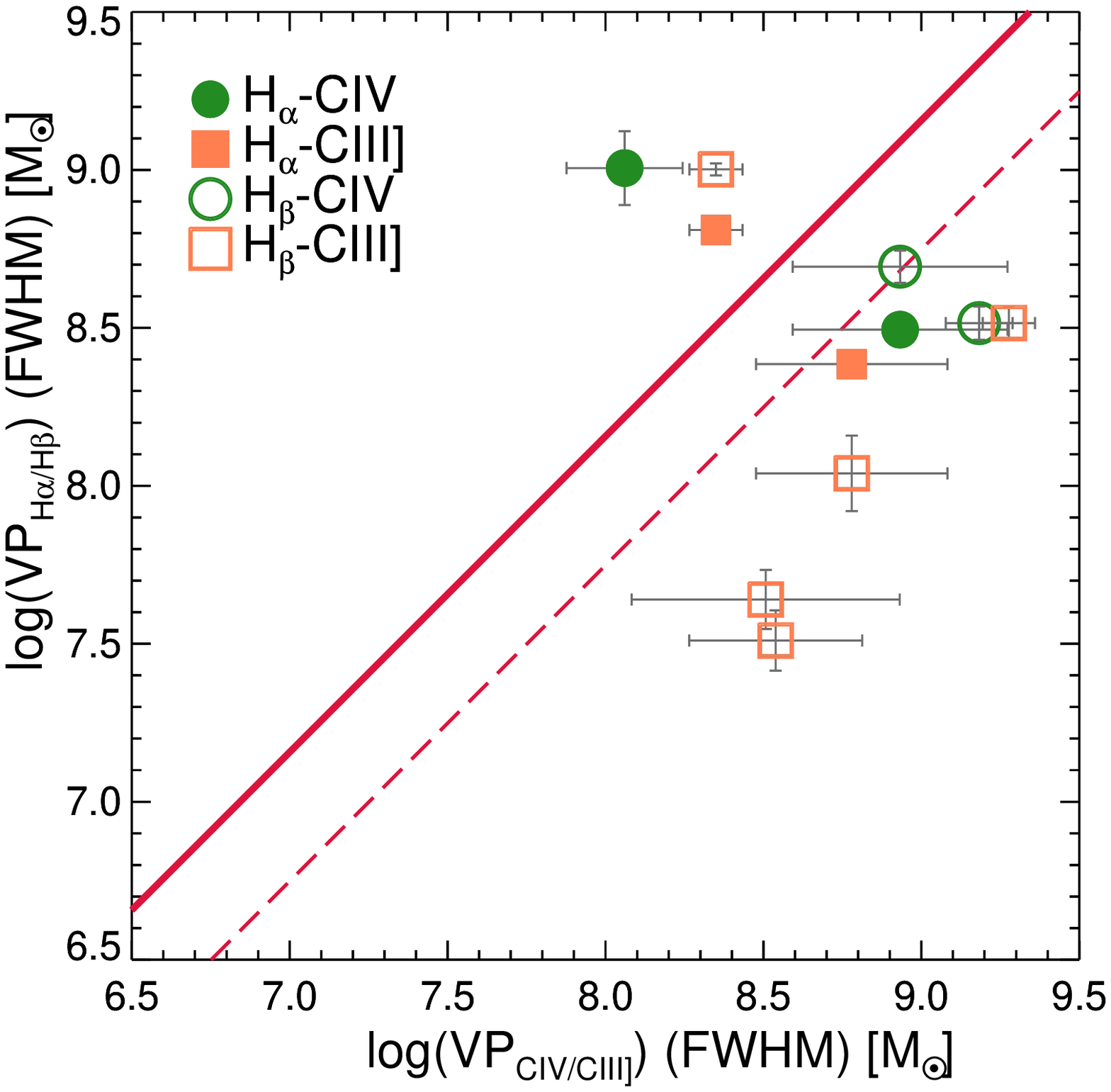}
\includegraphics[width=0.23\textwidth,angle=0]{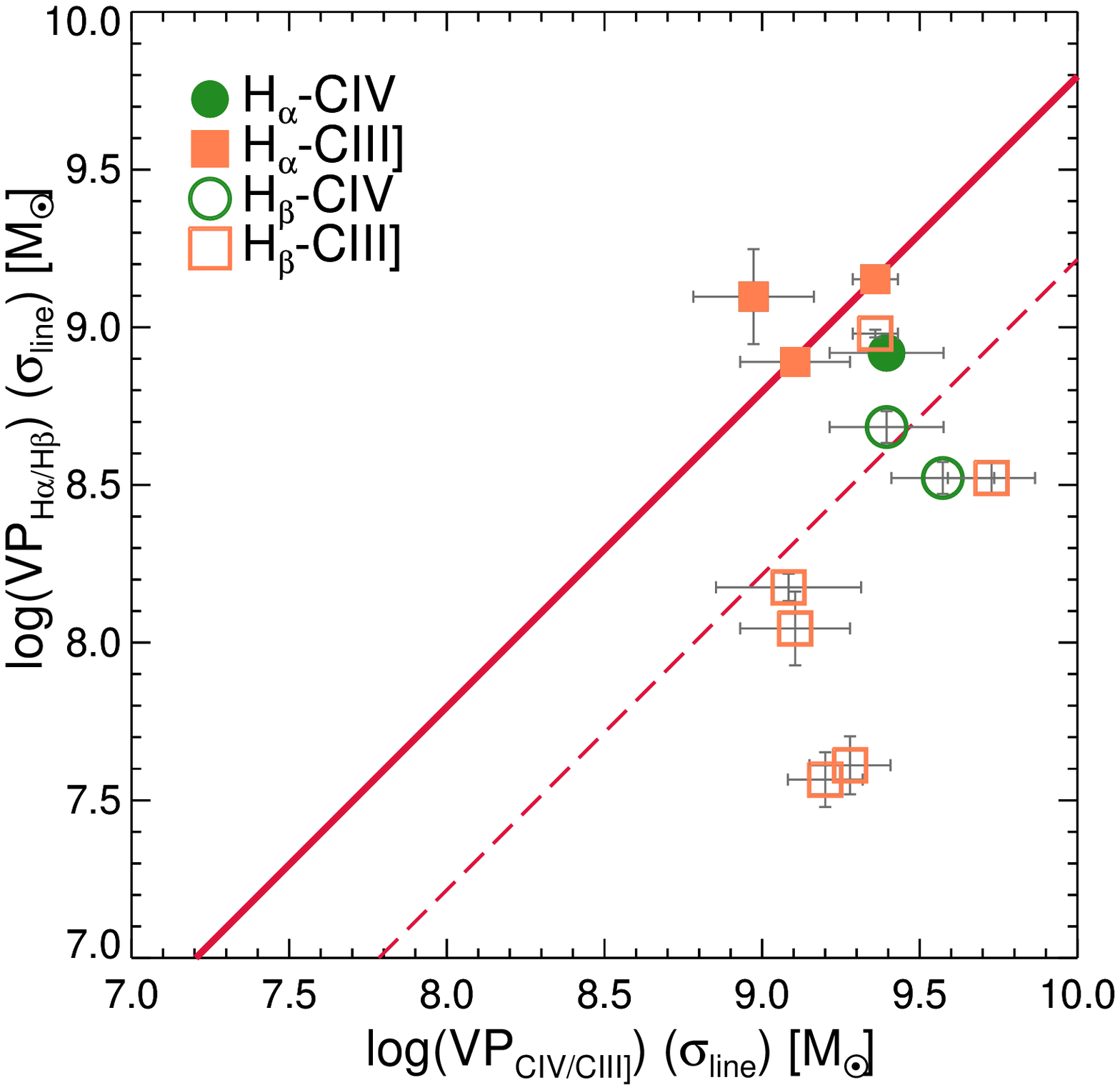}
\caption{Comparison between the VP using the FWHM (left) and $\sigma$ (right) of the H$\alpha$ (filled symbols) and H$\beta$ (open symbols) lines against the CIV (green circles) and CIII] (orange squares) lines. The dashed red line shows the linear fit of fixed slope $\alpha$=1 to all points on the plot, while the solid red solid line is the linear fit of fixed slope $\alpha$=1 to just the filled symbols (H$\alpha$-carbon). All CIV sources have visual quality flags B, while for CIII] one source has a visual quality flag A, one C, and the rest have visual quality flags B.}
\label{fig:virCIVCIIIHaHb}
\end{center}
\end{figure}

Next we investigate the consistency of the VPs derived from the Balmer lines and the carbon lines.
First, we directly compare the VPs of our sample in Fig. \ref{fig:virCIVCIIIHaHb}. {Since the BH mass must be the same, we expect a slope of unity between the Balmer and carbon line VPs,
with a non-zero intercept, reflecting the differences in the virial factor.
In the case of FWHM-based VPs (left panel of Fig. \ref{fig:virCIVCIIIHaHb}), we find no significant correlation based on Kendall's $\tau$, which is due to
the limited dynamical range and uncertain measurements from weak emission lines.} Linear fits with a fixed slope of $\alpha$=1 to all the VPs (dashed red line) and to only those involving H$\alpha$ (solid red line) are also shown. The dashed red line has an intercept of $-0.25\pm0.19$ dex with an intrinsic scatter $0.71\pm0.14$ dex. The solid red line has a smaller positive intercept of $0.16\pm0.29$ dex but an increased intrinsic scatter $1.14\pm0.34$ dex. 

In Fig \ref{fig:virCIVCIIIHaHb} (right), we show the same comparison for the $\sigma$-based VPs.  {Again we see no obvious correlation between the two VPs, which is confirmed by the Kendall's $\tau$ statistic}. From the H$\alpha$-carbon VPs we obtain a negative intercept of $-0.20\pm0.09$ dex (compared to an intercept  of $-0.78\pm0.16$ dex for all VPs; dashed red line) and an intrinsic scatter of $0.37\pm0.14$ ($0.61\pm0.10$) dex around the 1:1 relation. It is also worth noting that the $\sigma$-based VPs calculated from both CIV and CIII] are 0.5-1 dex systematically higher {than those based on the FWHM}.

\begin{figure}[tbp]
\begin{center}
\includegraphics[width=0.23\textwidth,angle=0]{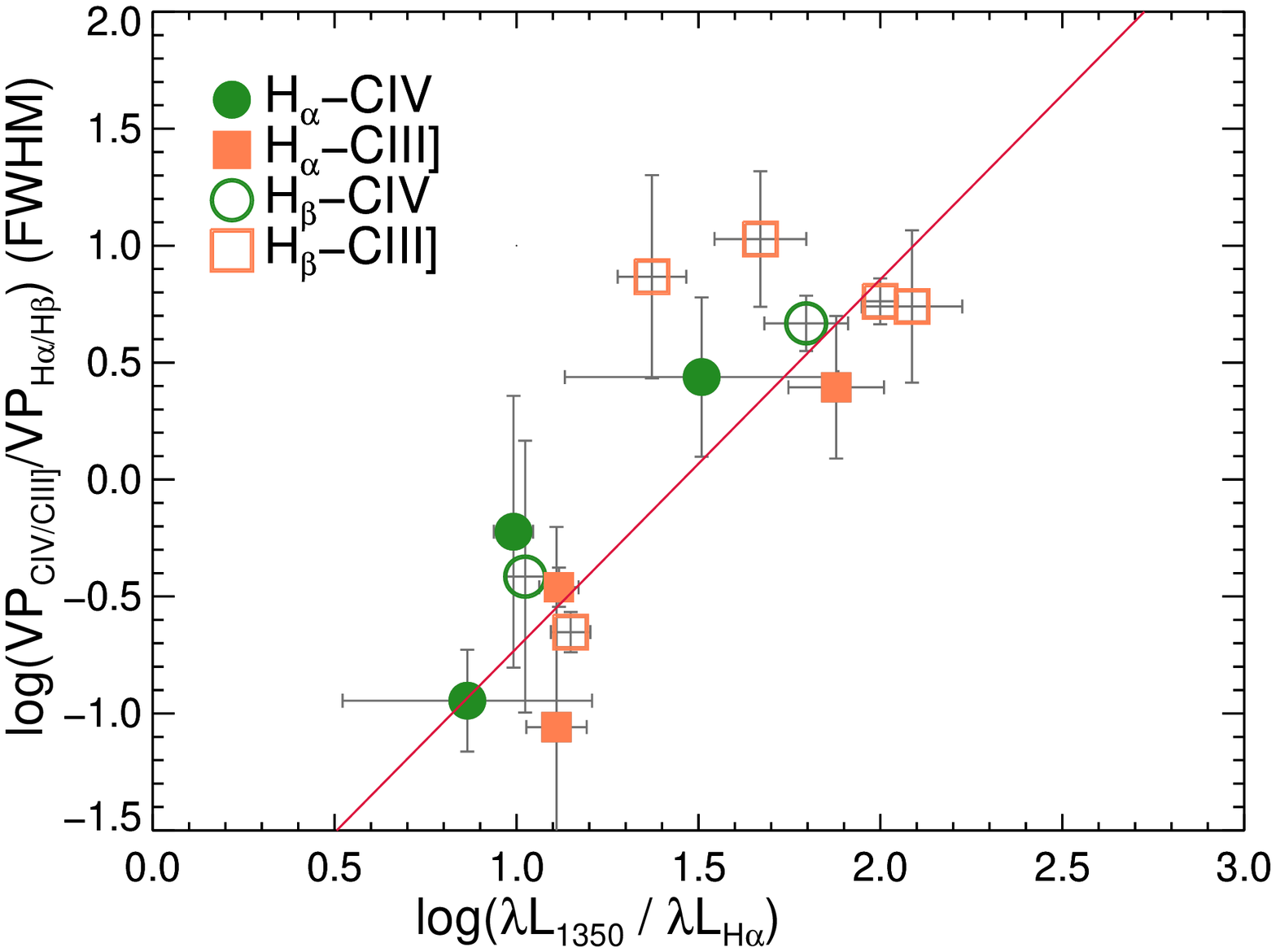}
\includegraphics[width=0.23\textwidth,angle=0]{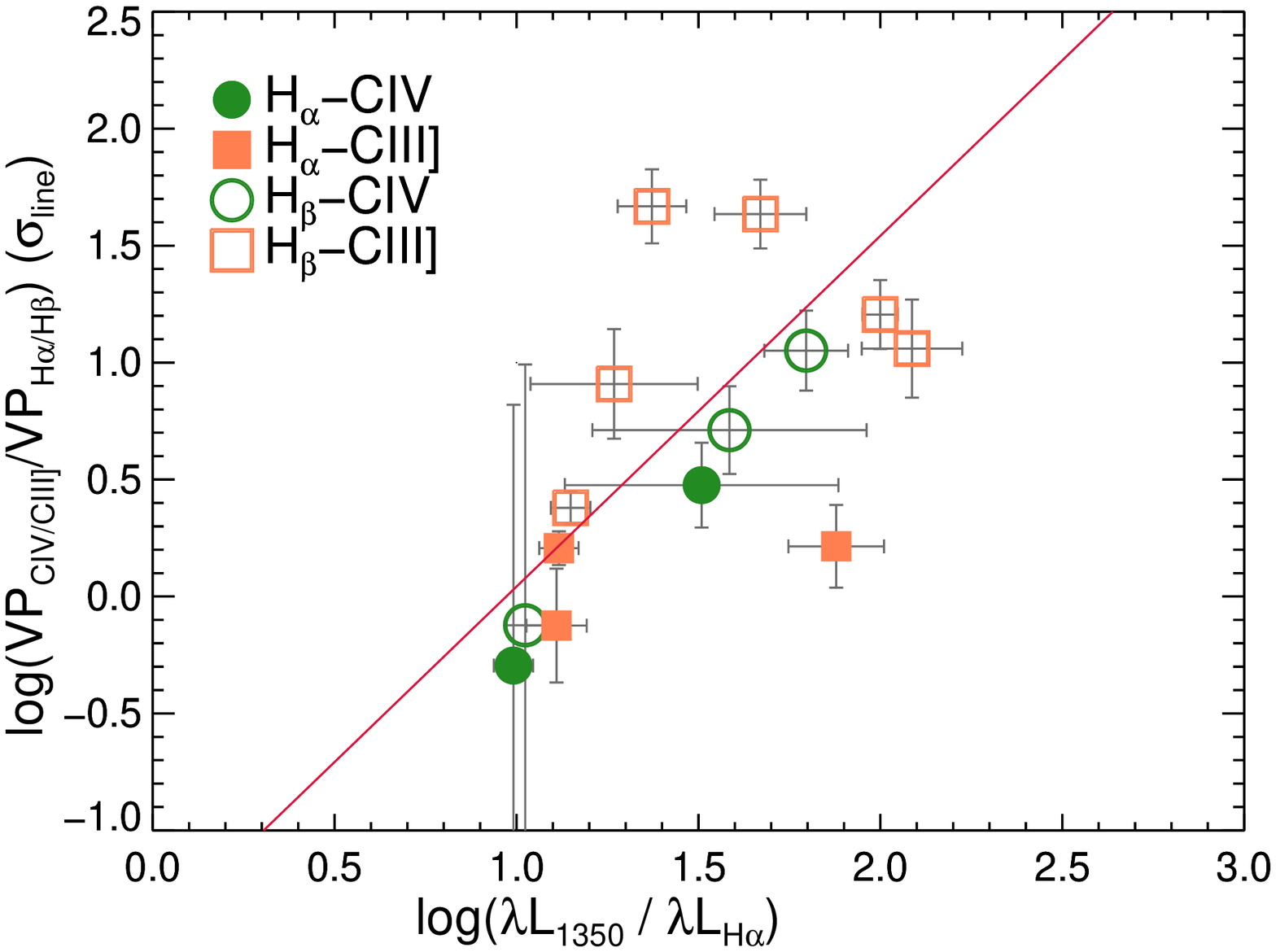}
\caption{CIV/CIII] to H$\alpha$/H$\beta$ VP ratio as a function of UV (1350\AA\, or 1800\AA\,) to optical luminosity ($L_{H\alpha}$) ratio. Solid red lines show the linear regression fits to the combined H$\alpha$ and H$\beta$ VP residuals. The symbol notation and visual flag information is the same as in Fig. \ref{fig:virCIVCIIIHaHb}.}
\label{fig:UVOPTcomp}
\end{center}
\end{figure}

Second, we investigate whether there is a systematic trend of the differences between the two VPs by investigating the residuals as a function of UV-to-optical flux ratios, as suggested by previous studies (e.g., \citealt{Assef2011}). In Fig. \ref{fig:UVOPTcomp} we compare the ratio of carbon and Balmer VPs with the luminosity ratio of UV continuum 
to H$\alpha$ emission line. We find a well defined correlation and calculate Kendall's $\tau$ to be 0.46 and 0.56 with significance 0.02 and 0.005 for FWHM-based and $\sigma$-based VPs, respectively. The best-fit slopes and intercepts of the correlations are ($\alpha$,$\beta)=(1.58\pm0.15$,$-2.30\pm0.23$) and ($1.50\pm0.17$,$-1.46\pm0.24$) for the FWHM-based and $\sigma$-based VPs, consistent with each other. They are considerably steeper than slopes calculated in \citet{Assef2011} (best-fit slopes $\sim$ 0.5-1.0). We note, however, that for asynchronous observations like the ones considered here, variability can affect the continuum luminosity and color, especially given the known wavelength dependence of the AGN variability power spectrum (e.g., \citealt{Ulrich1997}).

Based on the correlation between the VP residuals and the UV-to-H$\alpha$ flux ratios, 
we derive a correction formula for the carbon VPs of
\begin{equation}
\log{\mathrm{VP}}^{\mathrm{cor}}=\log{\mathrm{VP}}-\beta-\alpha\cdot\log{\frac{\lambda_{\mathrm{UV}}L_{\mathrm{UV}}}{\lambda_{\mathrm{opt}}L_{\mathrm{opt}}}},
\end{equation}
where L$_{UV}$ refers to the monochromatic luminosity measured at 1350\AA\ or 1800\AA, while 
L$_{opt}$ refers to the measured H$\alpha$ luminosity or the H$\alpha$ luminosity calculated from H$\beta$.

\begin{figure}[tbp]
\begin{center}
\includegraphics[width=0.48\textwidth,angle=0]{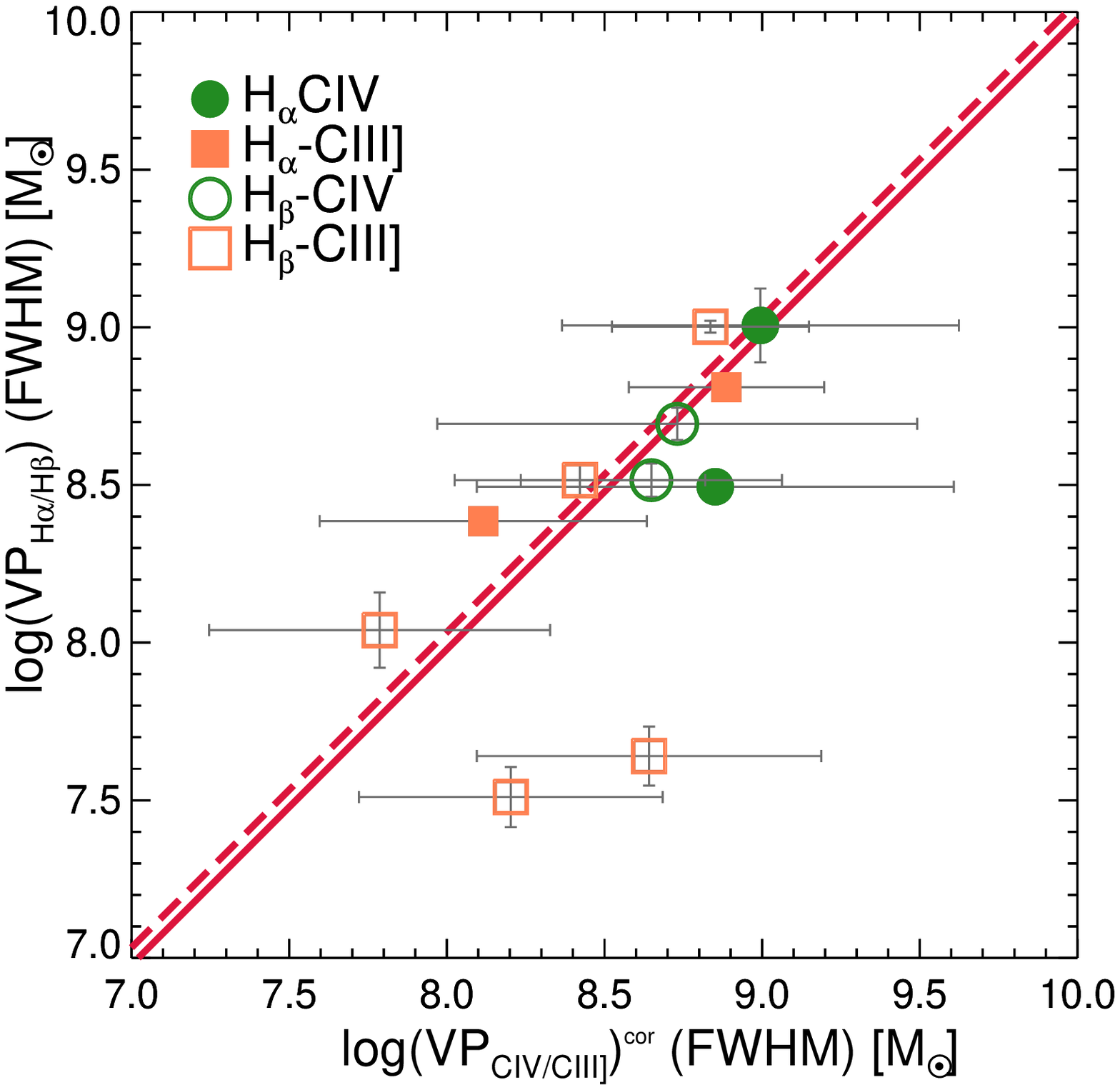}\\
\includegraphics[width=0.48\textwidth,angle=0]{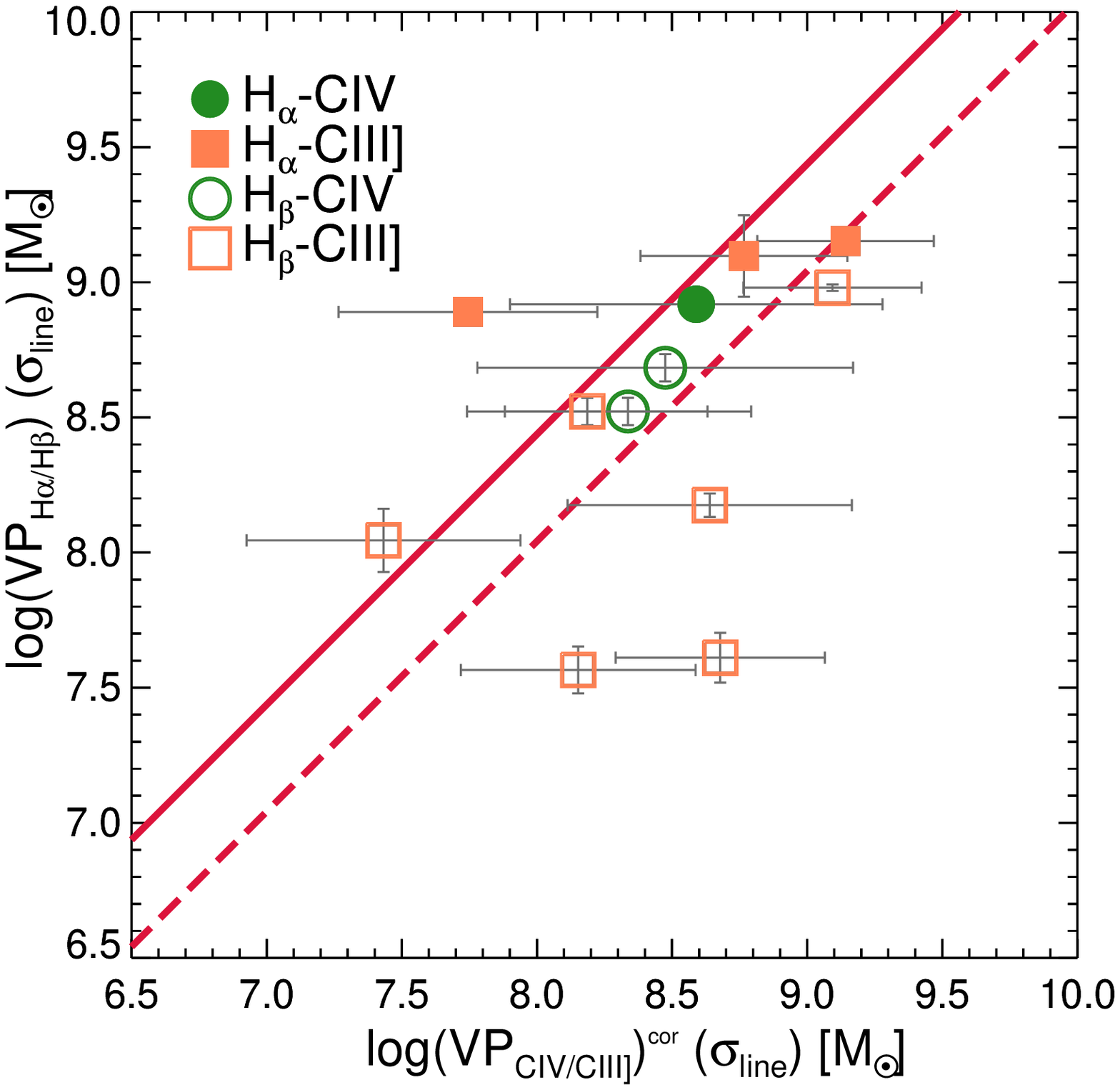}
\caption{Same as in Fig. \ref{fig:virCIVCIIIHaHb} but now using the corrected VPs for CIV and CIII] following the corrections derived from Fig. \ref{fig:UVOPTcomp}. The intercept  and intrinsic scatter of the fits are shown in Table \ref{tab:corVP}.}
\label{fig:cor_virCIVCIIIHaHb}
\end{center}
\end{figure}

{After applying the correction, the correlation between Balmer and carbon VPs becomes significantly stronger, as shown in Fig. \ref{fig:cor_virCIVCIIIHaHb}. {This is reflected in the $\tau$ values of 0.62 and 0.44 at a significance 0.007 and 0.04, respectively for the FWHM-based and $\sigma$-based VPs (compared to no correlation without the luminosity scaling), the improved intrinsic scatter, and the reduced intercept.} The intrinsic scatter and intercept are consistent with zero for the {FWHM-based VPs}. For the $\sigma$-based VPs, we obtain an intrinsic scatter $0.46\pm0.22$. {The reduced scatter compared to Fig. \ref{fig:virCIVCIIIHaHb} is in small part due to the increased uncertainties from the error propagation of the applied color correction\footnote{{If we do not consider the additional propagated errors of the color correction, we obtain intrinsic scatter 0.10 dex for the FWHM-based and 0.62 dex for the $\sigma$-based VPs (compared to 0.71 dex and 0.61 dex for the uncorrected VPs in Fig. \ref{fig:virCIVCIIIHaHb}).}}.} A summary of the measured scatter and intercept  values for Fig. \ref{fig:cor_virCIVCIIIHaHb} is given in Table \ref{tab:corVP}.}

{We note that there are two outliers that consistently show lower Balmer VPs than carbon VPs for both FWHM and $\sigma$. These are based on H$\beta$ measurements with low S/N and a visual quality flag C and B. As was noted for the MgII comparisons, our H$\beta$ measurements appear to be systematically underestimated compared to the UV lines.}

\begin{figure}[tbp]
\begin{center}
\includegraphics[width=0.48\textwidth,angle=0]{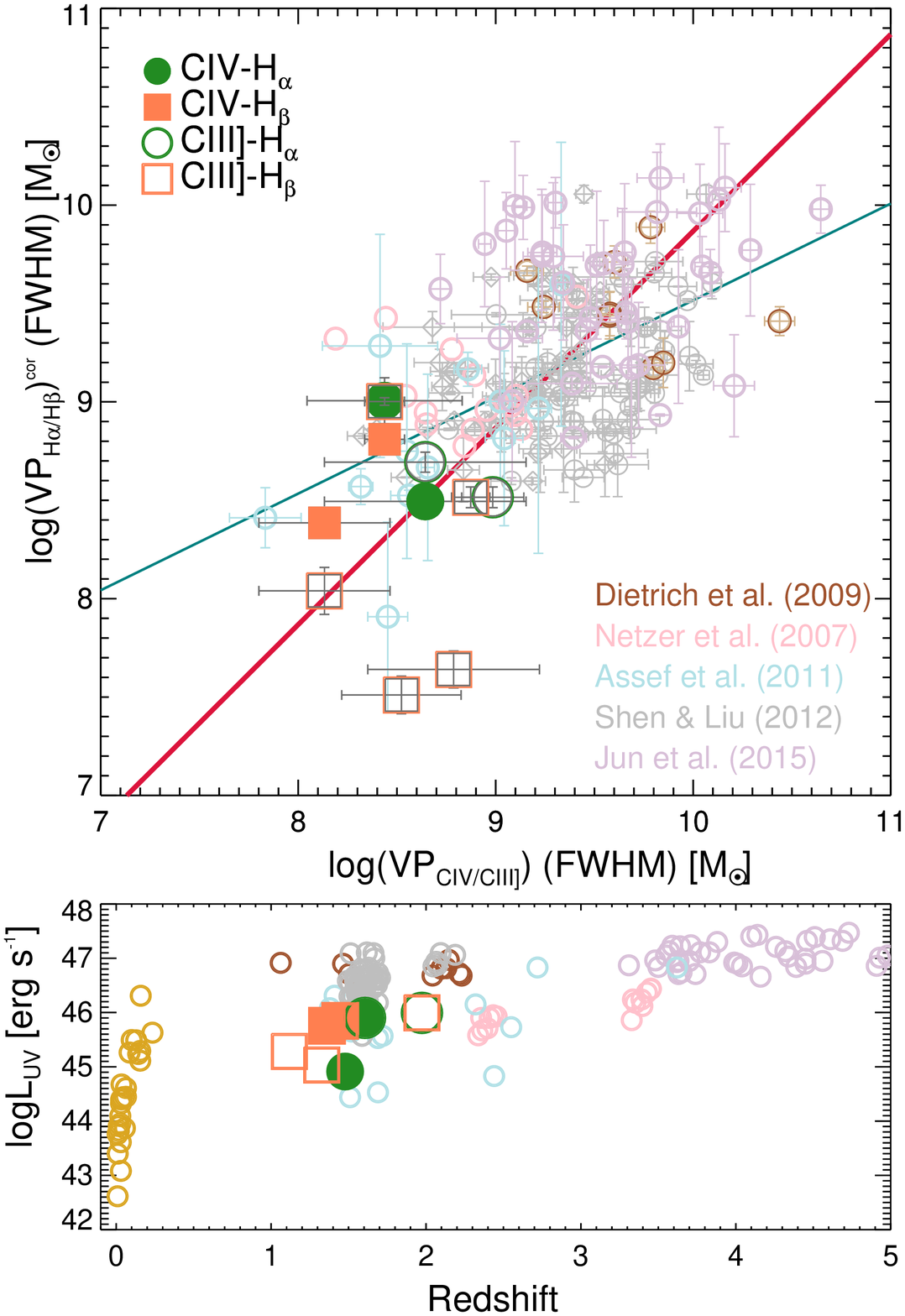}\\
\caption{Top: Carbon and Balmer {color-corrected} VP estimate comparison for our sources and sources from \citet{Dietrich2009} (brown), \citet{Netzer2007b} (pink), \citet{Assef2011} (powder blue), \citet{Shen2012} (gray), and \citet{Jun2015} (purple). The \citet{Shen2012} sample has both CIV (open diamonds) and CIII] (open circles) measurements, which are compared to H$\beta$ VPs. The other samples compare CIV and H$\alpha$ VPs. The notation for our sources is the same as in previous figures. The solid cyan line shows the forward linear regression fit with both the slope and intercept as free parameters. The solid red line shows a linear fit of fixed slope one. We find an intercept of $-0.13\pm0.03$ dex and an intrinsic scatter $0.38\pm0.02$ dex. Bottom: Continuum luminosity at 1350\AA\, as a function of redshift for the different samples presented in the top panel. The notation is as in the top panel. Also the redshift and luminosity range of the local reverberation-mapped AGN sample of \citet{Park2013} is shown with the golden circles at low redshift.}
\label{fig:CIVCIIIHaHb_lit}
\end{center}
\end{figure}

Next, we investigate the consistency of Balmer and carbon VPs for a much larger combined sample of Type 1 AGN from  \citet{Netzer2007b}, \citet{Dietrich2009}, \citet{Assef2011}, \citet{Shen2012}, and \citet{Jun2015}, in order to increase the parameter space we cover. Most of these studies investigated more luminous AGN than our AGN sample (Fig. \ref{fig:CIVCIIIHaHb_lit}, bottom panel). We convert our H$\alpha$ and H$\beta$ line luminosities using the L$_{\mathrm{H\alpha}}$-L$_{5100}$ relation from \citet{Jun2015} and the L$_{\mathrm{H\beta}}$-L$_{5100}$ relation from \citet{Greene2005}, since these studies typically adopted the monochromatic luminosity at 5100\AA\ as the optical luminosity measure.

\begin{deluxetable}{c c c c c}
\tabletypesize{\footnotesize}
\tablecolumns{5}
\tablewidth{0pt}
\tablecaption{Intrinsic scatter and intercept  values for Fig. \ref{fig:cor_virCIVCIIIHaHb}. \label{tab:corVP}}
\tablehead{	\colhead{}	&	\multicolumn{2}{c}{H$\alpha$+H$\beta$}	&	\multicolumn{2}{c}{H$\alpha$}\\
	\colhead{}	& \colhead{Scatter}	&	\colhead{Intercept }	& \colhead{Scatter}	&	\colhead{Intercept }\\
	\colhead{}	& \colhead{[dex]} & \colhead{[dex]} & \colhead{[dex]} & \colhead{[dex]}}
\startdata
$\sigma$	& 	$0.46\pm0.22$		&	$0.04\pm0.16$		&	$0.71\pm0.36$		&	$0.44\pm0.23$		\\
FWHM	& 	$0.00\pm0.09$		&	$-0.08\pm0.12$		&	$0.00\pm0.00$		&	$-0.02\pm0.09$		\\
\enddata
\end{deluxetable}%

A linear fit to the color terms in the combined data finds a flatter slope ($\alpha=1.01\pm0.03$) than the one derived from our sample only ($\alpha=1.58\pm0.15$) in Fig. \ref{fig:UVOPTcomp}. This may imply that the effect of the UV-to-optical continuum slope on the determination of the UV VPs is stronger 
for lower luminosity AGNs. Alternatively, this difference may be driven by the uncertainties in the flux measurements, which increase as we go to AGNs with lower fluxes as in our sample.  

Finally, we compile available high-z AGNs with Balmer and carbon VPs, and combine them with our measurements (Fig. \ref{fig:CIVCIIIHaHb_lit}, top). For consistency, we have used the color correction derived from the complete combined sample to correct both our measurements and the literature data. Our sources have lower luminosity than most samples presented in Fig. \ref{fig:CIVCIIIHaHb_lit} (bottom)\footnote{The sample of \citet{Assef2011} has a few sources with similar 1350\AA\, luminosities (and VP values) as our sample but on average is more luminous.}. For the combined sample, we obtain {$\tau$=0.34 at a very high significance (p$<$0.00001),} an intrinsic scatter of $0.37\pm0.02$ dex, and an intercept of $-0.13\pm0.03$ dex. {The derived $\tau$ is low and implies a modest to weak correlation between the carbon and Balmer VPs. The low $\tau$ value can be understood as a result of the significant scatter observed in Fig. \ref{fig:CIVCIIIHaHb_lit} and the fact that correlation coefficients, by definition, do not consider measurement uncertainties.} The intrinsic scatter is comparable to, but better than the scatter calculated in Fig. \ref{fig:cor_virCIVCIIIHaHb}. A linear regression provides a slope of $0.42\pm0.05$, significantly flatter than unity (solid cyan line in Fig. \ref{fig:CIVCIIIHaHb_lit}). {This is mainly due to the large scatter at VPs $>10^9$ M$_\sun$ and the relatively few measured VPs below $10^9$ M$_\sun$ (also see \citealt{Kelly2007} on how data with large measurement errors in both axes lead to flatter best-fit slopes).}

\subsection{The CIV/CIII] virial factor: From VPs to BH mass estimates}
In this section we determine the virial factor for the VPs derived from CIV/CIII] lines combined with UV continuum luminosity,
by calibrating with the best-studied virial factor of the H$\beta$ VPs. 
For this process, we convert the H$\beta$ and H$\alpha$ line luminosities to the 5100\AA\, luminosity using the correlations from \citet{Greene2005} and \citet{Jun2015}. Then we utilize the luminosity-size relation from \citet{Bentz2013} to calculate the BLR size, R$_{BLR}$. In this process, we derive the following BH mass equations based on the H$\alpha$ and H$\beta$ emission line properties,
\begin{equation}
M_{\mathrm{BH}}^{H\alpha}=f\times10^{6.58}\Bigg(\frac{\mathrm{FWHM}_{H\alpha}}{1000}\Bigg)^{2.12}\Bigg(\frac{L_{H\alpha}}{10^{42}}\Bigg)^{0.51}M_{\odot}
\label{eq:MBHHa}
\end{equation}
\begin{equation}
M_{\mathrm{BH}}^{H\beta}=f\times10^{6.74}\Bigg(\frac{\mathrm{FWHM}_{H\beta}}{1000 }\Bigg)^{2.00}\Bigg(\frac{L_{H\beta}}{10^{42}}\Bigg)^{0.47}M_{\odot}
\label{eq:MBHHb}
\end{equation}
where FWHM is measured in km s$^{-1}$ and luminosities are in erg s$^{-1}$. We use the updated virial factor $f_{H\beta}$ of $1.12\pm0.31$ derived for the FWHM-based VPs by \citet{Woo2015}.

\begin{figure}[bp]
\begin{center}
\includegraphics[width=0.48\textwidth,angle=0]{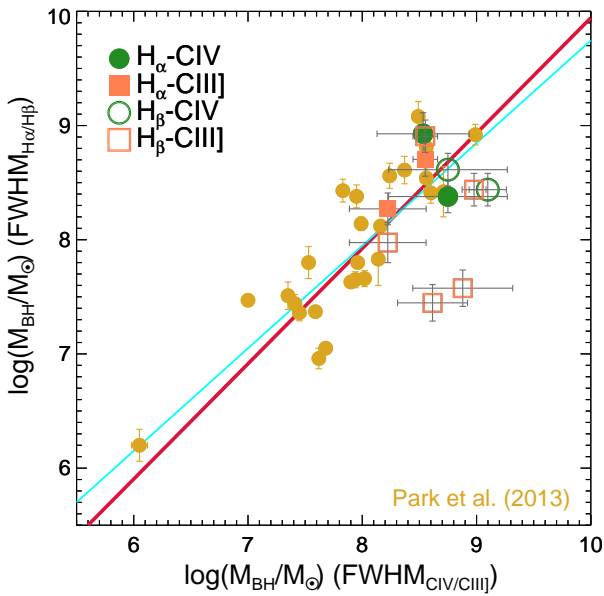}\\
\caption{Balmer and carbon virial BH mass estimates comparison for our sample (notation as in previous plots) and the updated local reverberation sample from \citet{Park2013} (shown in yellow filled circles). The fit to the combined sample with a fixed slope of $\alpha=1$ is shown with the solid red line, while the solid cyan line shows the free-slope forward linear regression fit to the combined data. {We find an offset consistent with zero and an intrinsic scatter $0.38\pm0.05$ dex around the 1:1 relation.}}
\label{fig:CIVCIIIHaHb_local}
\end{center}
\end{figure}

We calibrate the CIV/CIII] VPs {from the combined sample (FMOS-AGES and literature data, Fig. \ref{fig:CIVCIIIHaHb_lit})} by matching them to the H$\alpha$/H$\beta$ BH mass estimates and find a positive intercept  of $0.87\pm0.03$ dex for a fixed slope of unity\footnote{{The calculated Balmer BH masses for the FMOS-AGES sample, as well as the re-calculated BH masses for the literature samples, include the propagated uncertainty of the f$_{\mathrm{H\beta}}$ from \citet{Woo2015} and the uncertainty of the size-luminosity relation normalization and power index from \citet{Bentz2013}.}}. {We derive an intrinsic scatter of $0.40\pm0.02$ dex. An inverse regression fit gives consistent results.} Assuming that the mass based on the UV and optical lines should be the same, we require a normalization factor (i.e., virial factor) of the CIV/CIII] VPs of $\log\,f_{\mathrm{CIV/CIII]}}=0.87$ (f=7.45). This value is roughly consistent (within $<2\sigma$) with the value derived by \citet{Park2013} for the case of a fixed size-luminosity power index of 0.53 and a velocity power index of 2 (see Table 3 of \citealt{Park2013}). The derived \textit{f} may be affected by the systematic H$\alpha$ line width underestimation described in Section \ref{sec:mc}. However, we do not expect this to significantly impact our result since \textit{f} is derived based on the combined dataset.

{The uncertainty of f$_{\mathrm{CIV/CIII]}}$ is difficult to assess properly. The formal statistical error is very small (0.03) and thus is not the dominant source of uncertainty. \citet{ParkD2012}, by calibrating the virial factor based on the M$_{\mathrm{BH}}$-$\sigma_{*}$ relation, showed that there are differences of the order of 0.26 dex between different estimations of the virial factor, mainly due to sample selection effects. Additionally, differences in terms of forward and inverse linear regressions to the data were shown to lead to up to 0.2 dex differences in the estimated virial factor. As discussed previously, the H$\beta$ virial factor itself has systematic uncertainties of $\sim0.12$ dex (e.g., \citealt{Woo2013}). Combining these in quadrature results in a systematic uncertainty of $\delta f_{\mathrm{CIV/CIII]}}\sim0.4$ dex.} 


As a consistency check, we compare our results with those for the local low-luminosity reverberation sample from the updated analysis of \citet{Park2013} in Fig. \ref{fig:CIVCIIIHaHb_local}. We find a remarkable agreement, a linear regression fit to the combined sample giving a zero offset and a scatter of $0.38\pm0.05$ dex (for a fixed slope of 1). If we allow the slope to vary, we obtain a slightly sub-linear relation ($0.9\pm0.1$) that is, however, consistent with unity.\footnote{The plotted CIV BH mass estimates from \citet{Park2013} are calculated based on the virial factor and CIV BH mass relation proposed in that paper.}

\section{Comparison with Other Studies}
\label{sec:discuss}
\subsection{MgII versus H$\alpha$ and H$\beta$}
Locally, the most comprehensive sample of single epoch BH mass estimates was provided by the SDSS sample (\citealt{Shen2011,JGWang2009,Rafiee2011}). \citet{Shen2012} found a scatter of $0.25$ dex between the two BH mass estimators, using the recipes from \citet{Vestergaard2006}, consistent with the results shown in Fig. \ref{fig:MgIIHa_VP_lit}. At intermediate redshifts, \citet{McGill2008} presented a comprehensive comparison between MgII and the Balmer line virial mass estimators. The authors found an intrinsic scatter of $\sim0.24$ dex when using the MgII FWHM and the Balmer line luminosities, also consistent with our results.	 

\citet{Shen2012} used intermediate redshift ($1.5<z<2$) SDSS quasars to compare MgII and H$\beta$ virial masses. This yielded an intrinsic scatter of 0.16 dex again with a negligible offset. \citet{Matsuoka2013} provided the same comparison for a lower redshift ($0.5<z<1.6$) and lower luminosity (see the lower panel of Fig. \ref{fig:MgIIHa_VP_lit}) sample of Type 1 AGN. The authors found an intrinsic scatter of $\sim0.3$ dex with an offset of 0.17 dex, comparable to the one observed for the combined sample shown in Fig. \ref{fig:MgIIHa_VP_lit}. At even higher redshifts, \citet{Zuo2015} compared the two emission lines, finding an intrinsic scatter of $\sim0.3$ dex when using the FWHM.

Our data fit very well with previous similar studies of the MgII and {Balmer} BH mass estimates and extend the comparison of the two lines by at least half an order of magnitude in luminosity and roughly an order of magnitude in BH mass. 

\subsection{CIV and CIII] versus H$\alpha$ and H$\beta$}

In an initial comparison with the local H$\beta$ reverberation sample, \citet{Vestergaard2006} found an intrinsic scatter of 0.33 to 0.36 dex for the single epoch CIV BH mass calibration. Using a method of emission line fitting consistent with our study, \citet{Park2013} presented an updated comparison for the local Universe. They obtained intrinsic scatters of 0.29 and 0.35 dex for the $\sigma$ and FWHM-based virial BH mass estimates, respectively. These values are lower than the scatter we find in Fig. \ref{fig:cor_virCIVCIIIHaHb} and Table \ref{tab:corVP}, due to the difference in data quality. However, these local results are consistent with the values derived from the combined high redshift sample (shown in Fig. \ref{fig:CIVCIIIHaHb_lit}). Furthermore, our direct comparison of our measurements to those of \citet{Park2013} in Fig. \ref{fig:CIVCIIIHaHb_local} reveals no significant offsets or slope differences between the two. 

An intrinsic scatter of 0.18 dex was found by \citet{Assef2011} after correcting for the color dependency of the BH mass comparison residuals, corresponding to a factor of 2 improvement from the comparison of the uncorrected values. These values are significantly smaller than the ones we derive here, which most probably is a result of the low S/N of many of our carbon line measurements. \citet{Shen2012} found an ``irreducible'' scatter between the color-corrected CIV and CIII] FWHM to H$\beta$ FWHM of 0.13 and 0.15 dex, but did not provide a measurement of the resulting scatter in the VP or BH estimates. More recently, a number of studies demonstrated that both the S/N of the spectra, and a careful treatment of the non-virial component in the CIV emission profile can result in a substantial improvement in the agreement between CIV and H$\beta$ BH mass estimates (e.g., \citealt{Denney2013,Runnoe2013,Park2013}), resulting in an intrinsic scatter of $\lesssim0.3$ dex, similar to our results. 

The CIII] emission line has not been studied in depth in terms of its suitability as a virial estimator. \citet{Shen2012} found a strong correlation between the CIII] and CIV FWHM albeit with significant scatter and with CIII] FWHM measurements suffering from larger uncertainties due to the difficulty in properly deblending the CIII] emission complex. The authors also found a mild correlation between CIII] and H$\beta$ FWHM, with an intrinsic scatter of 0.15 dex, but did not provide BH mass estimates. Here we showed that by correcting for the luminosity ratio trends, our CIII] VPs are in broad agreement with both the CIV and Balmer VPs (Fig. \ref{fig:cor_virCIVCIIIHaHb}). We expect that higher S/N spectra would result in an even better agreement, since they would allow for a better treatment of possible non-virial components in the CIII] and CIV emission profiles. 

\section{Conclusions}
\label{sec:conc}

We performed near-infrared spectroscopy on a sample of low-luminosity Type 1 AGN at intermediate redshift. We measured the properties of  the rest-frame optical Balmer emission lines, H$\alpha$ and H$\beta$, and compared them to their rest-frame UV emission lines, including MgII, CIV, and CIII]. The main findings are summarized below:
\begin{itemize}

\item Based on detailed MC simulations for constraining measurement errors as well as systematic uncertainties induced by the emission line fitting method, we find that the exclusion of the H$\alpha$ narrow component does not significantly affect line width measurements, particularly when low resolution spectra are used (Fig. \ref{fig:sim_model}).

\item We find good agreement between MgII and H$\alpha$ and H$\beta$ VPs, with FWHM-based VPs showing slightly lower scatter and slopes closer to one than $\sigma$-based VPs. We extend previous high-redshift comparisons to lower BH masses, finding a scatter of $0.31\pm0.03$
(Fig. \ref{fig:MgIIHa_VP_lit}).

\item We find a strong dependence of the residual between Balmer and carbon VPs on the UV-to-optical continuum color. As previously found by \citet{Assef2011}, much of the scatter between Balmer and carbon VPs is due to the choice of luminosities (UV vs. optical) rather than any peculiarities of the carbon lines (Fig. \ref{fig:UVOPTcomp}).

\item By extending the comparison between Balmer and carbon VPs to lower BH mass scales, we find a good agreement between the two over $\sim3$ orders of magnitude in dynamical range. The scatter and intercept  of the comparison are $0.37\pm0.02$ 
(Fig. \ref{fig:CIVCIIIHaHb_lit}). The comparison with the local low luminosity AGN with reverberation measurements shows a good consistency with a negligible offset and intrinsic scatter $0.38\pm0.05$ dex (Fig. \ref{fig:CIVCIIIHaHb_local}).

\item Using the well calibrated virial factor for H$\beta$ BH masses, we derive a virial factor for CIV/CIII] BH mass estimates, as {$\log\,f_{\mathrm{CIV/CIII]}}=0.87\pm0.4$ (f = 7.45).}
\end{itemize}

By extending  the redshift and luminosity range, our comparisons between the two sets of lines (UV vs. Balmer) show good agreement with previous studies. We conclude that while both CIV and CIII] show larger scatter than MgII in comparison with the Balmer lines, they are viable virial BH mass estimators with a factor $\sim2$ uncertainty without a systematic offset. The derived virial factor for carbon line based VPs will be useful for black hole mass estimates for high-z AGN, although higher S/N data are necessary to further explore potential non-virial components in the CIV and CIII] emission lines for
more reliable  UV virial mass estimators. 

\acknowledgments{This research was supported by the National Research Foundation of Korea (NRF) grant funded by the Korea government (MEST) (No. 2010-0027910). We thank the referee for valuable comments, which improved the clarity of the manuscript. We thank Naoyuki Tamura for assistance in preparing the FMOS observations. We thank Roberto Assef for help in target selection. We acknowledge Phuong Thi Kim for her contribution to this study. Based [in part] on data collected at Subaru Telescope, which is operated by the National Astronomical Observatory of Japan. This research has made use of NASA's Astrophysics Data System Bibliographic Services. This research has made use of the VizieR catalogue access tool, CDS, Strasbourg, France. For this research, we have made extensive use of the TOPCAT software (\citealt{Taylor2005}), which is part of the suite of Virtual Observatory tools.}

\bibliographystyle{aa}
\bibliography{bibtex}

\begin{thebibliography}{81}
\expandafter\ifx\csname natexlab\endcsname\relax\def\natexlab#1{#1}\fi

\bibitem[{{Angl{\'e}s-Alc{\'a}zar} {et~al.}(2013){Angl{\'e}s-Alc{\'a}zar},
  {{\"O}zel}, \& {Dav{\'e}}}]{Angles2013}
{Angl{\'e}s-Alc{\'a}zar}, D., {{\"O}zel}, F., \& {Dav{\'e}}, R. 2013, \apj,
  770, 5

\bibitem[{{Assef} {et~al.}(2011){Assef}, {Denney}, {Kochanek}, {Peterson},
  {Koz{\l}owski}, {Ageorges}, {Barrows}, {Buschkamp}, {Dietrich}, {Falco},
  {Feiz}, {Gemperlein}, {Germeroth}, {Grier}, {Hofmann}, {Juette}, {Khan},
  {Kilic}, {Knierim}, {Laun}, {Lederer}, {Lehmitz}, {Lenzen}, {Mall}, {Madsen},
  {Mandel}, {Martini}, {Mathur}, {Mogren}, {Mueller}, {Naranjo}, {Pasquali},
  {Polsterer}, {Pogge}, {Quirrenbach}, {Seifert}, {Stern}, {Shappee}, {Storz},
  {Van Saders}, {Weiser}, \& {Zhang}}]{Assef2011}
{Assef}, R.~J., {Denney}, K.~D., {Kochanek}, C.~S., {et~al.} 2011, \apj, 742,
  93

\bibitem[{{Bahcall} {et~al.}(1972){Bahcall}, {Kozlovsky}, \&
  {Salpeter}}]{Bahcall1972}
{Bahcall}, J.~N., {Kozlovsky}, B.-Z., \& {Salpeter}, E.~E. 1972, \apj, 171, 467

\bibitem[{{Baskin} \& {Laor}(2005)}]{Baskin2005}
{Baskin}, A. \& {Laor}, A. 2005, \mnras, 356, 1029

\bibitem[{{Bennert} {et~al.}(2010){Bennert}, {Treu}, {Woo}, {Malkan}, {Le
  Bris}, {Auger}, {Gallagher}, \& {Blandford}}]{Bennert2010}
{Bennert}, V.~N., {Treu}, T., {Woo}, J.-H., {et~al.} 2010, \apj, 708, 1507

\bibitem[{{Bentz} {et~al.}(2013){Bentz}, {Denney}, {Grier}, {Barth},
  {Peterson}, {Vestergaard}, {Bennert}, {Canalizo}, {De Rosa}, {Filippenko},
  {Gates}, {Greene}, {Li}, {Malkan}, {Pogge}, {Stern}, {Treu}, \&
  {Woo}}]{Bentz2013}
{Bentz}, M.~C., {Denney}, K.~D., {Grier}, C.~J., {et~al.} 2013, \apj, 767, 149

\bibitem[{{Busch} {et~al.}(2014){Busch}, {Zuther}, {Valencia-S.}, {Moser},
  {Fischer}, {Eckart}, {Scharw{\"a}chter}, {Gadotti}, \&
  {Wisotzki}}]{Busch2014}
{Busch}, G., {Zuther}, J., {Valencia-S.}, M., {et~al.} 2014, \aap, 561, A140

\bibitem[{{Canalizo} \& {Stockton}(2013)}]{Canalizo2013}
{Canalizo}, G. \& {Stockton}, A. 2013, \apj, 772, 132

\bibitem[{{Cappellari}(2002)}]{Cappellari2002}
{Cappellari}, M. 2002, \mnras, 333, 400

\bibitem[{{Ciotti} \& {Ostriker}(2007)}]{Ciotti2007}
{Ciotti}, L. \& {Ostriker}, J.~P. 2007, \apj, 665, 1038

\bibitem[{{Croton} {et~al.}(2006){Croton}, {Springel}, {White}, {De Lucia},
  {Frenk}, {Gao}, {Jenkins}, {Kauffmann}, {Navarro}, \& {Yoshida}}]{Croton2006}
{Croton}, D.~J., {Springel}, V., {White}, S.~D.~M., {et~al.} 2006, \mnras, 365,
  11

\bibitem[{{Davis} {et~al.}(2007){Davis}, {Woo}, \& {Blaes}}]{Davis2007}
{Davis}, S.~W., {Woo}, J.-H., \& {Blaes}, O.~M. 2007, \apj, 668, 682

\bibitem[{{DeGraf} {et~al.}(2014){DeGraf}, {Di Matteo}, {Treu}, {Feng}, {Woo},
  \& {Park}}]{DeGraf2014}
{DeGraf}, C., {Di Matteo}, T., {Treu}, T., {et~al.} 2014, ArXiv e-prints

\bibitem[{{Denney}(2012)}]{Denney2012}
{Denney}, K.~D. 2012, \apj, 759, 44

\bibitem[{{Denney} {et~al.}(2009){Denney}, {Peterson}, {Dietrich},
  {Vestergaard}, \& {Bentz}}]{Denney2009}
{Denney}, K.~D., {Peterson}, B.~M., {Dietrich}, M., {Vestergaard}, M., \&
  {Bentz}, M.~C. 2009, \apj, 692, 246

\bibitem[{{Denney} {et~al.}(2013){Denney}, {Pogge}, {Assef}, {Kochanek},
  {Peterson}, \& {Vestergaard}}]{Denney2013}
{Denney}, K.~D., {Pogge}, R.~W., {Assef}, R.~J., {et~al.} 2013, \apj, 775, 60

\bibitem[{{Dietrich} {et~al.}(2009){Dietrich}, {Mathur}, {Grupe}, \&
  {Komossa}}]{Dietrich2009}
{Dietrich}, M., {Mathur}, S., {Grupe}, D., \& {Komossa}, S. 2009, \apj, 696,
  1998

\bibitem[{{Fabricant} {et~al.}(2005){Fabricant}, {Fata}, {Roll}, {Hertz},
  {Caldwell}, {Gauron}, {Geary}, {McLeod}, {Szentgyorgyi}, {Zajac}, {Kurtz},
  {Barberis}, {Bergner}, {Brown}, {Conroy}, {Eng}, {Geller}, {Goddard},
  {Honsa}, {Mueller}, {Mink}, {Ordway}, {Tokarz}, {Woods}, {Wyatt}, {Epps}, \&
  {Dell'Antonio}}]{Fabricant2005}
{Fabricant}, D., {Fata}, R., {Roll}, J., {et~al.} 2005, \pasp, 117, 1411

\bibitem[{{Fabricant} {et~al.}(1998){Fabricant}, {Hertz}, {Szentgyorgyi},
  {Fata}, {Roll}, \& {Zajac}}]{Fabricant1998}
{Fabricant}, D.~G., {Hertz}, E.~N., {Szentgyorgyi}, A.~H., {et~al.} 1998, in
  Society of Photo-Optical Instrumentation Engineers (SPIE) Conference Series,
  Vol. 3355, Optical Astronomical Instrumentation, ed. S.~{D'Odorico}, 285--296

\bibitem[{{Feng} {et~al.}(2014){Feng}, {Shen}, \& {Li}}]{Feng2014}
{Feng}, H., {Shen}, Y., \& {Li}, H. 2014, \apj, 794, 77

\bibitem[{{Ferrarese} \& {Merritt}(2000)}]{Ferrarese2000}
{Ferrarese}, L. \& {Merritt}, D. 2000, \apjl, 539, L9

\bibitem[{{Fine} {et~al.}(2010){Fine}, {Croom}, {Bland-Hawthorn}, {Pimbblet},
  {Ross}, {Schneider}, \& {Shanks}}]{Fine2010}
{Fine}, S., {Croom}, S.~M., {Bland-Hawthorn}, J., {et~al.} 2010, \mnras, 409,
  591

\bibitem[{{Gebhardt} {et~al.}(2000){Gebhardt}, {Bender}, {Bower}, {Dressler},
  {Faber}, {Filippenko}, {Green}, {Grillmair}, {Ho}, {Kormendy}, {Lauer},
  {Magorrian}, {Pinkney}, {Richstone}, \& {Tremaine}}]{Gebhardt2000}
{Gebhardt}, K., {Bender}, R., {Bower}, G., {et~al.} 2000, \apjl, 539, L13

\bibitem[{{Greene} \& {Ho}(2005)}]{Greene2005}
{Greene}, J.~E. \& {Ho}, L.~C. 2005, \apj, 630, 122

\bibitem[{{Ho} {et~al.}(2012){Ho}, {Goldoni}, {Dong}, {Greene}, \&
  {Ponti}}]{Ho2012}
{Ho}, L.~C., {Goldoni}, P., {Dong}, X.-B., {Greene}, J.~E., \& {Ponti}, G.
  2012, \apj, 754, 11

\bibitem[{{Iwamuro} {et~al.}(2012){Iwamuro}, {Moritani}, {Yabe}, {Sumiyoshi},
  {Kawate}, {Tamura}, {Akiyama}, {Kimura}, {Takato}, {Tait}, {Ohta}, {Totani},
  {Suzuki}, \& {Tonegawa}}]{Iwamuro2012}
{Iwamuro}, F., {Moritani}, Y., {Yabe}, K., {et~al.} 2012, \pasj, 64, 59

\bibitem[{{Jahnke} {et~al.}(2009){Jahnke}, {Bongiorno}, {Brusa}, {Capak},
  {Cappelluti}, {Cisternas}, {Civano}, {Colbert}, {Comastri}, {Elvis},
  {Hasinger}, {Ilbert}, {Impey}, {Inskip}, {Koekemoer}, {Lilly}, {Maier},
  {Merloni}, {Riechers}, {Salvato}, {Schinnerer}, {Scoville}, {Silverman},
  {Taniguchi}, {Trump}, \& {Yan}}]{Jahnke2009}
{Jahnke}, K., {Bongiorno}, A., {Brusa}, M., {et~al.} 2009, \apjl, 706, L215

\bibitem[{{Jannuzi} \& {Dey}(1999)}]{Jannuzi1999}
{Jannuzi}, B.~T. \& {Dey}, A. 1999, in Astronomical Society of the Pacific
  Conference Series, Vol. 191, Photometric Redshifts and the Detection of High
  Redshift Galaxies, ed. R.~{Weymann}, L.~{Storrie-Lombardi}, M.~{Sawicki}, \&
  R.~{Brunner}, 111

\bibitem[{{Jun} {et~al.}(2015){Jun}, {Im}, {Lee}, {Ohyama}, {Woo}, {Fan},
  {Goto}, {Kim}, {Kim}, {Kim}, {Lee}, {Nakagawa}, {Pearson}, \&
  {Serjeant}}]{Jun2015}
{Jun}, H.~D., {Im}, M., {Lee}, H.~M., {et~al.} 2015, ArXiv e-prints

\bibitem[{{Kaspi} {et~al.}(2000){Kaspi}, {Smith}, {Netzer}, {Maoz}, {Jannuzi},
  \& {Giveon}}]{Kaspi2000}
{Kaspi}, S., {Smith}, P.~S., {Netzer}, H., {et~al.} 2000, \apj, 533, 631

\bibitem[{{Kauffmann} \& {Haehnelt}(2000)}]{Kauffmann2000}
{Kauffmann}, G. \& {Haehnelt}, M. 2000, \mnras, 311, 576

\bibitem[{{Kelly} \& {Bechtold}(2007)}]{Kelly2007}
{Kelly}, B.~C. \& {Bechtold}, J. 2007, \apjs, 168, 1

\bibitem[{{Kelly} \& {Shen}(2013)}]{Kelly2013}
{Kelly}, B.~C. \& {Shen}, Y. 2013, \apj, 764, 45

\bibitem[{{Kimura} {et~al.}(2010){Kimura}, {Maihara}, {Iwamuro}, {Akiyama},
  {Tamura}, {Dalton}, {Takato}, {Tait}, {Ohta}, {Eto}, {Mochida}, {Elms},
  {Kawate}, {Kurakami}, {Moritani}, {Noumaru}, {Ohshima}, {Sumiyoshi}, {Yabe},
  {Brzeski}, {Farrell}, {Frost}, {Gillingham}, {Haynes}, {Moore}, {Muller},
  {Smedley}, {Smith}, {Bonfield}, {Brooks}, {Holmes}, {Curtis Lake}, {Lee},
  {Lewis}, {Froud}, {Tosh}, {Woodhouse}, {Blackburn}, {Content}, {Dipper},
  {Murray}, {Sharples}, \& {Robertson}}]{Kimura2010}
{Kimura}, M., {Maihara}, T., {Iwamuro}, F., {et~al.} 2010, \pasj, 62, 1135

\bibitem[{{Kochanek} {et~al.}(2012){Kochanek}, {Eisenstein}, {Cool},
  {Caldwell}, {Assef}, {Jannuzi}, {Jones}, {Murray}, {Forman}, {Dey}, {Brown},
  {Eisenhardt}, {Gonzalez}, {Green}, \& {Stern}}]{Kochanek2012}
{Kochanek}, C.~S., {Eisenstein}, D.~J., {Cool}, R.~J., {et~al.} 2012, \apjs,
  200, 8

\bibitem[{{Kollmeier} {et~al.}(2006){Kollmeier}, {Onken}, {Kochanek}, {Gould},
  {Weinberg}, {Dietrich}, {Cool}, {Dey}, {Eisenstein}, {Jannuzi}, {Le Floc'h},
  \& {Stern}}]{Kollmeier2006}
{Kollmeier}, J.~A., {Onken}, C.~A., {Kochanek}, C.~S., {et~al.} 2006, \apj,
  648, 128

\bibitem[{{Komatsu} {et~al.}(2011){Komatsu}, {Smith}, {Dunkley}, {Bennett},
  {Gold}, {Hinshaw}, {Jarosik}, {Larson}, {Nolta}, {Page}, {Spergel},
  {Halpern}, {Hill}, {Kogut}, {Limon}, {Meyer}, {Odegard}, {Tucker}, {Weiland},
  {Wollack}, \& {Wright}}]{Komatsu2011}
{Komatsu}, E., {Smith}, K.~M., {Dunkley}, J., {et~al.} 2011, \apjs, 192, 18

\bibitem[{{Kormendy} \& {Ho}(2013)}]{Kormendy2013}
{Kormendy}, J. \& {Ho}, L.~C. 2013, \araa, 51, 511

\bibitem[{{Markwardt}(2009)}]{Markwardt2009}
{Markwardt}, C.~B. 2009, in Astronomical Society of the Pacific Conference
  Series, Vol. 411, Astronomical Data Analysis Software and Systems XVIII, ed.
  D.~A. {Bohlender}, D.~{Durand}, \& P.~{Dowler}, 251

\bibitem[{Marquardt(1963)}]{Marquardt1963}
Marquardt, D.~W. 1963, Journal of the Society for Industrial and Applied
  Mathematics, 11, 431

\bibitem[{{Marziani} \& {Sulentic}(2012)}]{Marziani2012}
{Marziani}, P. \& {Sulentic}, J.~W. 2012, \nar, 56, 49

\bibitem[{{Marziani} {et~al.}(2013){Marziani}, {Sulentic}, {Plauchu-Frayn}, \&
  {del Olmo}}]{Marziani2013}
{Marziani}, P., {Sulentic}, J.~W., {Plauchu-Frayn}, I., \& {del Olmo}, A. 2013,
  \aap, 555, A89

\bibitem[{{Matsuoka} {et~al.}(2013){Matsuoka}, {Silverman}, {Schramm},
  {Steinhardt}, {Nagao}, {Kartaltepe}, {Sanders}, {Treister}, {Hasinger},
  {Akiyama}, {Ohta}, {Ueda}, {Bongiorno}, {Brandt}, {Brusa}, {Capak}, {Civano},
  {Comastri}, {Elvis}, {Lilly}, {Mainieri}, {Masters}, {Mignoli}, {Salvato},
  {Trump}, {Taniguchi}, {Zamorani}, {Alexander}, \&
  {Schawinski}}]{Matsuoka2013}
{Matsuoka}, K., {Silverman}, J.~D., {Schramm}, M., {et~al.} 2013, \apj, 771, 64

\bibitem[{{McGill} {et~al.}(2008){McGill}, {Woo}, {Treu}, \&
  {Malkan}}]{McGill2008}
{McGill}, K.~L., {Woo}, J.-H., {Treu}, T., \& {Malkan}, M.~A. 2008, \apj, 673,
  703

\bibitem[{{McLure} \& {Dunlop}(2002)}]{McLure2002}
{McLure}, R.~J. \& {Dunlop}, J.~S. 2002, \mnras, 331, 795

\bibitem[{{Merloni} {et~al.}(2010){Merloni}, {Bongiorno}, {Bolzonella},
  {Brusa}, {Civano}, {Comastri}, {Elvis}, {Fiore}, {Gilli}, {Hao}, {Jahnke},
  {Koekemoer}, {Lusso}, {Mainieri}, {Mignoli}, {Miyaji}, {Renzini}, {Salvato},
  {Silverman}, {Trump}, {Vignali}, {Zamorani}, {Capak}, {Lilly}, {Sanders},
  {Taniguchi}, {Bardelli}, {Carollo}, {Caputi}, {Contini}, {Coppa}, {Cucciati},
  {de la Torre}, {de Ravel}, {Franzetti}, {Garilli}, {Hasinger}, {Impey},
  {Iovino}, {Iwasawa}, {Kampczyk}, {Kneib}, {Knobel}, {Kova{\v c}},
  {Lamareille}, {Le Borgne}, {Le Brun}, {Le F{\`e}vre}, {Maier}, {Pello},
  {Peng}, {Perez Montero}, {Ricciardelli}, {Scodeggio}, {Tanaka}, {Tasca},
  {Tresse}, {Vergani}, \& {Zucca}}]{Merloni2010}
{Merloni}, A., {Bongiorno}, A., {Bolzonella}, M., {et~al.} 2010, \apj, 708, 137

\bibitem[{{Metzroth} {et~al.}(2006){Metzroth}, {Onken}, \&
  {Peterson}}]{Metzroth2006}
{Metzroth}, K.~G., {Onken}, C.~A., \& {Peterson}, B.~M. 2006, \apj, 647, 901

\bibitem[{Mor\'e(1978)}]{More1978}
Mor\'e, J. 1978, in Lecture Notes in Mathematics, Vol. 630, Numerical Analysis,
  ed. G.~Watson (Springer Berlin Heidelberg), 105--116

\bibitem[{{Nagao} {et~al.}(2006){Nagao}, {Marconi}, \& {Maiolino}}]{Nagao2006}
{Nagao}, T., {Marconi}, A., \& {Maiolino}, R. 2006, \aap, 447, 157

\bibitem[{{Netzer} {et~al.}(2007){Netzer}, {Lira}, {Trakhtenbrot}, {Shemmer},
  \& {Cury}}]{Netzer2007b}
{Netzer}, H., {Lira}, P., {Trakhtenbrot}, B., {Shemmer}, O., \& {Cury}, I.
  2007, \apj, 671, 1256

\bibitem[{{Park} {et~al.}(2012){Park}, {Kelly}, {Woo}, \& {Treu}}]{ParkD2012}
{Park}, D., {Kelly}, B.~C., {Woo}, J.-H., \& {Treu}, T. 2012, \apjs, 203, 6

\bibitem[{{Park} {et~al.}(2015){Park}, {Woo}, {Bennert}, {Treu}, {Auger}, \&
  {Malkan}}]{Park2015}
{Park}, D., {Woo}, J.-H., {Bennert}, V.~N., {et~al.} 2015, \apj, 799, 164

\bibitem[{{Park} {et~al.}(2013){Park}, {Woo}, {Denney}, \& {Shin}}]{Park2013}
{Park}, D., {Woo}, J.-H., {Denney}, K.~D., \& {Shin}, J. 2013, \apj, 770, 87

\bibitem[{{Peng} {et~al.}(2006){Peng}, {Impey}, {Rix}, {Kochanek}, {Keeton},
  {Falco}, {Leh{\'a}r}, \& {McLeod}}]{Peng2006}
{Peng}, C.~Y., {Impey}, C.~D., {Rix}, H.-W., {et~al.} 2006, \apj, 649, 616

\bibitem[{{Peterson}(1993)}]{Peterson1993}
{Peterson}, B.~M. 1993, \pasp, 105, 247

\bibitem[{{Peterson} \& {Wandel}(1999)}]{Peterson1999}
{Peterson}, B.~M. \& {Wandel}, A. 1999, \apjl, 521, L95

\bibitem[{{Rafiee} \& {Hall}(2011)}]{Rafiee2011}
{Rafiee}, A. \& {Hall}, P.~B. 2011, \apjs, 194, 42

\bibitem[{{Robertson} {et~al.}(2006){Robertson}, {Hernquist}, {Cox}, {Di
  Matteo}, {Hopkins}, {Martini}, \& {Springel}}]{Robertson2006}
{Robertson}, B., {Hernquist}, L., {Cox}, T.~J., {et~al.} 2006, \apj, 641, 90

\bibitem[{{Roll} {et~al.}(1998){Roll}, {Fabricant}, \& {McLeod}}]{Roll1998}
{Roll}, J.~B., {Fabricant}, D.~G., \& {McLeod}, B.~A. 1998, in Society of
  Photo-Optical Instrumentation Engineers (SPIE) Conference Series, Vol. 3355,
  Optical Astronomical Instrumentation, ed. S.~{D'Odorico}, 324--332

\bibitem[{{Runnoe} {et~al.}(2013){Runnoe}, {Brotherton}, {Shang}, \&
  {DiPompeo}}]{Runnoe2013}
{Runnoe}, J.~C., {Brotherton}, M.~S., {Shang}, Z., \& {DiPompeo}, M.~A. 2013,
  \mnras, 434, 848

\bibitem[{{Schramm} \& {Silverman}(2013)}]{Schramm2013}
{Schramm}, M. \& {Silverman}, J.~D. 2013, \apj, 767, 13

\bibitem[{{Shen} {et~al.}(2008){Shen}, {Greene}, {Strauss}, {Richards}, \&
  {Schneider}}]{Shen2008}
{Shen}, Y., {Greene}, J.~E., {Strauss}, M.~A., {Richards}, G.~T., \&
  {Schneider}, D.~P. 2008, \apj, 680, 169

\bibitem[{{Shen} \& {Liu}(2012)}]{Shen2012}
{Shen}, Y. \& {Liu}, X. 2012, \apj, 753, 125

\bibitem[{{Shen} {et~al.}(2011){Shen}, {Richards}, {Strauss}, {Hall},
  {Schneider}, {Snedden}, {Bizyaev}, {Brewington}, {Malanushenko},
  {Malanushenko}, {Oravetz}, {Pan}, \& {Simmons}}]{Shen2011}
{Shen}, Y., {Richards}, G.~T., {Strauss}, M.~A., {et~al.} 2011, \apjs, 194, 45

\bibitem[{{Sluse} {et~al.}(2011){Sluse}, {Schmidt}, {Courbin},
  {Hutsem{\'e}kers}, {Meylan}, {Eigenbrod}, {Anguita}, {Agol}, \&
  {Wambsganss}}]{Sluse2011}
{Sluse}, D., {Schmidt}, R., {Courbin}, F., {et~al.} 2011, \aap, 528, A100

\bibitem[{{Taylor}(2005)}]{Taylor2005}
{Taylor}, M.~B. 2005, in Astronomical Society of the Pacific Conference Series,
  Vol. 347, Astronomical Data Analysis Software and Systems XIV, ed.
  P.~{Shopbell}, M.~{Britton}, \& R.~{Ebert}, 29

\bibitem[{{Tremaine} {et~al.}(2002){Tremaine}, {Gebhardt}, {Bender}, {Bower},
  {Dressler}, {Faber}, {Filippenko}, {Green}, {Grillmair}, {Ho}, {Kormendy},
  {Lauer}, {Magorrian}, {Pinkney}, \& {Richstone}}]{Tremaine2002}
{Tremaine}, S., {Gebhardt}, K., {Bender}, R., {et~al.} 2002, \apj, 574, 740

\bibitem[{{Tsuzuki} {et~al.}(2006){Tsuzuki}, {Kawara}, {Yoshii}, {Oyabu},
  {Tanab{\'e}}, \& {Matsuoka}}]{Tsuzuki2006}
{Tsuzuki}, Y., {Kawara}, K., {Yoshii}, Y., {et~al.} 2006, \apj, 650, 57

\bibitem[{{Ulrich} {et~al.}(1997){Ulrich}, {Maraschi}, \& {Urry}}]{Ulrich1997}
{Ulrich}, M.-H., {Maraschi}, L., \& {Urry}, C.~M. 1997, \araa, 35, 445

\bibitem[{{van der Marel} \& {Franx}(1993)}]{vanderMarel1993}
{van der Marel}, R.~P. \& {Franx}, M. 1993, \apj, 407, 525

\bibitem[{{Vestergaard} \& {Osmer}(2009)}]{Vestergaard2009}
{Vestergaard}, M. \& {Osmer}, P.~S. 2009, \apj, 699, 800

\bibitem[{{Vestergaard} \& {Peterson}(2006)}]{Vestergaard2006}
{Vestergaard}, M. \& {Peterson}, B.~M. 2006, \apj, 641, 689

\bibitem[{{Wang} {et~al.}(2009{\natexlab{a}}){Wang}, {Dong}, {Wang}, {Ho},
  {Yuan}, {Wang}, {Zhang}, {Zhang}, \& {Zhou}}]{Wang2009}
{Wang}, J.-G., {Dong}, X.-B., {Wang}, T.-G., {et~al.} 2009{\natexlab{a}}, \apj,
  707, 1334

\bibitem[{{Wang} {et~al.}(2009{\natexlab{b}}){Wang}, {Dong}, {Wang}, {Ho},
  {Yuan}, {Wang}, {Zhang}, {Zhang}, \& {Zhou}}]{JGWang2009}
{Wang}, J.-G., {Dong}, X.-B., {Wang}, T.-G., {et~al.} 2009{\natexlab{b}}, \apj,
  707, 1334

\bibitem[{{Woo} {et~al.}(2013){Woo}, {Schulze}, {Park}, {Kang}, {Kim}, \&
  {Riechers}}]{Woo2013}
{Woo}, J.-H., {Schulze}, A., {Park}, D., {et~al.} 2013, \apj, 772, 49

\bibitem[{{Woo} {et~al.}(2010){Woo}, {Treu}, {Barth}, {Wright}, {Walsh},
  {Bentz}, {Martini}, {Bennert}, {Canalizo}, {Filippenko}, {Gates}, {Greene},
  {Li}, {Malkan}, {Stern}, \& {Minezaki}}]{Woo2010}
{Woo}, J.-H., {Treu}, T., {Barth}, A.~J., {et~al.} 2010, \apj, 716, 269

\bibitem[{{Woo} {et~al.}(2006){Woo}, {Treu}, {Malkan}, \&
  {Blandford}}]{Woo2006}
{Woo}, J.-H., {Treu}, T., {Malkan}, M.~A., \& {Blandford}, R.~D. 2006, \apj,
  645, 900

\bibitem[{{Woo} {et~al.}(2008){Woo}, {Treu}, {Malkan}, \&
  {Blandford}}]{Woo2008}
{Woo}, J.-H., {Treu}, T., {Malkan}, M.~A., \& {Blandford}, R.~D. 2008, \apj,
  681, 925

\bibitem[{{Woo} \& {Urry}(2002)}]{Woo2002}
{Woo}, J.-H. \& {Urry}, C.~M. 2002, \apj, 579, 530

\bibitem[{{Woo} {et~al.}(2015){Woo}, {Yoon}, {Park}, {Park}, \&
  {Kim}}]{Woo2015}
{Woo}, J.-H., {Yoon}, Y., {Park}, S., {Park}, D., \& {Kim}, S.~C. 2015, \apj,
  801, 38

\bibitem[{{Zuo} {et~al.}(2015){Zuo}, {Wu}, {Fan}, {Green}, {Wang}, \&
  {Bian}}]{Zuo2015}
{Zuo}, W., {Wu}, X.-B., {Fan}, X., {et~al.} 2015, \apj, 799, 189

\end{thebibliography}


\begin{appendix}

\section{FMOS near-IR spectra}
\label{sec:fmos_spec}
Below we present a compilation of all the FMOS spectra that are used in the analysis presented in this paper (i.e., 30 spectra, including 25 with $H\alpha$ and 8 with H$\beta$ broad emission, plus 1 spectrum showing a double-peaked H$\alpha$ profile and was therefore excluded from our analysis). Vertical dashed lines show the redshifted emission lines of interest (H$\alpha$ and H$\beta$). Redshift values for the identification of the lines come from AGES. The spectra have been smoothed with a 3 pixel boxcar filter for better visualization. Shaded areas mark the gap between the J and H bands and a small spectral gap around 12800\AA\, that suffers from high noise.
\begin{center}
\includegraphics[width=0.9\textwidth]{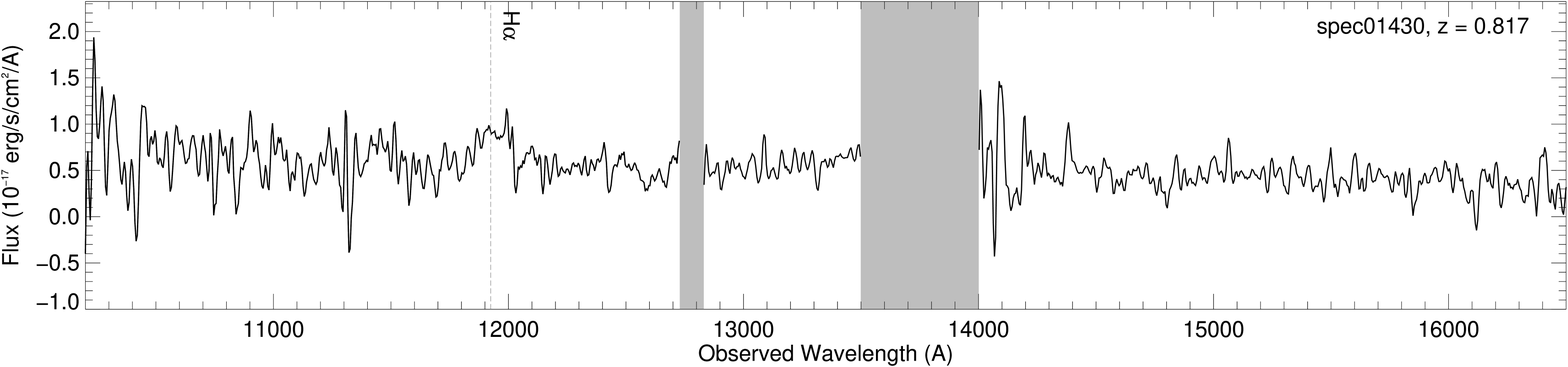}\\		
\includegraphics[width=0.9\textwidth]{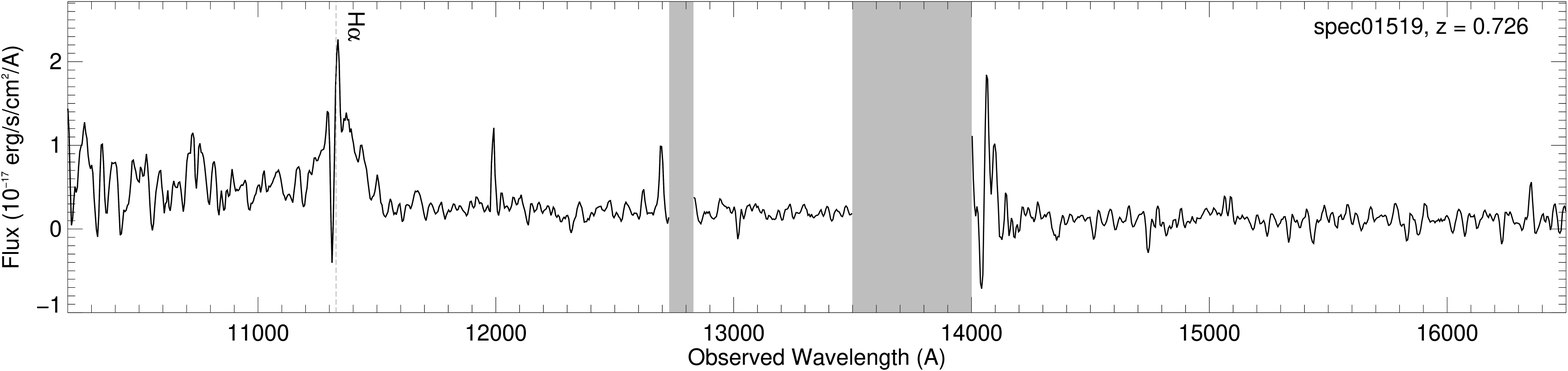}\\		
\includegraphics[width=0.9\textwidth]{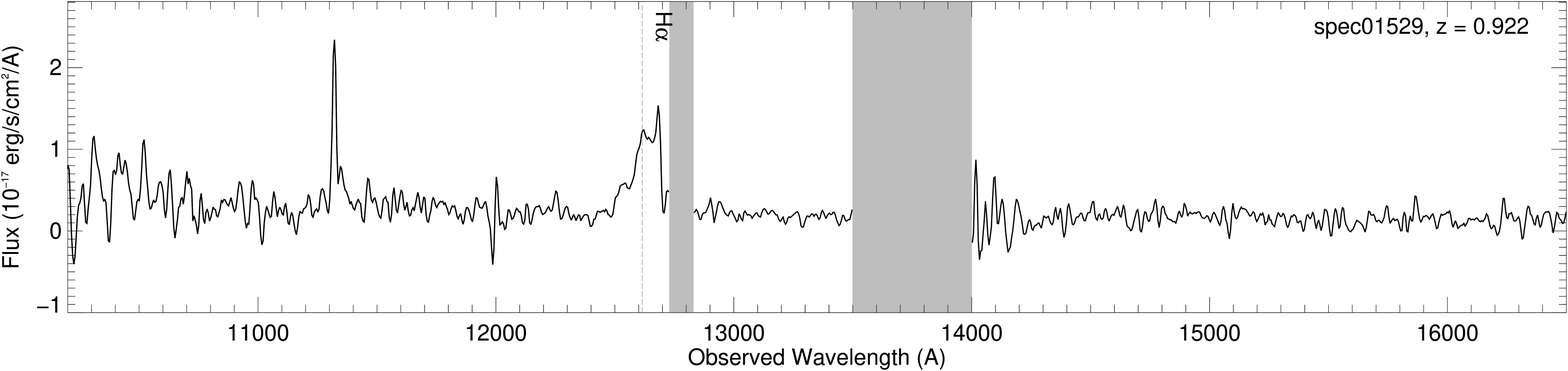}\\
\includegraphics[width=0.9\textwidth]{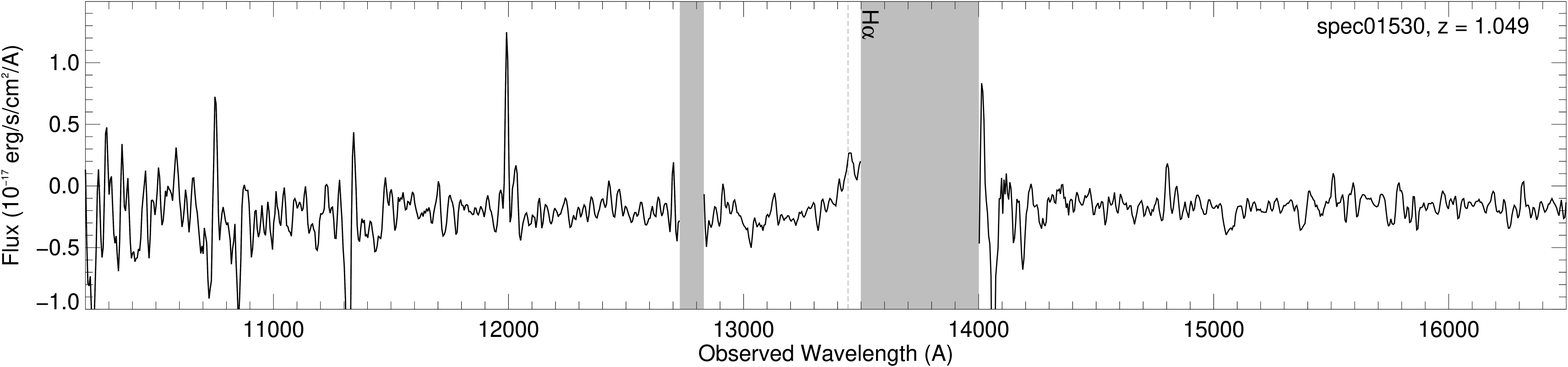}\\
\includegraphics[width=0.9\textwidth]{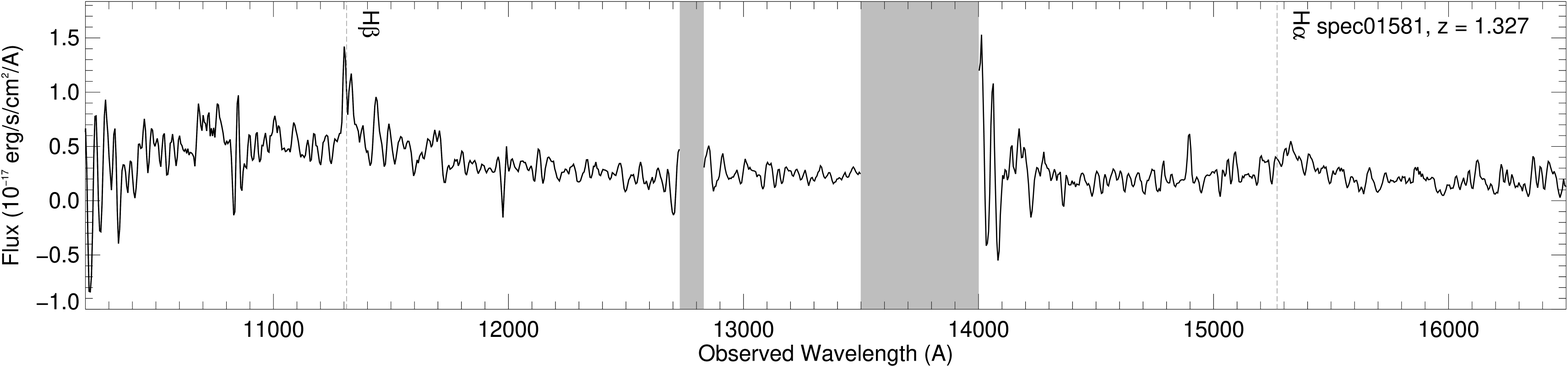}\\
\includegraphics[width=0.9\textwidth]{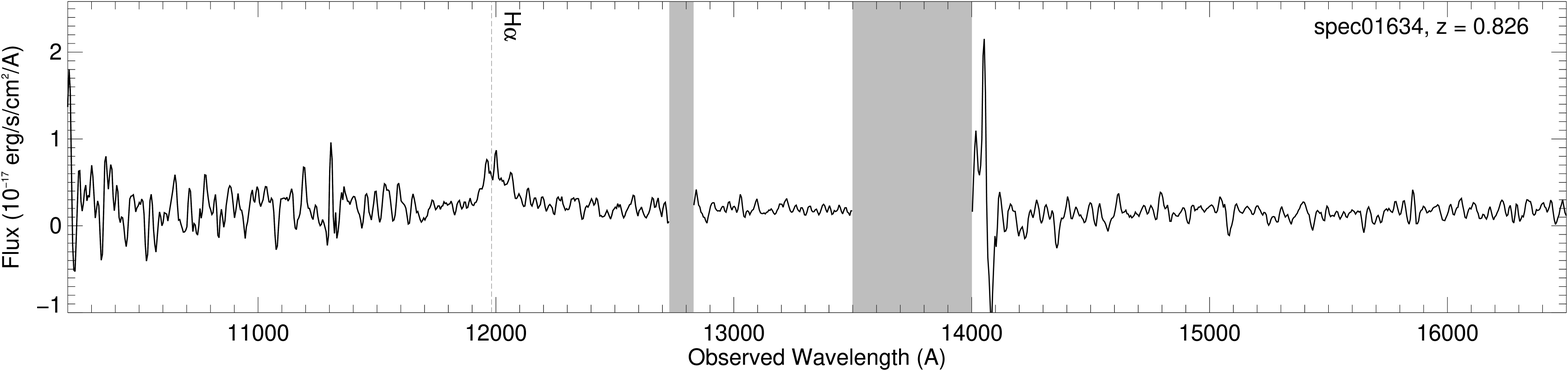}\\
\includegraphics[width=0.9\textwidth]{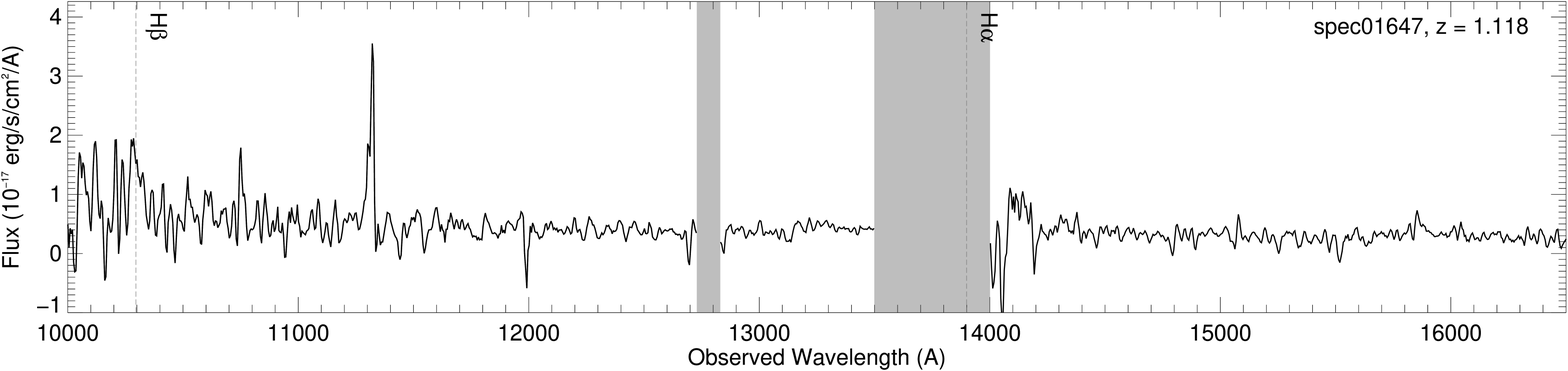}\\
\includegraphics[width=0.9\textwidth]{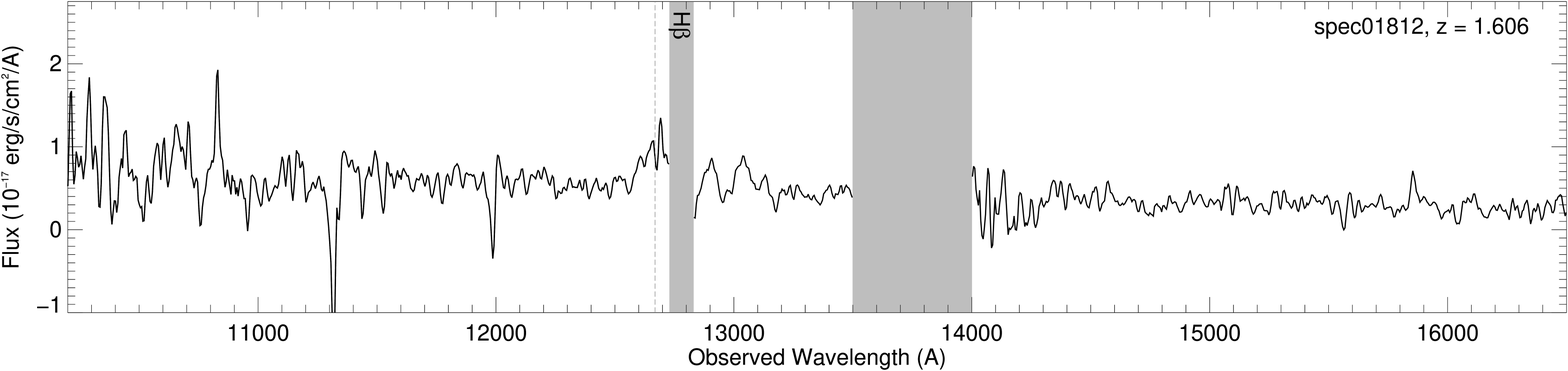}\\
\includegraphics[width=0.9\textwidth]{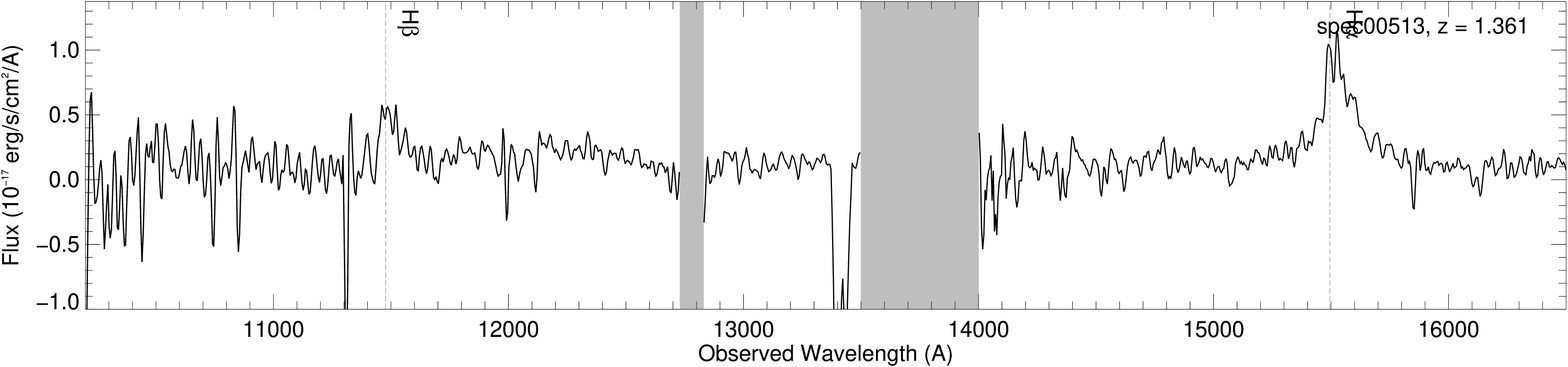}\\
\includegraphics[width=0.9\textwidth]{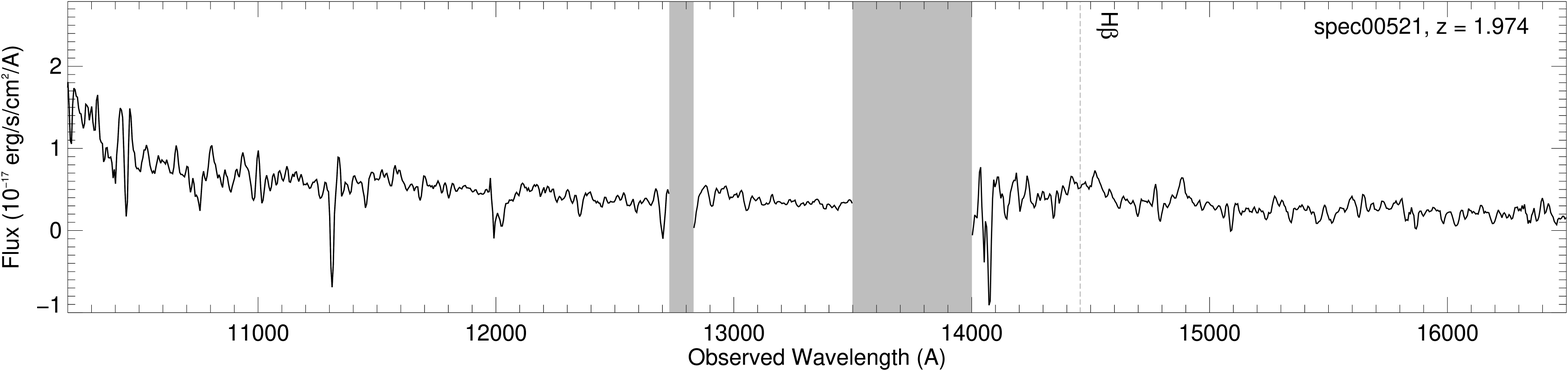}\\
\includegraphics[width=0.9\textwidth]{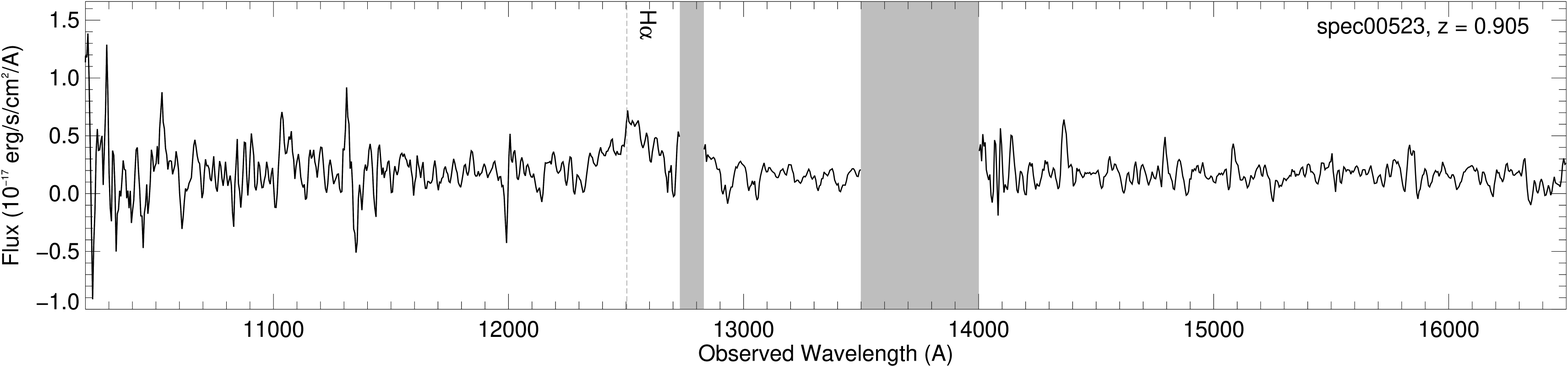}\\
\includegraphics[width=0.9\textwidth]{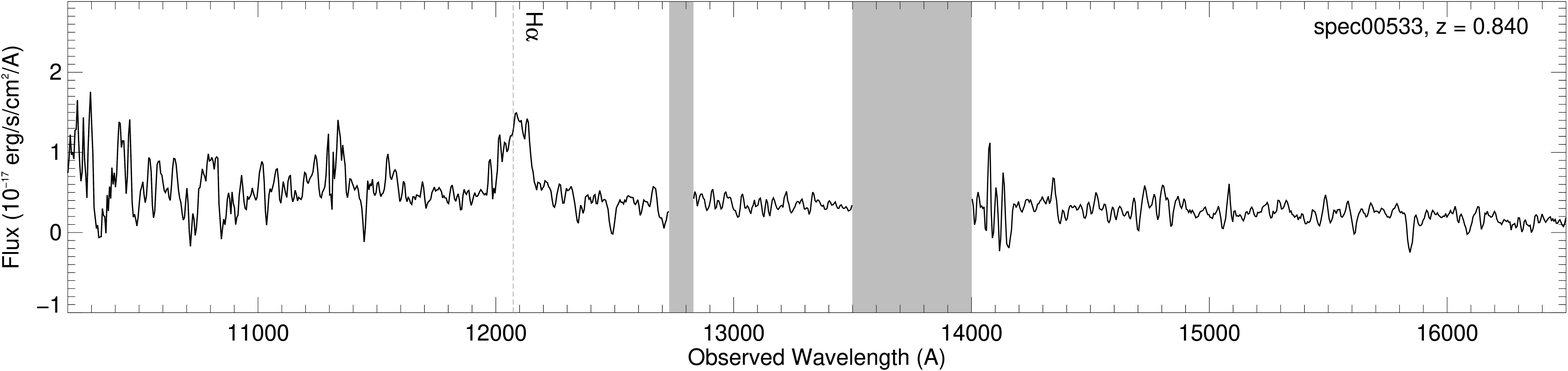}\\
\includegraphics[width=0.9\textwidth]{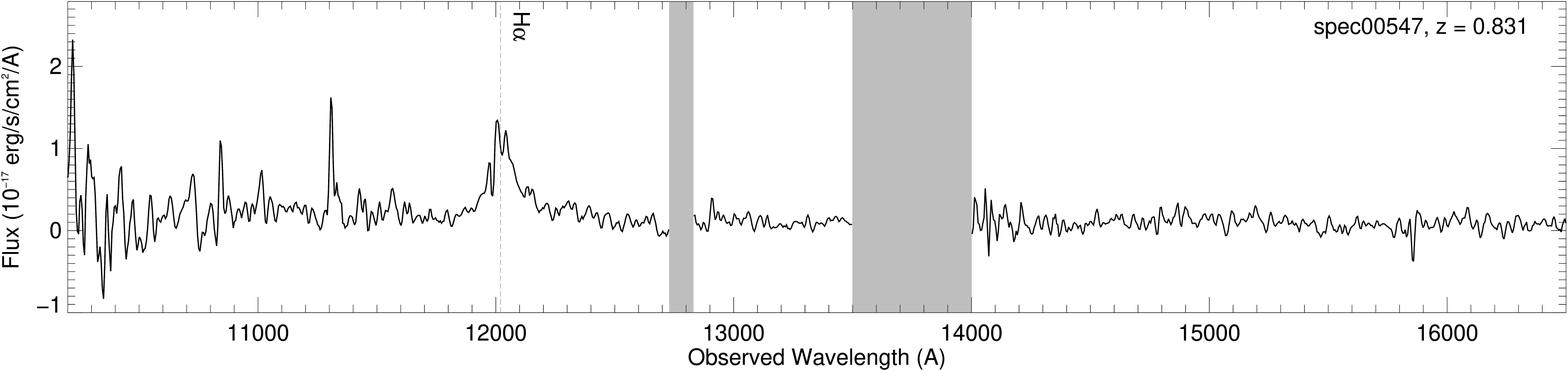}\\
\includegraphics[width=0.9\textwidth]{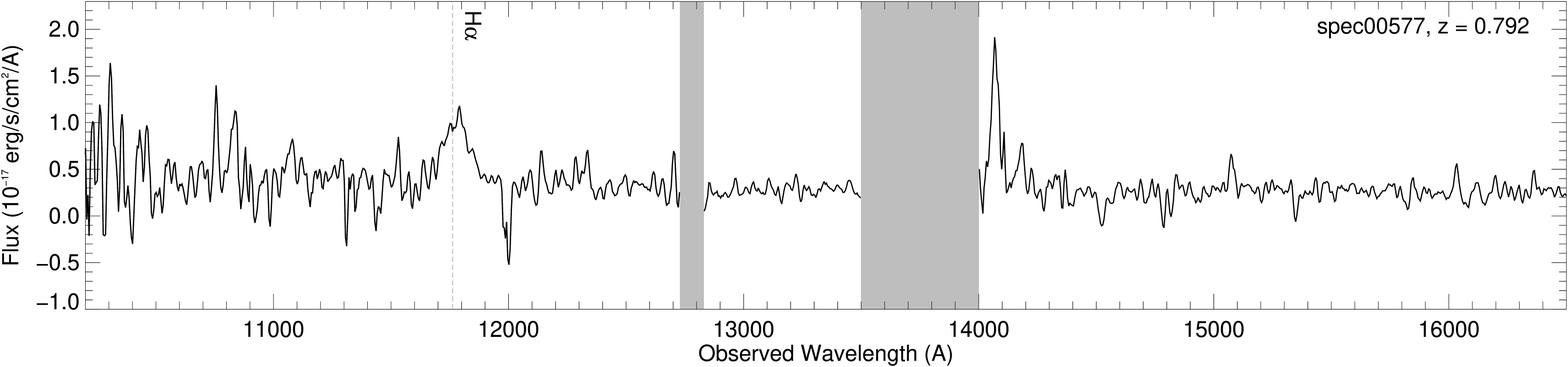}\\
\includegraphics[width=0.9\textwidth]{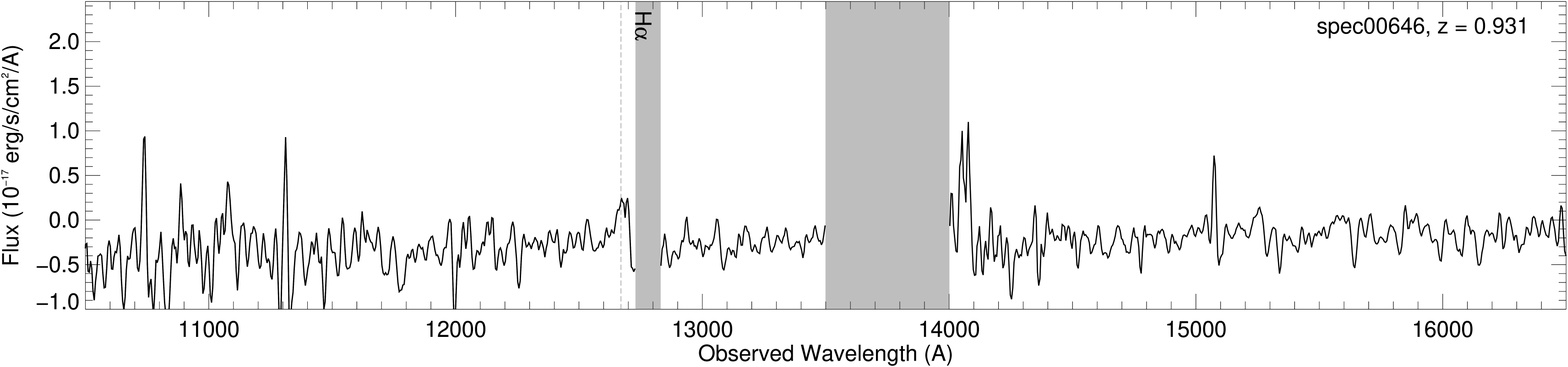}\\
\includegraphics[width=0.9\textwidth]{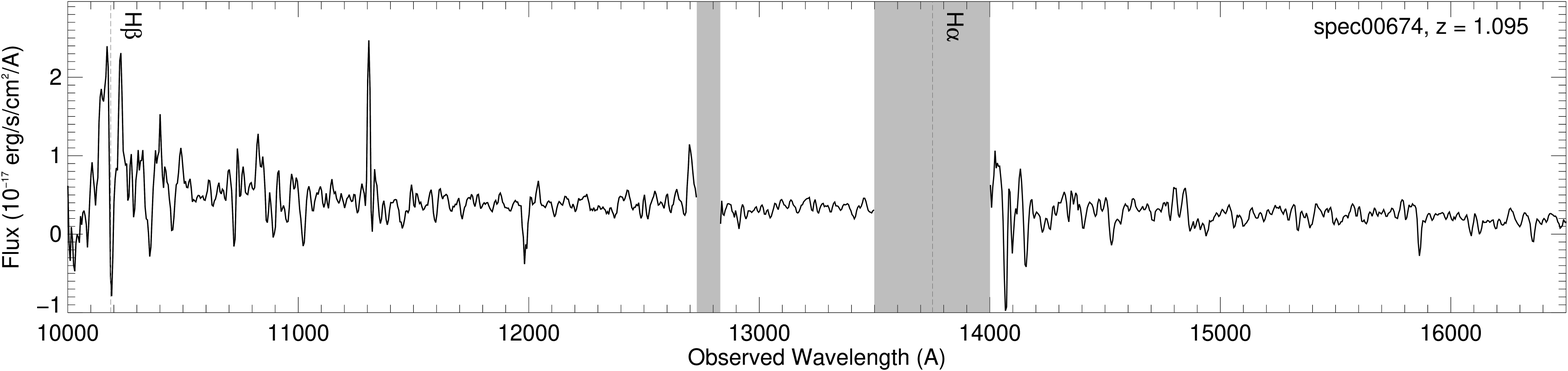}\\
\includegraphics[width=0.9\textwidth]{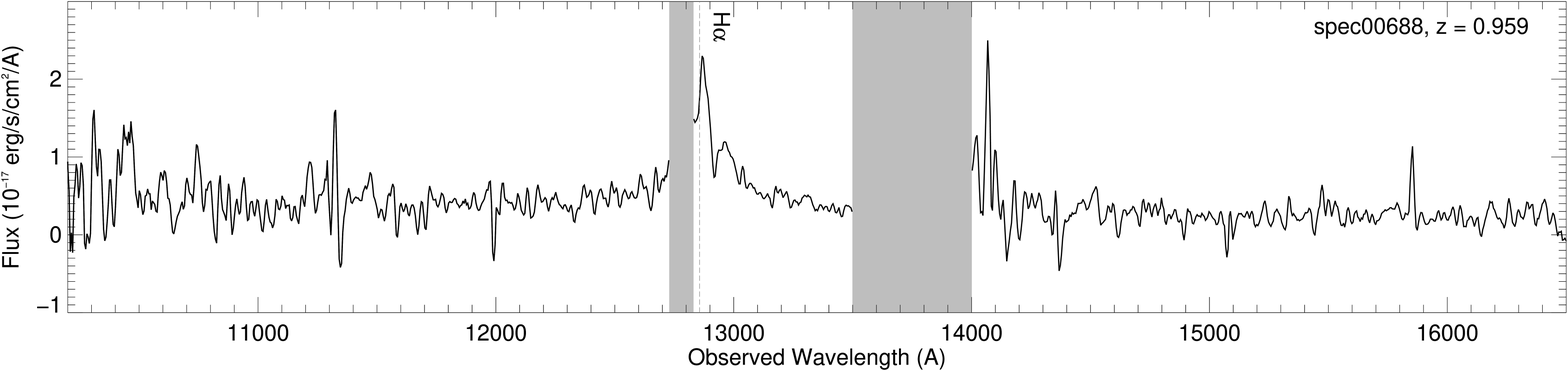}\\
\includegraphics[width=0.9\textwidth]{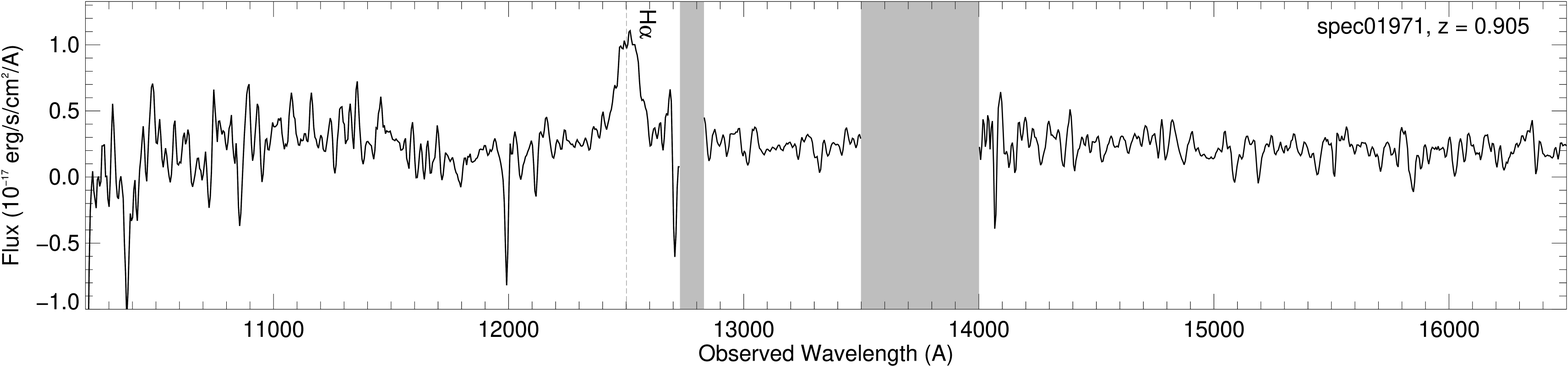}\\
\includegraphics[width=0.9\textwidth]{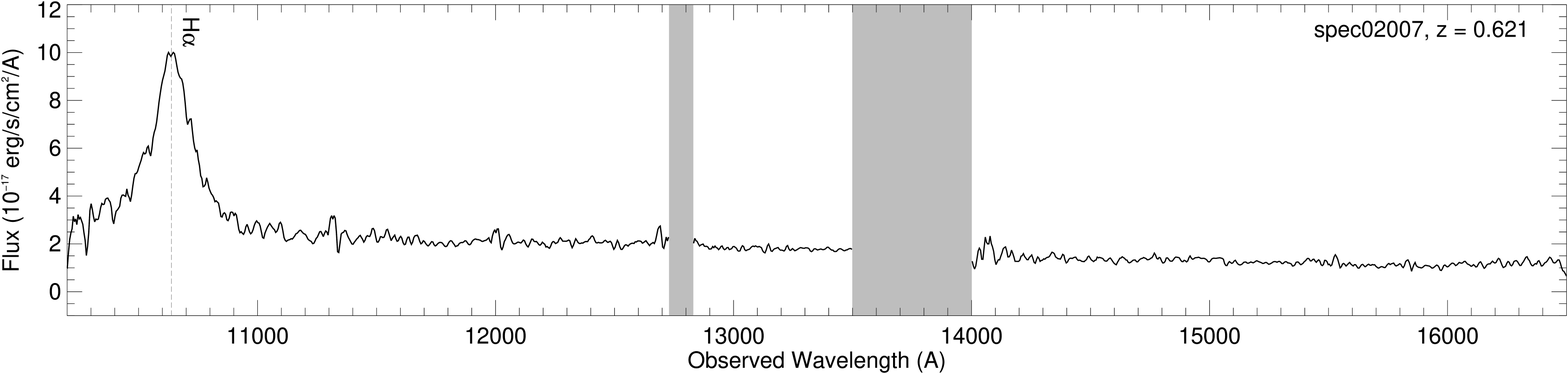}\\
\includegraphics[width=0.9\textwidth]{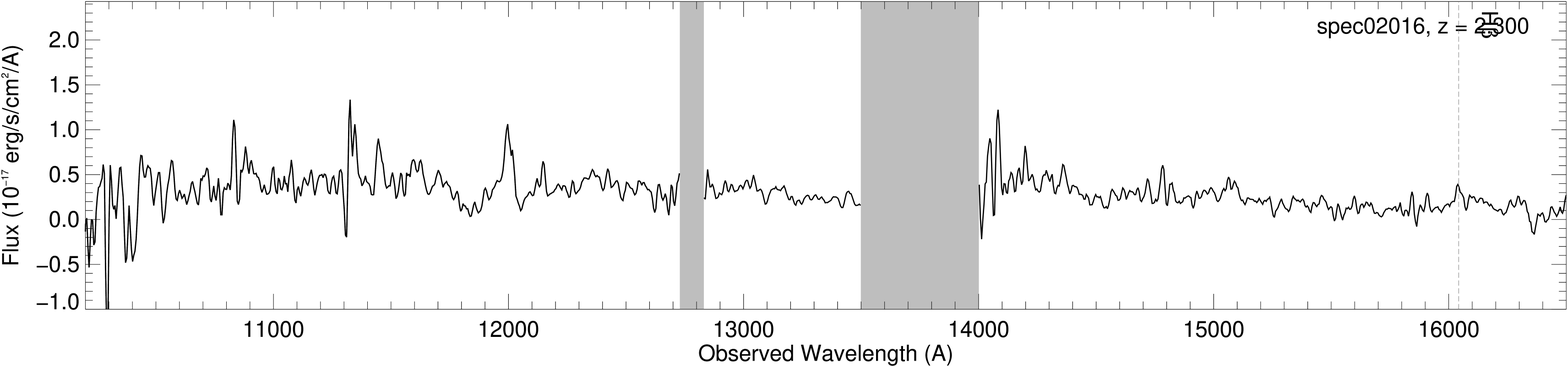}\\
\includegraphics[width=0.9\textwidth]{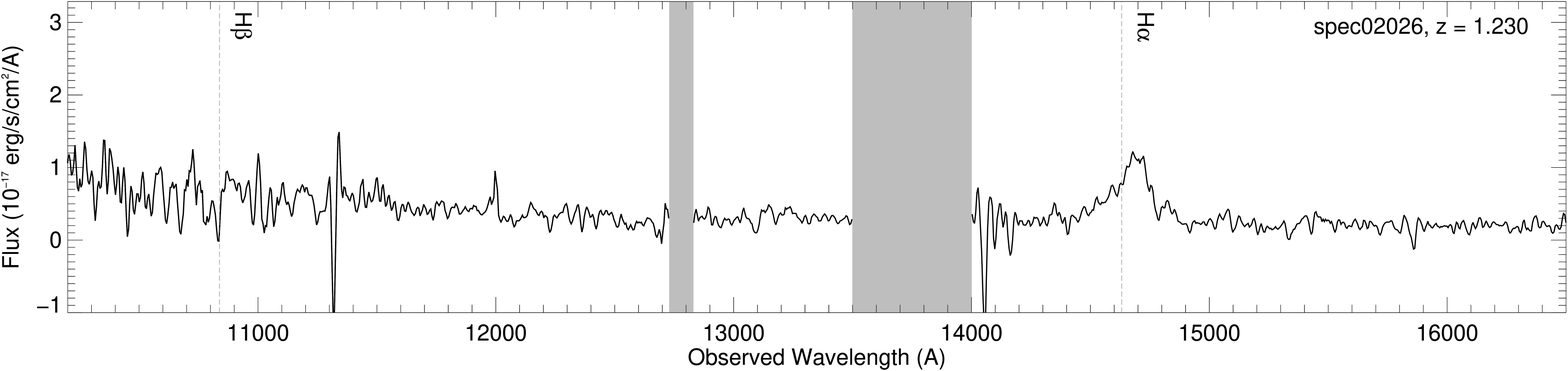}\\
\includegraphics[width=0.9\textwidth]{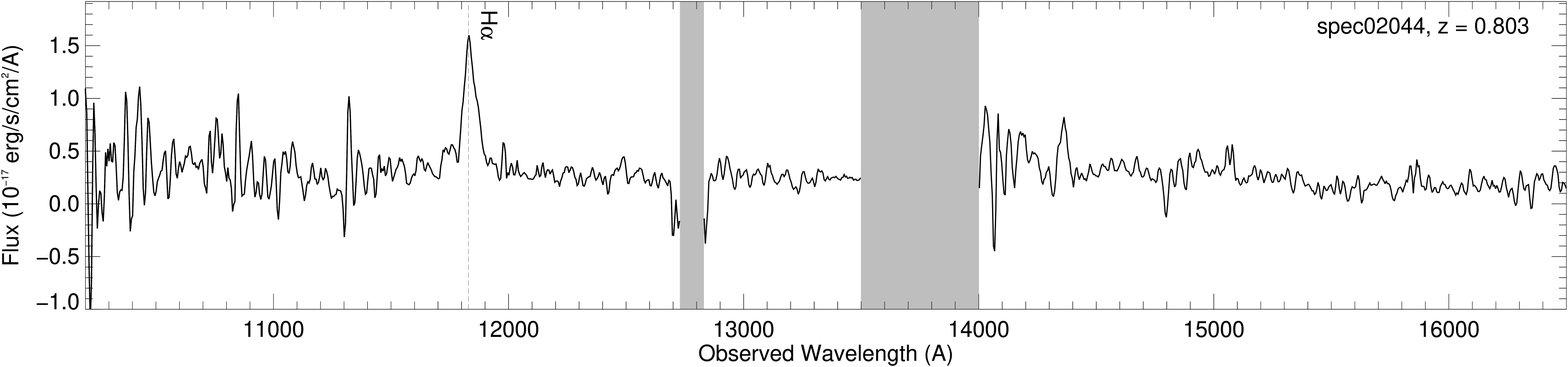}\\
\includegraphics[width=0.9\textwidth]{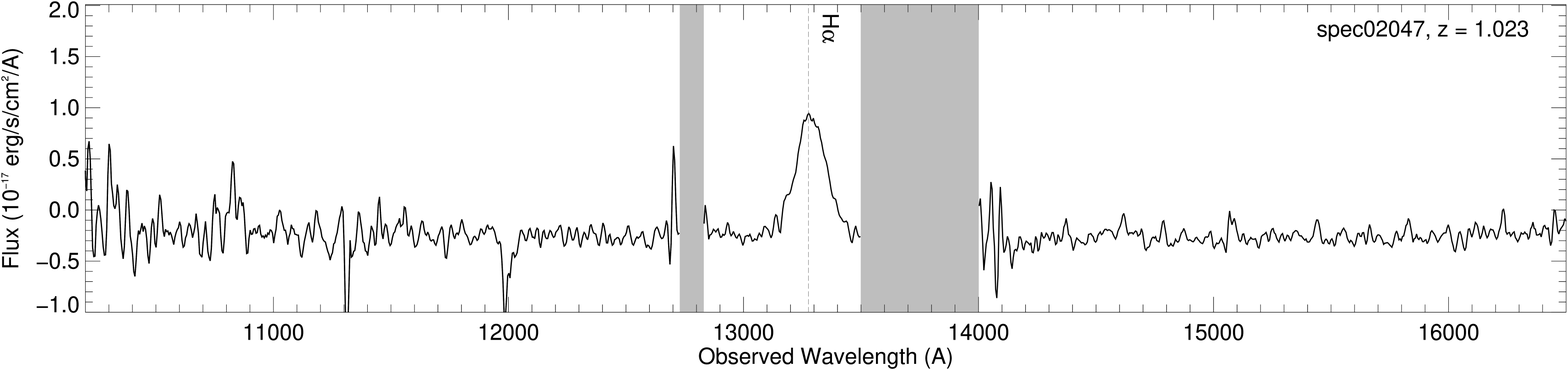}\\
\includegraphics[width=0.9\textwidth]{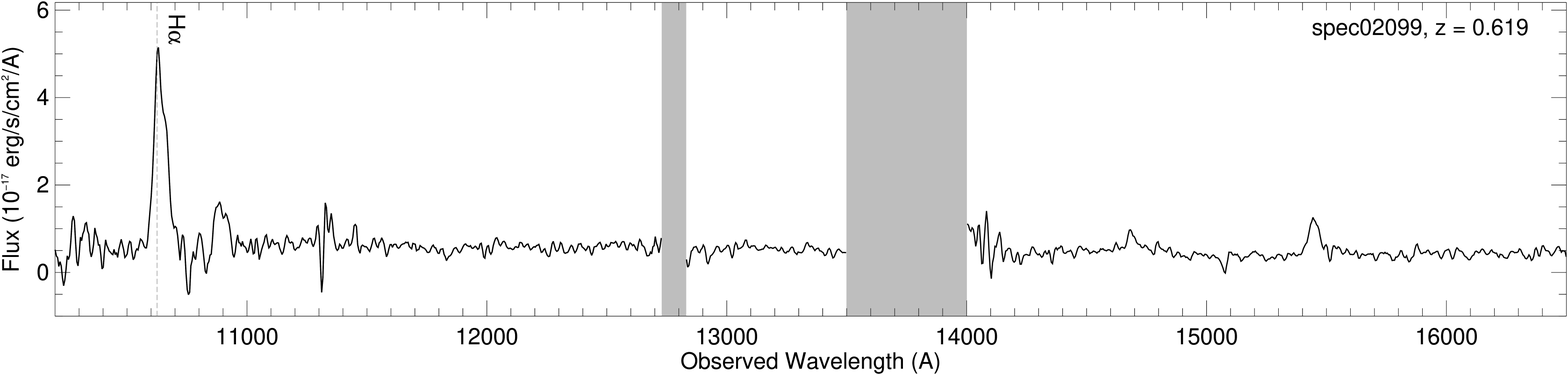}\\
\includegraphics[width=0.9\textwidth]{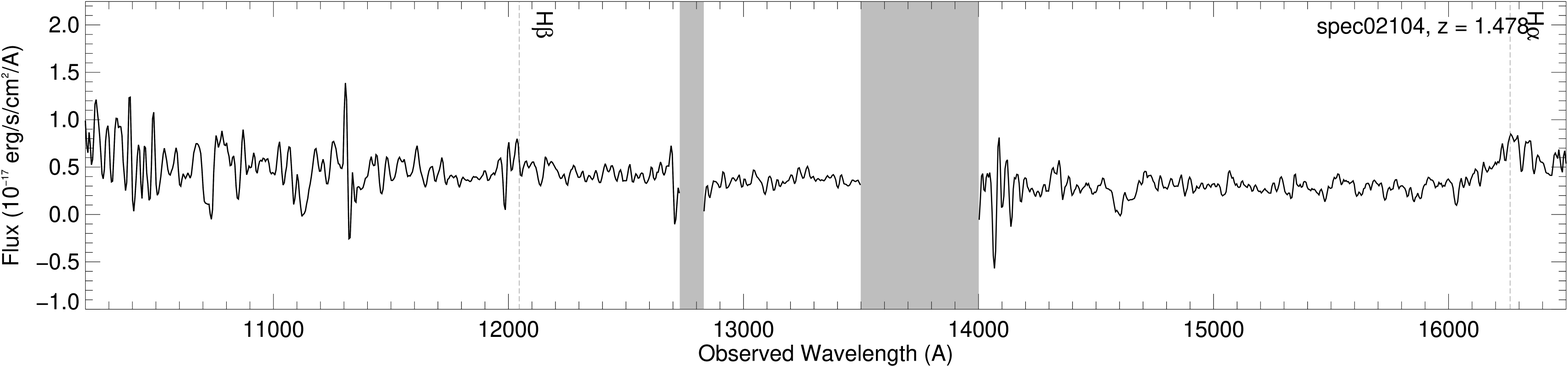}\\
\includegraphics[width=0.9\textwidth]{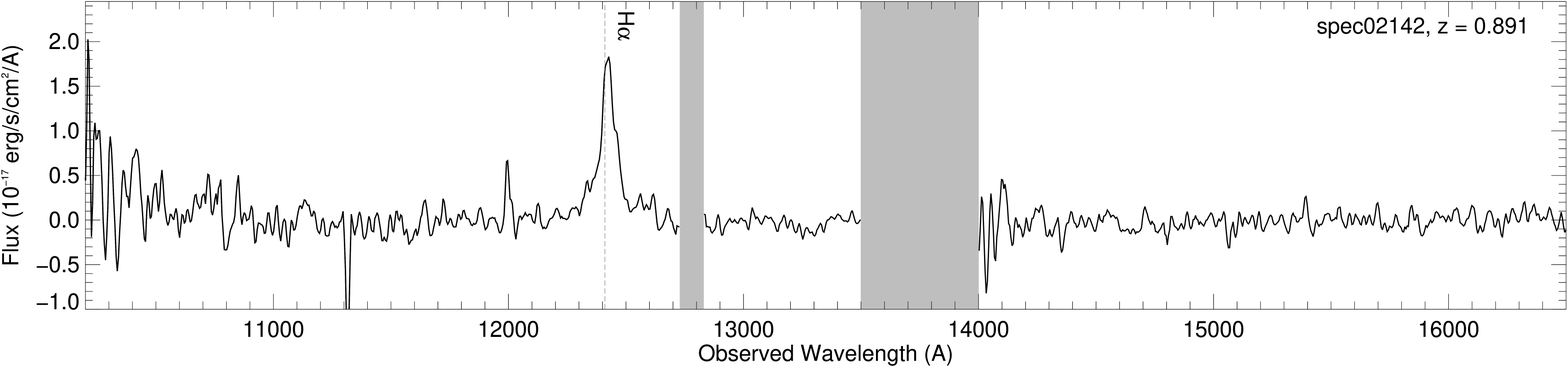}\\
\includegraphics[width=0.9\textwidth]{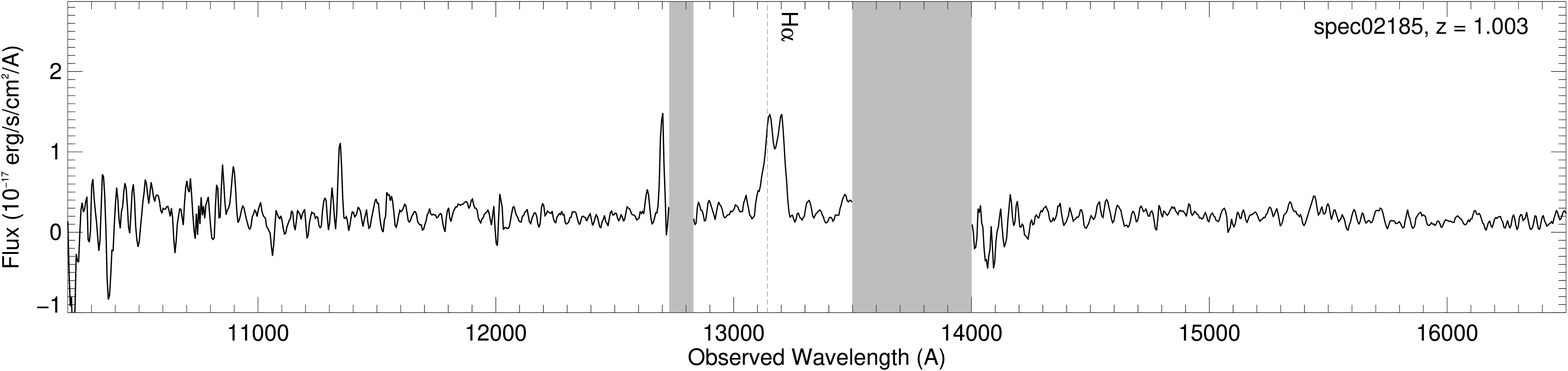}\\
\includegraphics[width=0.9\textwidth]{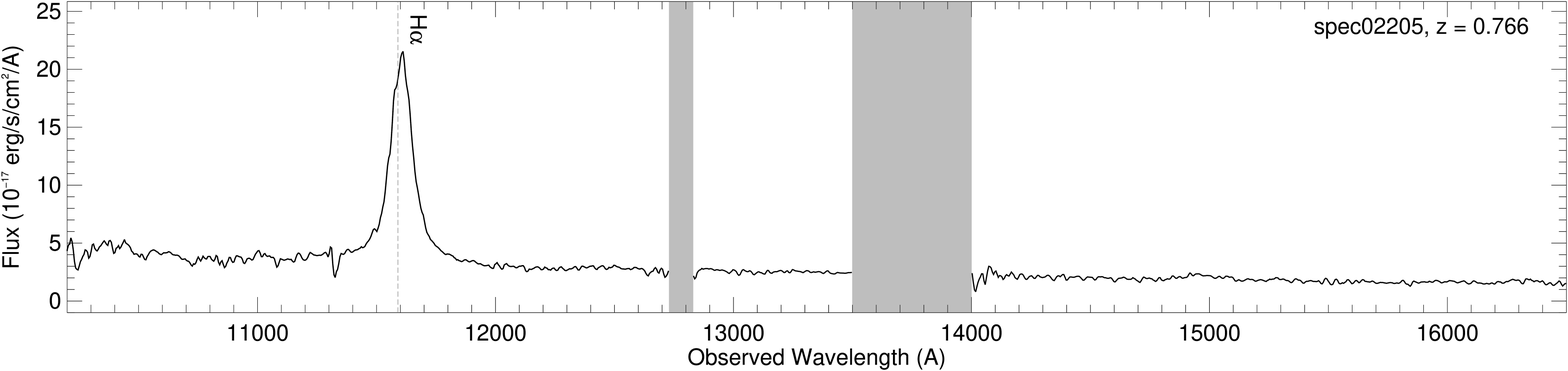}\\
\includegraphics[width=0.9\textwidth]{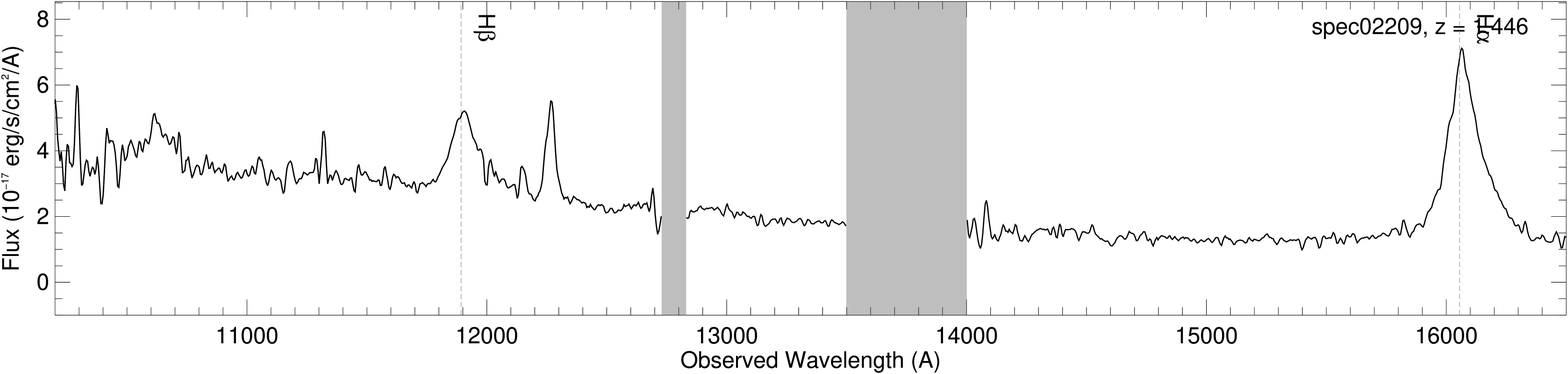}\\
\includegraphics[width=0.9\textwidth]{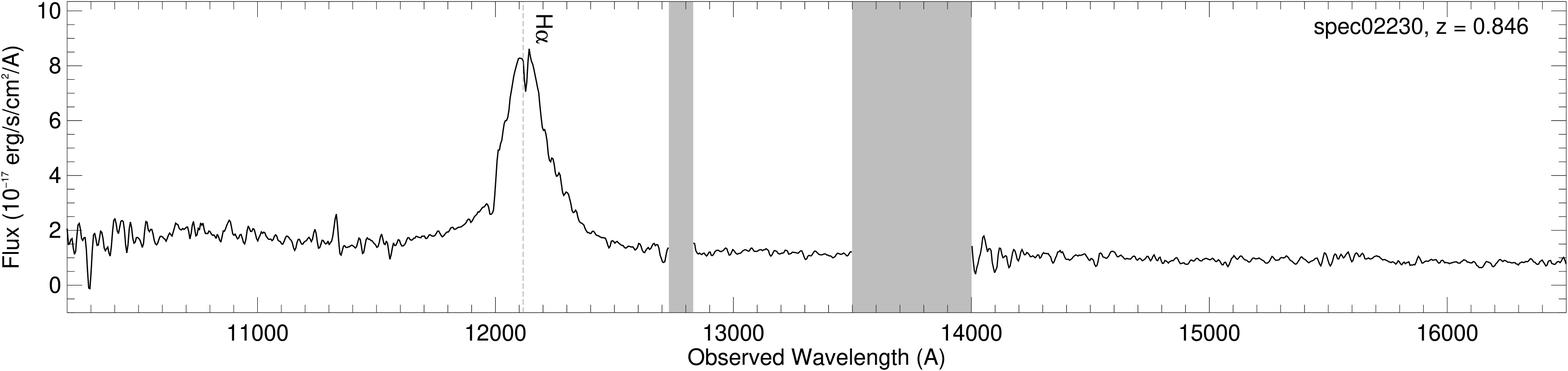}\\
\includegraphics[width=0.9\textwidth]{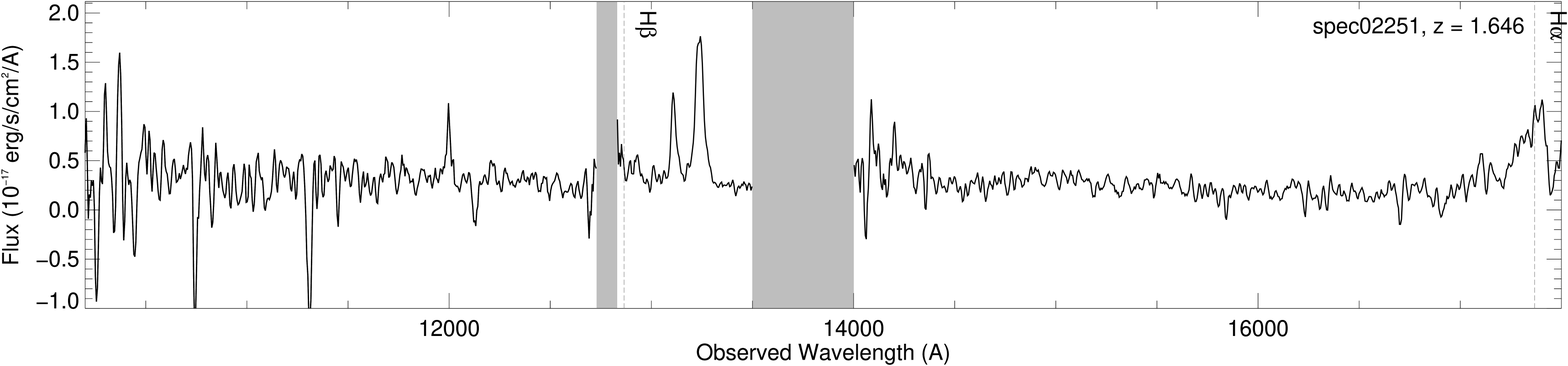}\\
\end{center}

\section{Continuum measurements}

\begin{center}
\begin{deluxetable*}{c c c c c c c}
\tablecolumns{7}
\tabletypesize{\footnotesize}
\tablewidth{0pt}
\tablecaption{Continuum and emission line luminosities in the UV and optical. \label{tab:flux}}
\tablehead{ \colhead{ID} & \colhead{$\log\lambda$ L$_{1350}$} & \colhead{$\log\lambda$ L$_{1800}$} & \colhead{$\log\lambda$ L$_{3000}$} & \colhead{$\log\lambda$ L$_{5100}$} & \colhead{$\log$L$_{H\alpha}$} & \colhead{$\log$L$_{H\beta}$}\\
\colhead{ } & \colhead{[erg s$^{-1}$]} & \colhead{[erg s$^{-1}$]} & \colhead{[erg s$^{-1}$]} & \colhead{[erg s$^{-1}$]} & \colhead{[erg s$^{-1}$]} & \colhead{[erg s$^{-1}$]}}
\startdata
spec01519 & \nodata & \nodata & $  44.27\pm   0.06$ & \nodata & $  42.66\pm   0.02$ & \nodata \\
spec01530 & $  44.24\pm   1.06$ & $  44.24\pm   1.06$ & $  44.19\pm   0.16$ & \nodata & $  42.49\pm   0.03$ & \nodata \\
spec01634 & \nodata & \nodata & $  46.12\pm   0.10$ & \nodata & $  42.31\pm   0.03$ & \nodata \\
spec00513 & $  45.75\pm   0.13$ & $  45.75\pm   0.13$ & $  45.42\pm   0.07$ & $  44.29\pm   0.88$ & $  43.31\pm   0.01$ & $  42.63\pm   0.04$ \\
spec00533 & \nodata & \nodata & $  44.95\pm   0.06$ & \nodata & $  42.69\pm   0.02$ & \nodata \\
spec00547 & \nodata & \nodata & $  43.85\pm   0.27$ & $  44.26\pm   4.03$ & $  42.63\pm   0.02$ & \nodata \\
spec00577 & \nodata & \nodata & $  44.38\pm   0.09$ & $  45.06\pm   1.05$ & $  42.42\pm   0.02$ & \nodata \\
spec00688 & \nodata & \nodata & $  45.00\pm   0.05$ & $  44.55\pm   0.89$ & $  43.11\pm   0.04$ & \nodata \\
spec01971 & \nodata & \nodata & $  43.85\pm   0.26$ & \nodata & $  42.70\pm   0.02$ & \nodata \\
spec02007 & \nodata & \nodata & $  45.13\pm   0.05$ & \nodata & $  43.39\pm   0.00$ & \nodata \\
spec02044 & \nodata & \nodata & $  43.84\pm   0.25$ & $  45.00\pm   1.05$ & $  42.52\pm   0.02$ & \nodata \\
spec02047 & $  44.99\pm   0.42$ & $  44.99\pm   0.42$ & $  44.80\pm   0.08$ & \nodata & $  43.06\pm   0.01$ & \nodata \\
spec02099 & \nodata & \nodata & $  43.74\pm   0.12$ & \nodata & $  42.61\pm   0.01$ & \nodata \\
spec02205 & \nodata & \nodata & $  45.35\pm   0.04$ & \nodata & $  43.83\pm   0.00$ & \nodata \\
spec01430 & \nodata & \nodata & $  46.27\pm   0.07$ & \nodata & $  42.47\pm   0.10$ & \nodata \\
spec01529 & \nodata & \nodata & $  44.31\pm   0.10$ & $  44.58\pm   0.56$ & $  42.79\pm   0.01$ & \nodata \\
spec01812 & $  45.90\pm   0.38$ & $  45.94\pm   0.04$ & $  45.52\pm   0.14$ & $  45.00\pm   0.09$ & $  43.71\pm   0.01$ & $  43.15\pm   0.02$ \\
spec00523 & \nodata & \nodata & $  44.77\pm   0.06$ & $  44.20\pm   1.83$ & $  42.50\pm   0.03$ & \nodata \\
spec02026 & $  44.48\pm   0.24$ & $  44.48\pm   0.24$ & $  44.64\pm   0.12$ & $  44.74\pm   0.48$ & $  43.18\pm   0.08$ & \nodata \\
spec02104 & $  44.91\pm   0.34$ & $  45.03\pm   0.08$ & $  44.87\pm   0.23$ & $  44.87\pm   0.36$ & $  43.36\pm   0.03$ & \nodata \\
spec02142 & \nodata & \nodata & $  44.53\pm   0.08$ & $  44.90\pm   0.45$ & $  42.81\pm   0.01$ & \nodata \\
spec02209 & $  45.85\pm   0.05$ & $  45.85\pm   0.05$ & $  45.68\pm   0.07$ & $  45.60\pm   0.06$ & $  44.17\pm   0.00$ & $  43.66\pm   0.01$ \\
spec02230 & \nodata & \nodata & $  45.40\pm   0.05$ & $  44.37\pm   2.55$ & $  43.80\pm   0.00$ & \nodata \\
spec02251 & $  43.41\pm   6.64$ & $  44.39\pm   0.35$ & $  44.62\pm   0.69$ & $  44.80\pm   0.42$ & $  43.27\pm   0.01$ & \nodata \\
spec00646 & \nodata & \nodata & $  44.22\pm   0.14$ & \nodata & $  42.09\pm   0.43$ & \nodata \\
spec01501 & \nodata & \nodata & $  46.00\pm   0.13$ & \nodata & \nodata & \nodata \\
spec01547 & \nodata & \nodata & $  43.47\pm   0.37$ & \nodata & \nodata & \nodata \\
spec01555 & $  44.46\pm   0.89$ & $  44.32\pm   0.46$ & $  44.53\pm   0.55$ & \nodata & \nodata & \nodata \\
spec01557 & $  46.10\pm   0.27$ & $  46.65\pm   0.21$ & \nodata & \nodata & \nodata & \nodata \\
spec01562 & \nodata & \nodata & $  43.63\pm   0.17$ & \nodata & \nodata & \nodata \\
spec01581 & $  45.03\pm   0.09$ & $  45.03\pm   0.09$ & $  45.00\pm   0.09$ & $  44.60\pm   0.19$ & \nodata & $  42.62\pm   0.03$ \\
spec01597 & $  44.28\pm   0.40$ & $  44.28\pm   0.40$ & $  44.39\pm   0.31$ & \nodata & \nodata & \nodata \\
spec01599 & $  45.19\pm   0.16$ & $  45.19\pm   0.09$ & \nodata & \nodata & \nodata & \nodata \\
spec01637 & $  44.91\pm   0.13$ & $  44.91\pm   0.13$ & $  44.55\pm   0.10$ & \nodata & \nodata & \nodata \\
spec01647 & $  45.28\pm   0.12$ & $  45.28\pm   0.12$ & $  44.74\pm   0.12$ & $  44.68\pm   0.31$ & \nodata & $  42.57\pm   0.03$ \\
spec01651 & \nodata & \nodata & $  43.19\pm   0.27$ & \nodata & \nodata & \nodata \\
spec01652 & $  45.08\pm   0.25$ & $  45.13\pm   0.08$ & $  44.94\pm   0.20$ & \nodata & \nodata & \nodata \\
spec01670 & $  45.23\pm   0.42$ & $  45.24\pm   0.07$ & $  45.21\pm   0.22$ & \nodata & \nodata & \nodata \\
spec01678 & $  46.09\pm   0.08$ & $  45.95\pm   0.13$ & \nodata & \nodata & \nodata & \nodata \\
spec01680 & \nodata & \nodata & $  43.78\pm   0.12$ & \nodata & \nodata & \nodata \\
spec01713 & $  45.47\pm   0.72$ & $  45.56\pm   0.05$ & $  45.51\pm   0.21$ & \nodata & \nodata & \nodata \\
spec01716 & \nodata & \nodata & $  43.89\pm   0.14$ & \nodata & \nodata & \nodata \\
spec01717 & $  44.28\pm   0.47$ & $  44.57\pm   0.12$ & $  44.67\pm   0.35$ & \nodata & \nodata & \nodata \\
spec01723 & $  44.73\pm   0.13$ & $  44.73\pm   0.13$ & $  44.81\pm   0.10$ & \nodata & \nodata & \nodata \\
spec01729 & \nodata & \nodata & $  43.71\pm   0.11$ & \nodata & \nodata & \nodata \\
spec01731 & \nodata & \nodata & $  43.90\pm   0.12$ & \nodata & \nodata & \nodata \\
spec01745 & $  45.16\pm   0.16$ & $  45.16\pm   0.16$ & $  44.91\pm   0.04$ & $  44.29\pm   0.64$ & $  47.11\pm   0.43$ & \nodata \\
spec01752 & $  45.63\pm   0.05$ & $  45.63\pm   0.05$ & $  45.39\pm   0.04$ & \nodata & \nodata & \nodata \\
spec01754 & $  45.37\pm   0.18$ & $  45.47\pm   0.13$ & \nodata & \nodata & \nodata & \nodata \\
spec01788 & $  46.03\pm   0.08$ & $  45.99\pm   0.10$ & \nodata & \nodata & \nodata & \nodata \\
spec01805 & $  44.34\pm   0.47$ & $  44.27\pm   0.31$ & $  44.82\pm   0.69$ & \nodata & \nodata & \nodata \\
spec00409 & $  45.95\pm   0.06$ & $  45.84\pm   0.17$ & \nodata & \nodata & \nodata & \nodata \\
spec00424 & \nodata & \nodata & $  44.01\pm   0.07$ & \nodata & \nodata & \nodata \\
spec00427 & $  45.11\pm   0.39$ & $  45.03\pm   0.19$ & $  44.99\pm   1.38$ & \nodata & \nodata & \nodata \\
spec00467 & $  45.71\pm   0.16$ & $  45.70\pm   0.06$ & $  45.49\pm   0.19$ & \nodata & \nodata & \nodata \\
spec00521 & $  45.99\pm   0.11$ & $  46.07\pm   0.04$ & $  45.97\pm   0.18$ & $  45.06\pm   0.22$ & \nodata & $  43.03\pm   0.03$ \\
spec00528 & $  45.13\pm   0.13$ & $  45.13\pm   0.13$ & $  45.01\pm   0.12$ & \nodata & \nodata & \nodata \\
spec00540 & $  44.79\pm   0.24$ & $  45.03\pm   0.08$ & $  45.38\pm   0.14$ & \nodata & \nodata & \nodata \\
spec00571 & $  45.05\pm   0.19$ & $  45.10\pm   0.13$ & $  46.01\pm   0.35$ & \nodata & \nodata & \nodata \\
spec00584 & \nodata & \nodata & $  44.43\pm   0.12$ & \nodata & \nodata & \nodata \\
spec00588 & $  45.74\pm   0.08$ & $  45.88\pm   0.02$ & $  45.91\pm   0.23$ & \nodata & \nodata & \nodata \\
spec00591 & \nodata & \nodata & $  44.45\pm   0.21$ & \nodata & \nodata & \nodata \\
spec00598 & $  46.18\pm   0.06$ & $  46.16\pm   0.11$ & \nodata & \nodata & \nodata & \nodata \\
spec00600 & $  45.85\pm   0.06$ & $  45.84\pm   0.06$ & \nodata & \nodata & \nodata & \nodata \\
spec00601 & $  45.42\pm   0.07$ & $  45.51\pm   0.07$ & \nodata & \nodata & \nodata & \nodata \\
spec00602 & \nodata & \nodata & $  44.14\pm   0.07$ & \nodata & \nodata & \nodata \\
spec00658 & $  45.06\pm   0.19$ & $  45.10\pm   0.09$ & $  45.09\pm   0.21$ & \nodata & \nodata & \nodata \\
spec00674 & $  45.34\pm   0.23$ & $  45.34\pm   0.23$ & $  45.20\pm   0.05$ & $  44.60\pm   0.35$ & \nodata & $  43.03\pm   0.04$ \\
spec00679 & \nodata & \nodata & $  44.58\pm   0.04$ & \nodata & \nodata & \nodata \\
spec00716 & $  45.76\pm   0.06$ & $  45.77\pm   0.04$ & \nodata & \nodata & \nodata & \nodata \\
spec00724 & $  45.36\pm   0.24$ & $  45.41\pm   0.05$ & $  45.37\pm   0.42$ & \nodata & \nodata & \nodata \\
spec00732 & $  44.94\pm   0.25$ & $  45.10\pm   0.15$ & \nodata & \nodata & \nodata & \nodata \\
spec00739 & $  44.93\pm   0.36$ & $  44.93\pm   0.36$ & $  44.90\pm   0.53$ & \nodata & \nodata & \nodata \\
spec00745 & \nodata & \nodata & $  44.57\pm   0.09$ & \nodata & \nodata & \nodata \\
spec00752 & $  46.20\pm   0.15$ & $  45.97\pm   0.03$ & $  45.87\pm   0.07$ & \nodata & \nodata & \nodata \\
spec01980 & \nodata & $  44.46\pm   0.44$ & $  44.56\pm   1.23$ & \nodata & \nodata & \nodata \\
spec02016 & $  44.26\pm   1.71$ & $  44.83\pm   0.32$ & \nodata & $  44.89\pm   0.72$ & \nodata & $  43.09\pm   0.04$ \\
spec02019 & $  45.54\pm   0.08$ & $  45.40\pm   0.11$ & \nodata & \nodata & \nodata & \nodata \\
spec02037 & $  45.28\pm   0.24$ & $  45.37\pm   0.48$ & \nodata & \nodata & \nodata & \nodata \\
spec02080 & \nodata & \nodata & $  43.23\pm   0.58$ & \nodata & \nodata & \nodata \\
spec02091 & $  45.83\pm   0.12$ & $  45.51\pm   0.72$ & \nodata & \nodata & \nodata & \nodata \\
spec02138 & \nodata & \nodata & $  43.30\pm   0.17$ & \nodata & $  42.73\pm   0.43$ & \nodata \\
spec02153 & $  45.18\pm   0.15$ & $  45.18\pm   0.14$ & \nodata & \nodata & \nodata & \nodata \\
spec02170 & $  46.01\pm   0.07$ & $  45.87\pm   0.07$ & \nodata & \nodata & \nodata & \nodata \\
spec02171 & \nodata & \nodata & $  43.87\pm   0.08$ & \nodata & \nodata & \nodata \\
spec02192 & $  45.05\pm   0.21$ & $  45.11\pm   0.06$ & $  44.94\pm   0.44$ & \nodata & \nodata & \nodata \\
spec02247 & $  45.58\pm   0.10$ & $  45.40\pm   0.12$ & $  45.92\pm   0.43$ & \nodata & \nodata & \nodata \\
spec02277 & $  45.27\pm   0.13$ & $  45.32\pm   0.10$ & \nodata & \nodata & \nodata & \nodata \\
spec02185 & $  43.85\pm   2.13$ & $  43.85\pm   2.13$ & $  44.52\pm   0.09$ & $  43.69\pm   4.10$ & $  42.73\pm   0.03$ & \nodata \\
\enddata
\tablecomments{We provide the FMOS ID (Column 1, as in Table \ref{tab:sample}) and {the logarithmic monochromatic luminosities at 1350, 1800, 3000, 5100 and their uncertainties (Columns 2-5), as well as the integrated luminosities and uncertainties of the H$\alpha$ and H$\beta$ emission lines (Columns 6 and 7).}}
\end{deluxetable*}
\end{center}

\section{UV rest-frame broad emission line fitting results (AGES)}
\LongTables
\begin{deluxetable*}{c c c c c c c c c c}
\tabletypesize{\footnotesize}
\tablecolumns{10}
\tablewidth{0pt}
\tablecaption{Best-fit profile parameters for rest-frame UV emission lines. \label{tab:UV}}
\tablehead{\colhead{ID}	&	\multicolumn{3}{c}{CIII}	&	\multicolumn{3}{c}{CIV}	&	\multicolumn{3}{c}{MgII}	\\
 \colhead{ } & \colhead{$\sigma$} & \colhead{FWHM} & \colhead{Flag} & \colhead{$\sigma$} & \colhead{FWHM} & \colhead{Flag} & \colhead{$\sigma$} & \colhead{FWHM} & \colhead{Flag} \\
 \colhead{ } & \colhead{[km s$^{-1}$]} & \colhead{[km s$^{-1}$]} & \colhead{ } & \colhead{[km s$^{-1}$]} & \colhead{[km s$^{-1}$]} & \colhead{ } & \colhead{[km s$^{-1}$]} & \colhead{[km s$^{-1}$]} & \colhead{ }}
\startdata
spec01519 & \nodata & \nodata & F & \nodata & \nodata & F & $     3.41\pm     0.12$ & $     3.61\pm     0.09$ & A \\
spec01530 & \nodata & \nodata & F & \nodata & \nodata & F & $     3.14\pm     0.25$ & $     3.52\pm     0.17$ & B \\
spec01634 & \nodata & \nodata & F & \nodata & \nodata & F & $     3.19\pm     0.06$ & $     3.40\pm     0.06$ & B \\
spec00513 & $     3.33\pm     0.08$ & $     3.54\pm     0.15$ & A & \nodata & \nodata & F & $     3.18\pm     0.03$ & $     3.46\pm     0.02$ & A \\
spec00533 & \nodata & \nodata & F & \nodata & \nodata & F & $     3.14\pm     0.02$ & $     3.48\pm     0.01$ & A \\
spec00577 & \nodata & \nodata & F & \nodata & \nodata & F & $     3.23\pm     0.02$ & $     3.61\pm     0.02$ & B \\
spec00688 & $     3.48\pm     0.25$ & $     3.64\pm     0.77$ & B & \nodata & \nodata & F & $     3.40\pm     0.03$ & $     3.82\pm     0.02$ & A \\
spec01971 & \nodata & \nodata & F & \nodata & \nodata & F & $     3.25\pm     0.15$ & $     3.56\pm     0.10$ & A \\
spec02007 & \nodata & \nodata & F & \nodata & \nodata & F & $     3.28\pm     0.20$ & $     3.56\pm     0.14$ & A \\
spec02044 & \nodata & \nodata & F & \nodata & \nodata & F & $     3.15\pm     0.11$ & $     3.37\pm     0.16$ & C \\
spec02047 & \nodata & \nodata & F & \nodata & \nodata & F & $     3.20\pm     0.02$ & $     3.49\pm     0.03$ & A \\
spec02205 & \nodata & \nodata & F & \nodata & \nodata & F & $     3.17\pm     0.02$ & $     3.46\pm     0.02$ & A \\
spec01430 & \nodata & \nodata & F & \nodata & \nodata & F & $     3.29\pm     0.03$ & $     3.60\pm     0.01$ & A \\
spec01529 & \nodata & \nodata & F & \nodata & \nodata & F & $     3.29\pm     0.01$ & $     3.59\pm     0.01$ & A \\
spec01812 & $     3.49\pm     0.40$ & $     3.60\pm     1.27$ & A & $     3.43\pm     0.09$ & $     3.57\pm     0.17$ & A & $     3.25\pm     0.02$ & $     3.55\pm     0.02$ & A \\
spec00523 & \nodata & \nodata & F & \nodata & \nodata & F & $     3.23\pm     0.02$ & $     3.51\pm     0.01$ & A \\
spec02026 & $     3.47\pm     0.63$ & $     3.84\pm     1.03$ & B & \nodata & \nodata & F & $     3.31\pm     0.09$ & $     3.69\pm     0.07$ & A \\
spec02104 & $     3.48\pm     0.04$ & $     3.34\pm     0.42$ & B & $     3.53\pm     0.39$ & $     3.36\pm     0.09$ & B & $     3.73\pm     0.06$ & $     3.90\pm     0.18$ & C \\
spec02142 & \nodata & \nodata & F & \nodata & \nodata & F & $     2.91\pm     0.02$ & $     3.27\pm     0.02$ & A \\
spec02209 & $     3.44\pm     0.03$ & $     3.30\pm     0.04$ & A & $     3.19\pm     0.56$ & $     3.42\pm     0.29$ & B & $     3.24\pm     0.01$ & $     3.46\pm     0.01$ & B \\
spec02230 & \nodata & \nodata & F & \nodata & \nodata & F & $     3.41\pm     0.01$ & $     3.68\pm     0.02$ & A \\
spec02251 & $     3.50\pm     0.05$ & $     3.87\pm     0.04$ & C & \nodata & \nodata & F & \nodata & \nodata & F \\
spec00646 & \nodata & \nodata & F & \nodata & \nodata & F & $     2.93\pm     0.02$ & $     3.30\pm     0.03$ & B \\
spec01501 & \nodata & \nodata & F & \nodata & \nodata & F & $     3.16\pm     0.12$ & $     3.40\pm     0.05$ & A \\
spec01547 & \nodata & \nodata & F & \nodata & \nodata & F & $     3.01\pm     0.05$ & $     3.38\pm     0.05$ & C \\
spec01555 & $     3.59\pm     0.10$ & $     3.71\pm     0.22$ & C & \nodata & \nodata & F & $     3.72\pm     0.23$ & $     4.09\pm     0.27$ & C \\
spec01557 & $     3.36\pm     0.61$ & $     3.45\pm     0.65$ & B & $     3.55\pm     0.07$ & $     3.27\pm     0.03$ & A & \nodata & \nodata & F \\
spec01581 & $     3.60\pm     0.06$ & $     3.59\pm     0.21$ & B & \nodata & \nodata & F & $     3.32\pm     0.48$ & $     3.70\pm     0.15$ & A \\
spec01597 & $     3.63\pm     0.12$ & $     3.26\pm     0.84$ & C & \nodata & \nodata & F & $     3.20\pm     0.41$ & $     3.71\pm     0.17$ & B \\
spec01599 & $     3.53\pm     0.27$ & $     3.54\pm     0.11$ & B & $     3.48\pm     0.07$ & $     3.52\pm     0.15$ & B & \nodata & \nodata & F \\
spec01637 & $     3.74\pm     0.38$ & $     3.47\pm     2.05$ & B & \nodata & \nodata & F & $     3.14\pm     0.02$ & $     3.51\pm     0.02$ & A \\
spec01647 & $     3.50\pm     0.05$ & $     3.54\pm     0.13$ & A & \nodata & \nodata & F & $     3.41\pm     0.05$ & $     3.61\pm     0.06$ & A \\
spec01652 & $     3.61\pm     0.32$ & $     3.67\pm     1.20$ & A & $     3.54\pm     0.28$ & $     3.67\pm     0.06$ & A & $     3.31\pm     0.09$ & $     3.61\pm     0.03$ & A \\
spec01670 & $     3.42\pm     0.08$ & $     3.79\pm     0.09$ & B & $     3.60\pm     0.08$ & $     3.68\pm     0.06$ & A & $     3.64\pm     0.11$ & $     4.01\pm     0.11$ & A \\
spec01678 & $     3.28\pm     0.46$ & $     3.52\pm     0.18$ & C & $     3.40\pm     0.09$ & $     3.17\pm     0.02$ & B & \nodata & \nodata & F \\
spec01680 & \nodata & \nodata & F & \nodata & \nodata & F & $     3.68\pm     0.10$ & $     4.05\pm     0.10$ & C \\
spec01713 & $     3.62\pm     0.06$ & $     3.57\pm     0.17$ & A & $     3.30\pm     0.16$ & $     3.67\pm     0.21$ & A & $     3.27\pm     0.04$ & $     3.51\pm     0.05$ & B \\
spec01716 & \nodata & \nodata & F & \nodata & \nodata & F & $     3.31\pm     0.21$ & $     3.76\pm     0.09$ & A \\
spec01717 & $     3.64\pm     0.25$ & $     3.57\pm     1.20$ & B & \nodata & \nodata & F & $     3.14\pm     0.09$ & $     3.51\pm     0.09$ & A \\
spec01723 & $     3.75\pm     0.14$ & $     3.75\pm     0.13$ & B & \nodata & \nodata & F & $     3.42\pm     0.05$ & $     3.70\pm     0.10$ & B \\
spec01731 & \nodata & \nodata & F & \nodata & \nodata & F & $     3.27\pm     0.06$ & $     3.64\pm     0.06$ & B \\
spec01745 & $     3.31\pm     0.62$ & $     3.11\pm     0.48$ & B & \nodata & \nodata & F & $     3.52\pm     0.18$ & $     3.91\pm     0.07$ & A \\
spec01752 & $     3.62\pm     0.07$ & $     3.66\pm     0.09$ & A & \nodata & \nodata & F & $     3.19\pm     0.03$ & $     3.52\pm     0.02$ & A \\
spec01754 & $     3.71\pm     0.21$ & $     4.08\pm     0.44$ & C & \nodata & \nodata & F & \nodata & \nodata & F \\
spec01788 & $     3.46\pm     0.23$ & $     3.43\pm     0.12$ & A & $     3.36\pm     0.07$ & $     3.52\pm     0.02$ & A & \nodata & \nodata & F \\
spec01805 & $     3.82\pm     0.16$ & $     3.43\pm     0.05$ & C & $     2.64\pm     0.07$ & $     2.92\pm     0.06$ & A & \nodata & \nodata & F \\
spec00409 & \nodata & \nodata & F & $     3.58\pm     0.26$ & $     3.58\pm     1.16$ & A & \nodata & \nodata & F \\
spec00424 & \nodata & \nodata & F & \nodata & \nodata & F & $     3.21\pm     0.14$ & $     3.41\pm     0.07$ & A \\
spec00427 & $     3.03\pm     1.74$ & $     3.41\pm     0.15$ & C & $     3.63\pm     0.11$ & $     3.47\pm     0.07$ & B & $     3.33\pm     0.37$ & $     3.70\pm     0.37$ & C \\
spec00467 & $     3.68\pm     0.21$ & $     3.77\pm     0.76$ & A & $     3.41\pm     0.04$ & $     3.78\pm     0.03$ & A & $     3.23\pm     0.06$ & $     3.56\pm     0.05$ & A \\
spec00521 & $     4.18\pm     0.06$ & $     3.73\pm     0.03$ & B & $     3.49\pm     0.08$ & $     3.66\pm     0.05$ & A & $     3.37\pm     0.06$ & $     3.61\pm     0.03$ & B \\
spec00528 & $     3.69\pm     0.26$ & $     4.01\pm     0.53$ & C & $     3.46\pm     0.38$ & $     3.75\pm     0.12$ & B & $     3.51\pm     0.08$ & $     3.80\pm     0.05$ & A \\
spec00540 & $     3.62\pm     0.05$ & $     3.93\pm     0.05$ & B & $     3.11\pm     1.06$ & $     3.47\pm     0.03$ & A & $     2.99\pm     0.48$ & $     3.36\pm     0.48$ & C \\
spec00571 & $     3.92\pm     0.03$ & $     3.88\pm     0.10$ & B & $     3.54\pm     0.12$ & $     3.54\pm     0.14$ & A & \nodata & \nodata & F \\
spec00584 & $     3.38\pm     0.48$ & $     3.18\pm     0.72$ & C & \nodata & \nodata & F & $     3.40\pm     0.07$ & $     3.81\pm     0.03$ & B \\
spec00588 & $     4.07\pm     0.07$ & $     3.64\pm     0.08$ & B & $     3.59\pm     0.38$ & $     3.79\pm     0.27$ & A & $     3.17\pm     0.02$ & $     3.55\pm     0.02$ & C \\
spec00591 & $     3.39\pm     0.38$ & $     3.65\pm     0.12$ & C & \nodata & \nodata & F & $     3.21\pm     0.11$ & $     3.66\pm     0.03$ & B \\
spec00598 & $     3.47\pm     0.17$ & $     3.84\pm     0.32$ & B & $     3.23\pm     0.26$ & $     3.40\pm     0.10$ & C & \nodata & \nodata & F \\
spec00600 & $     3.41\pm     0.12$ & $     3.78\pm     0.07$ & B & $     3.35\pm     0.03$ & $     3.59\pm     0.03$ & A & \nodata & \nodata & F \\
spec00601 & $     3.61\pm     0.30$ & $     3.98\pm     0.49$ & C & \nodata & \nodata & F & \nodata & \nodata & F \\
spec00602 & \nodata & \nodata & F & \nodata & \nodata & F & $     3.83\pm     0.16$ & $     3.75\pm     0.14$ & C \\
spec00658 & $     3.46\pm     0.04$ & $     3.50\pm     0.23$ & B & $     3.50\pm     0.29$ & $     3.69\pm     0.04$ & A & $     3.11\pm     0.02$ & $     3.48\pm     0.02$ & B \\
spec00674 & $     3.43\pm     0.10$ & $     3.43\pm     0.35$ & C & \nodata & \nodata & F & $     3.24\pm     0.03$ & $     3.53\pm     0.02$ & A \\
spec00679 & \nodata & \nodata & F & \nodata & \nodata & F & $     3.36\pm     0.05$ & $     3.66\pm     0.05$ & A \\
spec00716 & $     3.67\pm     0.17$ & $     3.40\pm     0.25$ & B & $     3.13\pm     0.42$ & $     3.50\pm     0.07$ & B & \nodata & \nodata & F \\
spec00724 & $     3.59\pm     0.25$ & $     3.54\pm     0.17$ & B & $     3.54\pm     0.31$ & $     3.64\pm     0.08$ & A & $     3.39\pm     0.16$ & $     3.58\pm     0.08$ & A \\
spec00732 & $     3.22\pm     0.08$ & $     3.30\pm     0.17$ & B & \nodata & \nodata & F & \nodata & \nodata & F \\
spec00739 & $     3.66\pm     0.24$ & $     3.84\pm     0.67$ & C & \nodata & \nodata & F & $     3.65\pm     0.07$ & $     4.03\pm     0.07$ & C \\
spec00745 & \nodata & \nodata & F & \nodata & \nodata & F & $     3.25\pm     0.03$ & $     3.64\pm     0.03$ & A \\
spec00752 & $     3.48\pm     0.40$ & $     3.57\pm     1.35$ & A & $     3.47\pm     0.39$ & $     3.62\pm     0.38$ & A & $     3.18\pm     0.03$ & $     3.49\pm     0.03$ & A \\
spec01980 & $     3.55\pm     0.08$ & $     3.92\pm     0.14$ & C & $     3.79\pm     0.06$ & $     3.91\pm     0.24$ & B & \nodata & \nodata & F \\
spec02016 & $     3.48\pm     0.43$ & $     3.86\pm     0.61$ & C & \nodata & \nodata & F & \nodata & \nodata & F \\
spec02019 & $     3.69\pm     0.15$ & $     3.29\pm     0.48$ & B & $     3.85\pm     0.15$ & $     3.35\pm     0.15$ & A & \nodata & \nodata & F \\
spec02037 & \nodata & \nodata & F & $     3.29\pm     0.08$ & $     3.57\pm     0.09$ & B & \nodata & \nodata & F \\
spec02091 & \nodata & \nodata & F & $     3.55\pm     0.32$ & $     3.54\pm     0.23$ & A & \nodata & \nodata & F \\
spec02153 & $     3.31\pm     0.08$ & $     3.68\pm     0.02$ & A & $     3.17\pm     0.10$ & $     3.35\pm     0.07$ & A & \nodata & \nodata & F \\
spec02170 & $     3.61\pm     0.04$ & $     3.80\pm     0.08$ & A & $     3.66\pm     0.20$ & $     3.75\pm     0.10$ & A & \nodata & \nodata & F \\
spec02171 & \nodata & \nodata & F & \nodata & \nodata & F & $     3.15\pm     0.31$ & $     3.43\pm     0.07$ & A \\
spec02192 & $     3.75\pm     0.15$ & $     3.65\pm     0.93$ & B & $     3.69\pm     0.29$ & $     4.02\pm     0.18$ & A & $     3.40\pm     0.02$ & $     3.77\pm     0.02$ & C \\
spec02247 & $     3.64\pm     0.17$ & $     3.84\pm     0.12$ & B & $     3.48\pm     0.62$ & $     3.63\pm     1.97$ & A & \nodata & \nodata & F \\
spec02277 & \nodata & \nodata & F & $     3.79\pm     0.18$ & $     3.93\pm     0.33$ & B & \nodata & \nodata & F \\
spec02185 & $     3.76\pm     0.17$ & $     3.53\pm     0.52$ & B & \nodata & \nodata & F & $     3.44\pm     0.07$ & $     3.62\pm     0.03$ & A \\
\enddata
\tablecomments{We provide the FMOS ID (Column 1, as in Table \ref{tab:sample}), the second moment and FWHM and their uncertainties in logarithmic scale, and the visual flag for CIII], CIV, and MgII (Columns 2-4, 5-7, 8-10, respectively).}
\end{deluxetable*}

\section{Optical rest-frame broad emission line fitting results (FMOS)}

\begin{center}
\begin{deluxetable*}{c c c c  c c c }
\tablecolumns{7}
\tabletypesize{\footnotesize}
\tablewidth{0pt}
\tablecaption{Best-fit profile parameters for the H$\alpha$ and H$\beta$ lines. \label{tab:balmer}}
\tablehead{\multirow{2}{*}{ID}	&	\multicolumn{3}{c}{H$\beta$}	&	\multicolumn{3}{c}{H$\alpha$} \\
\colhead{} & \colhead{$\sigma$} & \colhead{FWHM} & \colhead{Flag} & \colhead{$\sigma$} & \colhead{FWHM} & \colhead{Flag} \\
\colhead{} & \colhead{[km s$^{-1}$]} & \colhead{[km s$^{-1}$]} & \colhead{}  & \colhead{[km s$^{-1}$]} & \colhead{[km s$^{-1}$]} & \colhead{} }
\startdata
spec01519 & \nodata & \nodata & F & $     3.25\pm     0.02$ & $     2.64\pm     1.88$ & A   \\
spec01530 & \nodata & \nodata & F & $     3.08\pm     0.04$ & $     3.34\pm     0.04$ & B   \\
spec01634 & \nodata & \nodata & F & $     3.13\pm     0.05$ & $     3.49\pm     0.04$ & A   \\
spec00513 & $     2.97\pm     0.06$ & $     3.33\pm     0.06$ & B & $     3.34\pm     0.02$ & $     3.46\pm     0.02$ & A   \\
spec00533 & \nodata & \nodata & F & $     3.16\pm     0.03$ & $     3.49\pm     0.02$ & A   \\
spec00547 & \nodata & \nodata & F & $     3.21\pm     0.02$ & $     3.41\pm     0.02$ & A   \\
spec00577 & \nodata & \nodata & F & $     3.09\pm     0.04$ & $     3.51\pm     0.02$ & A   \\
spec00688 & \nodata & \nodata & F & $     3.41\pm     0.03$ & $     3.12\pm     0.48$ & A   \\
spec01971 & \nodata & \nodata & F & $     3.20\pm     0.04$ & $     3.41\pm     0.02$ & A   \\
spec02007 & \nodata & \nodata & F & $     3.46\pm     0.00$ & $     3.70\pm     0.01$ & A   \\
spec02044 & \nodata & \nodata & F & $     3.08\pm     0.04$ & $     3.20\pm     0.03$ & A   \\
spec02047 & \nodata & \nodata & F & $     3.19\pm     0.02$ & $     3.52\pm     0.01$ & A   \\
spec02099 & \nodata & \nodata & F & $     2.78\pm     0.02$ & $     3.21\pm     0.01$ & A   \\
spec02205 & \nodata & \nodata & F & $     3.34\pm     0.01$ & $     3.45\pm     0.00$ & A   \\
spec01430 & \nodata & \nodata & F & $     3.36\pm     0.26$ & $     3.60\pm     0.07$ & B   \\
spec01529 & \nodata & \nodata & F & $     3.13\pm     0.02$ & $     3.45\pm     0.04$ & B   \\
spec01812 & $     3.15\pm     0.02$ & $     3.53\pm     0.02$ & B & $     3.25\pm     0.01$ & $     3.41\pm     0.01$ & B   \\
spec00523 & \nodata & \nodata & F & $     3.28\pm     0.08$ & $     3.61\pm     0.03$ & B   \\
spec02026 & \nodata & \nodata & F & $     3.27\pm     0.25$ & $     3.40\pm     0.20$ & B   \\
spec02104 & \nodata & \nodata & F & $     3.43\pm     0.07$ & $     3.75\pm     0.06$ & B   \\
spec02142 & \nodata & \nodata & F & $     3.05\pm     0.02$ & $     3.17\pm     0.02$ & B   \\
spec02209 & $     3.18\pm     0.01$ & $     3.56\pm     0.01$ & B & $     3.25\pm     0.00$ & $     3.45\pm     0.00$ & B   \\
spec02230 & \nodata & \nodata & F & $     3.49\pm     0.00$ & $     3.70\pm     0.00$ & B   \\
spec02251 & \nodata & \nodata & F & $     2.99\pm     0.02$ & $     3.17\pm     0.12$ & B   \\
spec00646 & \nodata & \nodata & F & $     2.83\pm  -999.00$ & $     2.97\pm  -999.00$ & C   \\
spec01581 & $     2.75\pm     0.05$ & $     3.14\pm     0.05$ & B & \nodata & \nodata & F   \\
spec01647 & $     2.74\pm     0.04$ & $     3.08\pm     0.05$ & C & \nodata & \nodata & F   \\
spec00521 & $     3.10\pm     0.02$ & $     3.47\pm     0.03$ & B & \nodata & \nodata & F   \\
spec00674 & $     2.93\pm     0.02$ & $     3.30\pm     0.02$ & C & \nodata & \nodata & F   \\
spec02016 & $     3.12\pm     0.05$ & $     3.50\pm     0.05$ & C & \nodata & \nodata & F   \\
spec02185 & \nodata & \nodata & F & $     2.83\pm     0.06$ & $     3.06\pm     0.15$ & dbl \\
\enddata
\tablecomments{Same as in Table \ref{tab:UV} but for the H$\beta$ (Columns 2-4) and H$\alpha$ (Columns (5-7) emission lines. Source spec02185 shows a double-peaked H$\alpha$ profile and is thus conservatively not included in our analysis.}
\end{deluxetable*}
\end{center}

\end{appendix}
\end{document}